\documentclass[apjl]{emulateapj}
\usepackage{amsfonts,amsmath,graphicx,natbib,apjfonts,subfigure,textcomp}

\def\ra#1#2#3{#1$^{\rm h}$#2$^{\rm m}$#3$^{\rm s}$}
\def\dec#1#2#3{#1$^\circ$#2$'$#3$''$}

\def\swift{{\it Swift}}

\def\cfa{1}
\def\princeton{2}
\def\ferrara{3}
\def\mullard{4}
\def\albanova{5}
\def\casa{6}
\def\india{7}
\def\nyu{8}
\def\texasagain{9}
\def\texas{10}
\def\naval{11}
\def\seoul{12}
\def\michigan{13}
\def\socorro{14}
\def\russia{15}
\def\australia{16}
\def\dartmouth{17}
\def\mit{18}
\def\dark{19}
\def\nasa{20}
\def\chile{21}
\def\sweden{22}
\def\illinois{23}
\def\chileagain{24}
\def\az{25}
\def\stanford{26}
\def\baltimore{27}
\def\denmark{28}
\def\hungary{29}
\def\azagain{30}
\def\brera{31}
\def\virginia{32}
\def\texasagainagain{33}
\def\pennstate{34}

\shorttitle{SN2009ip}
\shortauthors{Margutti et al.}

\begin{document}
\title{A panchromatic view of the restless SN\,2009ip reveals the explosive ejection of a
massive star envelope} 
\author{R.~Margutti\altaffilmark{\cfa}, D. Milisavljevic\altaffilmark{\cfa}, A.~M. Soderberg\altaffilmark{\cfa}, R. Chornock\altaffilmark{\cfa}, 
B.~A. Zauderer\altaffilmark{\cfa}, K. Murase\altaffilmark{\princeton}, C. Guidorzi\altaffilmark{\ferrara}, 
N.~E. Sanders\altaffilmark{\cfa}, P. Kuin\altaffilmark{\mullard},
C. Fransson\altaffilmark{\albanova}, E.~M. Levesque\altaffilmark{\casa}, P. Chandra\altaffilmark{\india}, 
E. Berger\altaffilmark{\cfa}, F.~B.  Bianco\altaffilmark{\nyu}, P. J. Brown\altaffilmark{\texasagain}, P. Challis\altaffilmark{\india}, E. Chatzopoulos\altaffilmark{\texas},
C.~C. Cheung\altaffilmark{\naval}, C. Choi\altaffilmark{\seoul}, L. Chomiuk\altaffilmark{\michigan,\socorro}, N. Chugai\altaffilmark{\russia}, C. Contreras\altaffilmark{\australia},
M.~R. Drout\altaffilmark{\cfa}, R. Fesen\altaffilmark{\dartmouth}, R.~J. Foley\altaffilmark{\cfa}, W. Fong\altaffilmark{\cfa}, 
A.~S. Friedman\altaffilmark{\cfa,\mit}, C. Gall\altaffilmark{\dark,\nasa}, 
N. Gehrels\altaffilmark{\nasa}, J. Hjorth\altaffilmark{\dark},  E. Hsiao\altaffilmark{\chile}, R. Kirshner\altaffilmark{\cfa}, M. Im\altaffilmark{\seoul}, G. Leloudas\altaffilmark{\sweden,\dark}, R. Lunnan\altaffilmark{\cfa}, G.~H. Marion\altaffilmark{\cfa}, J. Martin\altaffilmark{\illinois},
 N. Morrell\altaffilmark{\chileagain}, K. ~F. Neugent\altaffilmark{\az}, N. Omodei\altaffilmark{\stanford},  M.~M. Phillips\altaffilmark{\chileagain}, 
A. Rest\altaffilmark{\baltimore}, J.~M. Silverman\altaffilmark{\texas},  J. Strader\altaffilmark{\michigan}, 
M.~D. Stritzinger\altaffilmark{\denmark}, T. Szalai\altaffilmark{\hungary}, N.~B. Utterback\altaffilmark{\dartmouth}, 
J. Vinko\altaffilmark{\hungary,\texas}, J.~C. Wheeler\altaffilmark{\texas}, 
D. Arnett\altaffilmark{\azagain},  S. Campana\altaffilmark{\brera}, R. Chevalier\altaffilmark{\virginia},  A. Ginsburg\altaffilmark{\casa}, A. Kamble\altaffilmark{\cfa}, 
P.~W. ~A.  Roming\altaffilmark{\texasagainagain,\pennstate}, T. Pritchard\altaffilmark{\pennstate}, G. Stringfellow\altaffilmark{\casa}}

\altaffiltext{\cfa}{Harvard-Smithsonian Center for Astrophysics, 60 Garden St., Cambridge, MA 02138, USA.}
\altaffiltext{\princeton}{Institute for Advanced Study, Princeton, New Jersey 08540, USA.}
\altaffiltext{\ferrara}{Department of Physics, University of Ferrara, via Saragat 1, I-44122 Ferrara, Italy.}
\altaffiltext{\mullard}{University College London, MSSL, Holmbury St. Mary, Dorking, Surrey RH5 6NT, UK.}
\altaffiltext{\albanova}{Department of Astronomy and the Oskar Klein Centre, Stockholm University, AlbaNova, SE-106 91 Stockholm, Sweden.}
\altaffiltext{\casa}{CASA, Department of Astrophysical and Planetary Sciences, University of Colorado, 389-UCB, Boulder, CO 80309, USA.}
\altaffiltext{\india}{National Centre for Radio Astrophysics, Tata Institute of Fundamental Research, Pune University Campus, Ganeshkhind, Pune 411007, India.}
\altaffiltext{\nyu}{Center for Cosmology and Particle Physics, New York University, 4 Washington Place, New York, NY 10003.}
\altaffiltext{\texasagain}{George P. and Cynthia Woods Mitchell Institute for Fundamental Physics \& Astronomy, Texas A. \& M. University, Department of Physics and Astronomy,
4242 TAMU, College Station, TX 77843, USA.}
\altaffiltext{\texas}{Department of Astronomy, University of Texas at Austin, Austin, TX 78712-1205, USA.}
\altaffiltext{\naval}{Space Science Division, Naval Research Laboratory, Washington, DC 20375-5352, USA.}
\altaffiltext{\seoul}{CEOU/Department of Physics and Astronomy, Seoul National University, Seoul 151-742, Republic of Korea.}
\altaffiltext{\michigan}{Department of Physics and Astronomy, Michigan State University, East Lansing, MI 48824, USA.}
\altaffiltext{\socorro}{National Radio Astronomy Observatory, P.O. Box O, Socorro, NM 87801.}
\altaffiltext{\russia}{Institute of Astronomy, Russian Academy of Sciences, Pyatnitskaya 48, 119017, Moscow, Russian Federation.}
\altaffiltext{\australia}{Centre for Astrophysics \& Supercomputing, Swinburne University of Technology, PO Box 218, Hawthorn, VIC 3122, Australia.}

\altaffiltext{\dartmouth}{Department of Physics \& Astronomy, Dartmouth College, 6127 Wilder Lab, Hanover, NH 03755, USA.}
\altaffiltext{\mit}{Massachusetts Institute of Technology, 77 Massachusetts Ave., Bldg. E51-173, Cambridge, MA 02138, USA.}
\altaffiltext{\dark}{Dark Cosmology Centre, Niels Bohr Institute, University of Copenhagen, Juliane Maries Vej 30, DK-2100 Copenhagen, Denmark.}
\altaffiltext{\nasa}{NASA, Goddard Space Flight Center, 8800 Greenbelt Road, Greenbelt, MD 20771, USA.}
\altaffiltext{\chile}{Carnegie Observatories, Las Campanas Observatory, Colina El Pino, Casilla 601, Chile.}
\altaffiltext{\sweden}{The Oskar Klein Centre, Department of Physics, Stockholm University, SE-10691, Stockholm, Sweden.}
\altaffiltext{\illinois}{Astronomy/Physics MS HSB 314, One University Plaza Springfield, IL 62730, USA.}
\altaffiltext{\chileagain}{Carnegie Observatories, Las Campanas Observatory, Casilla 601, La Serena, Chile.}
\altaffiltext{\az}{Lowell Observatory, 1400 W Mars Hill Road, Flagstaff, AZ 86001, USA.}
\altaffiltext{\stanford}{W. W. Hansen Experimental Physics Laboratory, Kavli Institute for Particle Astrophysics and Cosmology, Department of Physics and SLAC National Accelerator Laboratory, Stanford University, Stanford, CA 94305, USA.}
\altaffiltext{\baltimore}{Space Telescope Science Institute, 3700 San Martin Dr., Baltimore, MD 21218, USA.}
\altaffiltext{\denmark}{Department of Physics and Astronomy, Aarhus University, Ny Munkegade, DK-8000 Aarhus C, Denmark.}
\altaffiltext{\hungary}{Department of Optics and Quantum Electronics, University of Szeged, D\'om t\'er 9., Szeged H-6720, Hungary.}
\altaffiltext{\azagain}{Department of Astronomy and Steward Observatory, University of Arizona, Tucson, AZ 85721, USA.}

\altaffiltext{\brera}{INAF/Brera Astronomical Observatory, via Bianchi 46, 23807, Merate (LC), Italy.}
\altaffiltext{\virginia}{Department of Astronomy, University of Virginia, P.O. Box 400325, Charlottesville, VA 22904-4325, USA.}
\altaffiltext{\texasagainagain}{Southwest Research Institute, Department of Space Science, 6220 Culebra Road, San Antonio, TX 78238, USA.}
\altaffiltext{\pennstate}{Department of Astronomy \& Astrophysics, Penn State University, 525 Davey Lab, University Park, PA 16802, USA.}

\begin{abstract}
The 2012 explosion of SN\,2009ip raises questions about our understanding of the late stages 
of massive star evolution. Here we present a comprehensive study of SN\,2009ip during its 
remarkable re-brightening(s). High-cadence photometric and spectroscopic observations from the GeV 
to the radio band obtained from a variety of ground-based and space facilities 
(including the VLA, Swift, Fermi, HST and XMM) constrain SN\,2009ip to be a low energy 
($E\sim 10^{50}$ erg for an ejecta mass $\sim 0.5 \,\rm{M_{\sun}}$) and likely asymmetric explosion 
in a complex medium shaped by multiple eruptions of the restless progenitor star. 
Most of the energy is radiated as a result of the shock breaking out through a dense shell of material 
located at $\sim5\times10^{14}\,\rm{cm}$ with $M\sim0.1\,\rm{M_{\sun}}$, ejected by the precursor 
outburst $\sim40$ days before the major explosion. 
We interpret the NIR excess of emission as signature of dust vaporization of material located further 
out ($R>4\times 10^{15}\,\rm{cm}$), the origin of which has to be connected with documented mass loss 
episodes in the previous years. Our modeling predicts bright neutrino emission associated with
the shock break-out if the cosmic ray energy is comparable to the radiated energy. 
We connect this phenomenology with the explosive ejection of the outer layers 
of the massive progenitor star, that later interacted with material deposited in the surroundings by 
previous eruptions. Future observations will reveal if the luminous blue variable (LBV) progenitor star survived. 
Irrespective of whether the explosion was terminal, SN\,2009ip brought to light the 
existence of new channels for sustained episodic mass-loss, the physical origin of which has yet to be
identified. 
\end{abstract}

\section{Introduction}
\label{Sec:Intro}

\begin{figure}
\vskip -0.0 true cm
\centering
\includegraphics[scale=0.65]{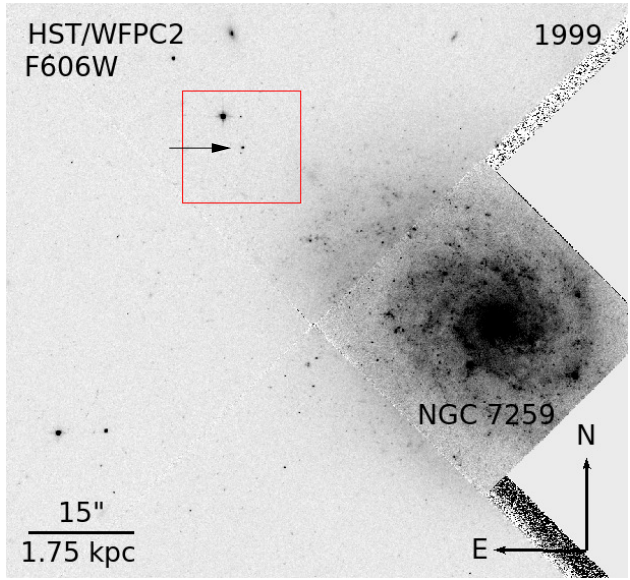}
\caption{Hubble Space Telescope (HST) pre-explosion image acquired in 1999 
(Program 6359; PI: Stiavelli). The location 
of the progenitor of SN\,2009ip is marked by an arrow. SN\,2009ip exploded in the outskirts
of its host galaxy NGC\,7259 at an angular distance of $\sim43.4''$ from the host center,
corresponding to $\sim 5\,\rm{kpc}$. }
\label{Fig:SN2009ip}
\end{figure}

Standard stellar evolutionary models predict massive stars with $M\gtrsim 40\, \rm{M_{\odot}}$ 
to spend half a Myr in the Wolf-Rayet (WR) phase before exploding
as supernovae (SNe, e.g. \citealt{Georgy12} and references therein).
As a result, massive stars are \emph{not} expected to be H rich at the 
time of explosion. Yet, recent observations have
questioned this picture, revealing the limitations of our
current understanding of the last stages of massive star evolution
and in particular the uncertainties in the commonly assumed 
mass loss prescriptions (e.g. \citealt{Smith06}). Here, we present observations from
an extensive, broad-band monitoring campaign
of SN\,2009ip (Fig. \ref{Fig:SN2009ip}) during its double explosion in 2012 that revealed extreme 
mass-loss properties, raising questions about our understanding of the late stages 
of massive star evolution.

An increasingly complex picture is emerging connecting SN progenitor stars with explosion
properties. The most direct link arguably comes from the detection of progenitor stars
in pre-explosion images. These efforts have been successful connecting 
Type IIP SNe with the death of red supergiants ($M\sim 8-15\,\rm{M_{\sun}}$, \citealt{Smartt09}).
However,  massive progenitor stars have proven
to be more elusive (e.g. \citealt{Kochanek08}): SN\,2005gl constitutes the first direct evidence for a massive 
($M>50\,M_{\sun}$) and H rich star to explode as a core-collapse SN, contrary to 
theoretical expectations (\citealt{GalYam07}; \citealt{GalYam09}).

SN\,2005gl belongs to the class of Type IIn SNe (\citealt{Schlegel90}). Their
spectra  show evidence for strong interaction between the explosion ejecta and
a dense circumstellar medium (CSM) previously enriched by  mass loss from the 
progenitor star.  In order for the SN to appear as a Type IIn explosion, the mass loss and the
core collapse have to be \emph{timed}, with mass loss occurring in the
decades to years before the collapse. This timing requirement constitutes a further challenge to current 
evolutionary models and emphasizes the importance of the progenitor mass loss 
in the years before the explosion in determining its observable properties.

Mass loss in massive stars can either occur through steady winds 
(on a typical time scale of $10^3$ yr) or episodic outbursts lasting months to years, 
reminiscent of Luminous Blue Variable (LBV) eruptions (see \citealt{Humphreys94} 
for a review). SN\,2005gl, with its LBV-like progenitor, established the first direct
observational connection between SNe IIn and LBVs. On the other hand, there are
controversial objects like SN\,1961V,  highlighting the present
difficulty in distinguishing between a giant LBV eruption and a genuine core-collapse
explosion even 50 years after the event (\citealt{VanDyk12},
\citealt{Kochanek11} and references therein).  The dividing line between SNe and
impostors can be ambiguous.

Here we report on our extensive multi-wavelength campaign
to monitor the evolution of SN\,2009ip, which offers an unparalleled opportunity to
study the effects and causes of significant  mass loss in massive stars in real time.
Discovered in 2009 \citep{Maza09} in NGC 7259
(spiral galaxy with brightness $M_{\rm{B}}\sim -18$ mag, \citealt{Lauberts89}),
it was first mistaken as a faint SN candidate (hence the name SN\,2009ip). Later
observations (\citealt{Miller09}; \citealt{Li09}; \citealt{Berger09c}) 
showed the behavior of  SN\,2009ip to be consistent instead with that of LBVs. 
Pre-explosion Hubble Space Telescope (HST) 
images constrain the progenitor to be a massive star with $M\gtrsim 60\, \rm{M_{\odot}}$
(\citealt{Smith10}, \citealt{Foley11}), consistent with an LBV nature. 
The studies by \cite{Smith10} and \cite{Foley11} showed that SN\,2009ip
underwent multiple explosions in rapid succession in 2009.
Indeed, a number of LBV-like eruptions were also observed in  2010 
and 2011: a detailed summary can be found in \cite{Levesque12} and a
historic light-curve is presented by \cite{Pastorello12}. Among the most 
important findings is the presence of blue-shifted absorption lines corresponding
to ejecta traveling at a velocity of $2000-7000\,\rm{km\,s^{-1}}$ during the
2009 outbursts (\citealt{Smith10}, \citealt{Foley11}), 
extending to $v\sim13000\,\rm{km\,s^{-1}}$ in Semptember 2011 \citep{Pastorello12}.
Velocities this large have never been associated with LBV outbursts to date.

SN\,2009ip re-brightened again on 2012 July 24 (\citealt{Drake12}), only to dim considerably
$\sim40$ days afterwards (hereafter referred to as the 2012a outburst). The appearance
of high-velocity spectral features was first noted by \cite{Smith12} on 2012 September 22.
This was shortly followed  by the major 2012 re-brightening on September 23 
(2012b explosion hereafter,
\citealt{Brimacombe12}, \citealt{Margutti12}).
SN\,2009ip reached $M_{\rm{V}}<-18\,\rm{mag}$ at this time, consequently questioning the actual survival of the
progenitor star: SN or impostor? (\citealt{Pastorello12}; \citealt{Prieto13}; \citealt{Mauerhan12};
\citealt{Fraser13}; \citealt{Soker13}).

Here we present a comprehensive study of SN\,2009ip during its remarkable
evolution in 2012. Using observations spanning more than 15 decades in wavelength,
from the GeV to the radio band, we
constrain the properties of the explosion and its complex environment, identify 
characteristic time scales that regulate the mass loss history of the progenitor star,
and study the process of dust vaporization in the progenitor surroundings.
We further predict the neutrino emission associated with this transient.


This paper is organized as follows. In Section \ref{Sec:Obs}-\ref{Sec:Energetics} 
we describe our follow up campaign and derive the observables that can be
\emph{directly} constrained by our data. In Section \ref{Sec:source} we present the
properties of the explosion and environment that can be \emph{inferred} from 
the data under reasonable assumptions. In Section \ref{Sec:Discussion}
we address the major questions raised by this explosion and speculate
about answers. Conclusions are drawn in Section \ref{Sec:Conclusions}.

Uncertainties are $1\sigma$ unless stated otherwise. 
Following \cite{Foley11} we adopt a distance modulus of $\mu=32.05\,\rm{mag}$
corresponding to a distance $d_{\rm{L}}=24$ Mpc and a Milky Way 
extinction $E(B-V)=0.019$ mag \citep{Schlegel98} with no additional host galaxy or
circumstellar extinction. From VLT-Xshooter high-resolution spectroscopy
our best estimate for the redshift of the explosion is  $z=0.005720$,
which we adopt through out the paper. We use $U$, $B$ and $V$ for the Johnson filters.
u, b and v refer to \emph{Swift-}UVOT filters.
Standard cosmological quantities have been adopted: 
$H_{0}=71\,\rm{km\,s^{-1}\,Mpc^{-1}}$, $\Omega_{\Lambda }=0.73$, 
$\Omega_{\rm{M}}=0.27$. All dates are in UT and are reported with respect to MJD 56203
(2012 October 3) which corresponds to the UV peak ($t_{\rm{pk}}$).
\section{Observations and data analysis}
\label{Sec:Obs}
Our campaign includes data from the radio band to the GeV range. We first describe
the data acquisition and reduction in the UV, optical and NIR bands (which are dominated
by thermal emission processes) and then describe the radio, X-ray and GeV observations, 
which sample the portion of the spectrum where non-thermal processes are likely to dominate.
\subsection{UV photometry}
\label{SubSec:UVphot}

\begin{figure*}
\vskip -0.0 true cm
\centering
\includegraphics[scale=0.8]{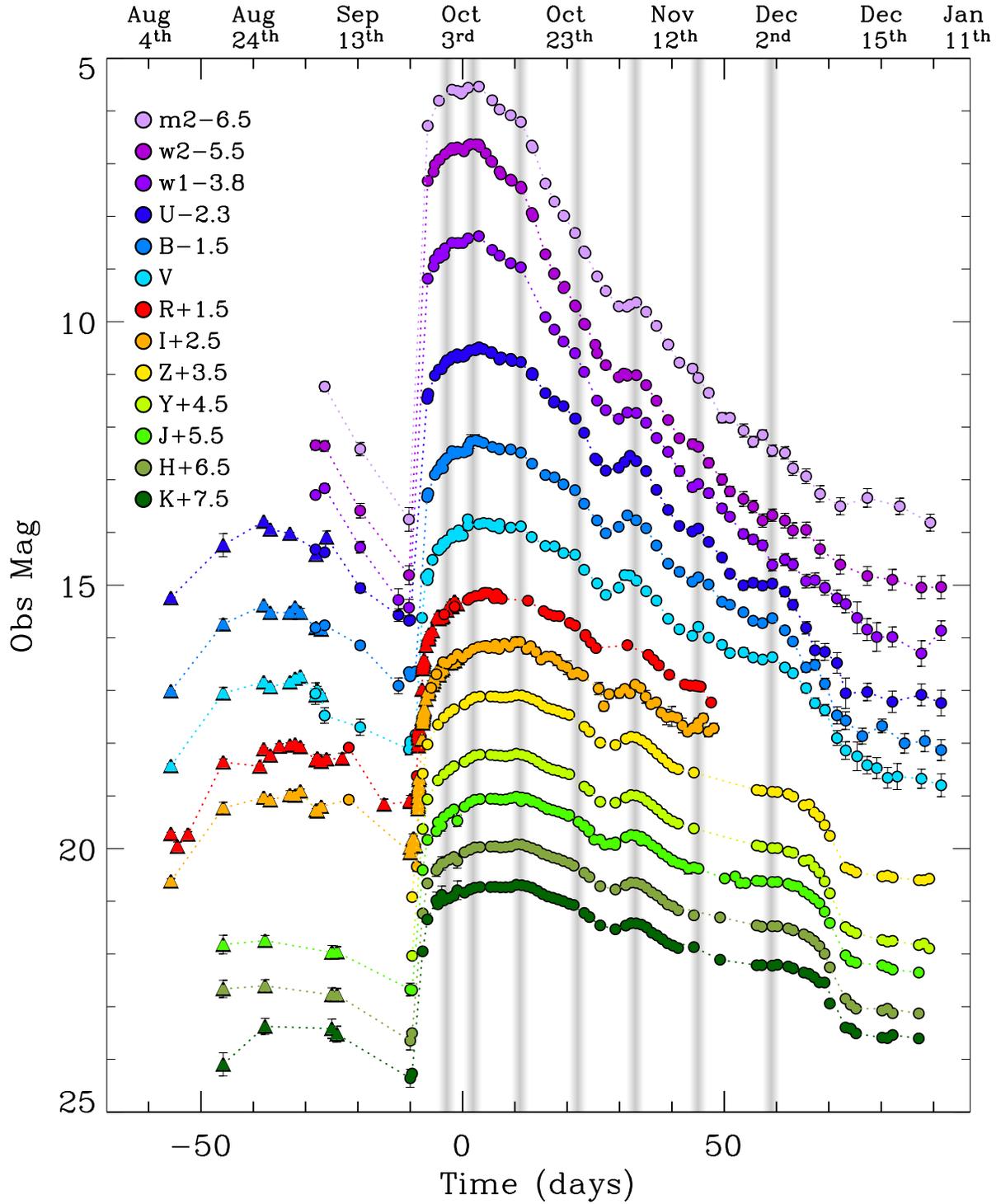}
\caption{Photometric evolution of SN\,2009ip in the UV, optical and NIR (filled circles).  We add
 NIR observations of the 2012a outburst published by \cite{Pastorello12} for 
$t<-10\,\rm{days}$ (triangles) together with R and I-band photometry from \cite{Prieto13} obtained during the rise-time (triangles). 
The shaded gray vertical bands mark the time of observed bumps in the light-curve. 
Our late time UVOT photometry from April 2013 is not shown here.}
\label{Fig:OptPhot}
\end{figure*}

We initiated our \emph{Swift}-UVOT \citep{Roming05} photometric 
campaign on 2009 September 10 and followed the evolution of SN\,2009ip
in the 6 UVOT filters up until April 2013. \emph{Swift}-UVOT observations
span the wavelength range $\lambda_c=1928$ \AA\, (w2 filter) - $\lambda_c=5468$ 
\AA\, (v filter, central wavelength listed, see \citealt{Poole08} for details).
Data have been analyzed following the prescriptions by \cite{Brown09}.
We used different apertures during the fading of the 2012a outburst to maximize
the signal-to-noise ratio and limit the contamination by a nearby star. 
For the 2012a event we used  a  $3''$ aperture for the b and v filters; $4''$ for 
the u filter and $5''$ for the UV filters. We correct for PSF losses following standard 
prescriptions. At peak SN\,2009ip reaches u$\sim 12.5$ mag potentially at risk for
major coincidence losses. For this reason we requested a smaller readout 
region around maximum light.
Our final photometry is reported in Table \ref{Tab:UVOTphot} and shown in Fig. \ref{Fig:OptPhot}.
The photometry is based on the UVOT photometric system (\citealt{Poole08}) and
the revised zero-points of \cite{Breeveld11}.

\subsection{UV spectroscopy: Swift-UVOT and HST}
\label{SubSec:UVspec}

\begin{figure}
\vskip -0.0 true cm
\centering
\includegraphics[scale=0.45]{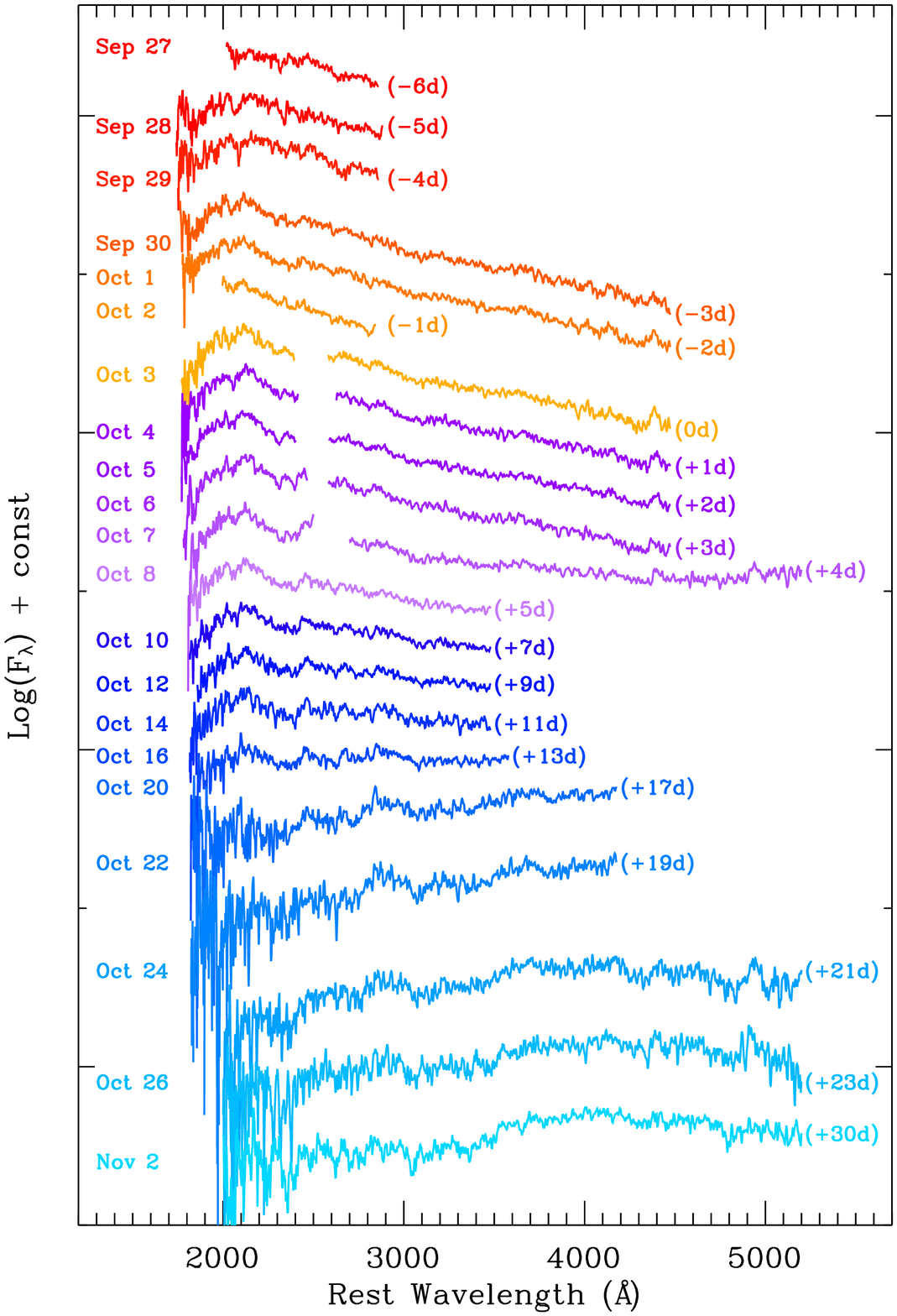}
\caption{Sequence of \emph{Swift}-UVOT spectra of SN\,2009ip covering
the rise time (red to orange), peak time (shades of purple) and decay time (shades of blue) of the 2012b explosion.}
\label{Fig:UVOTspec}
\end{figure}

\begin{figure}
\vskip -0.0 true cm
\centering
\includegraphics[scale=0.35]{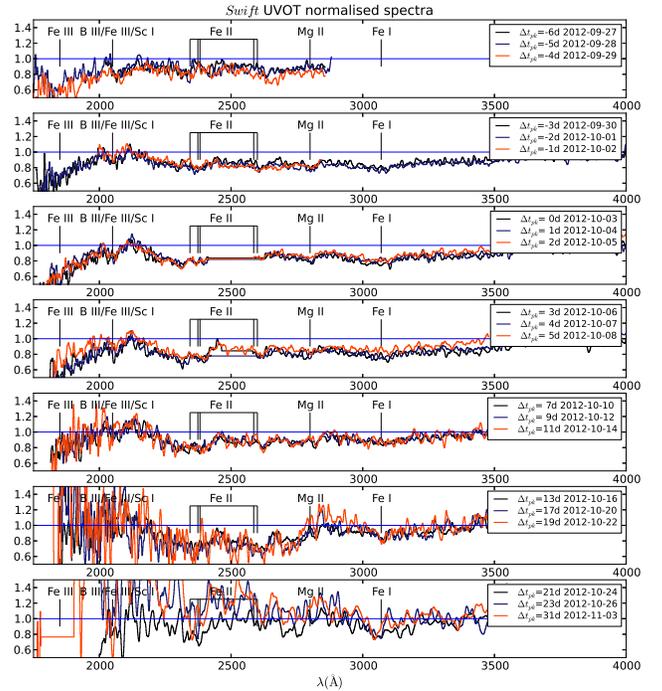}
\caption{UV portion of the \emph{Swift}-UVOT spectra re-normalized using the black-body fits of Sect. \ref{Sec:SED},
with identifications. As time proceeds  \ion{Fe}{3} absorption features become weaker while \ion{Fe}{2} develops stronger
absorption features, consistent with the progressive decrease of the black-body temperature with time (Fig. \ref{Fig:Lbol}).}
\label{Fig:UVOTspecnormalised}
\end{figure}

\begin{figure}
\vskip -0.0 true cm
\centering
\includegraphics[scale=0.44]{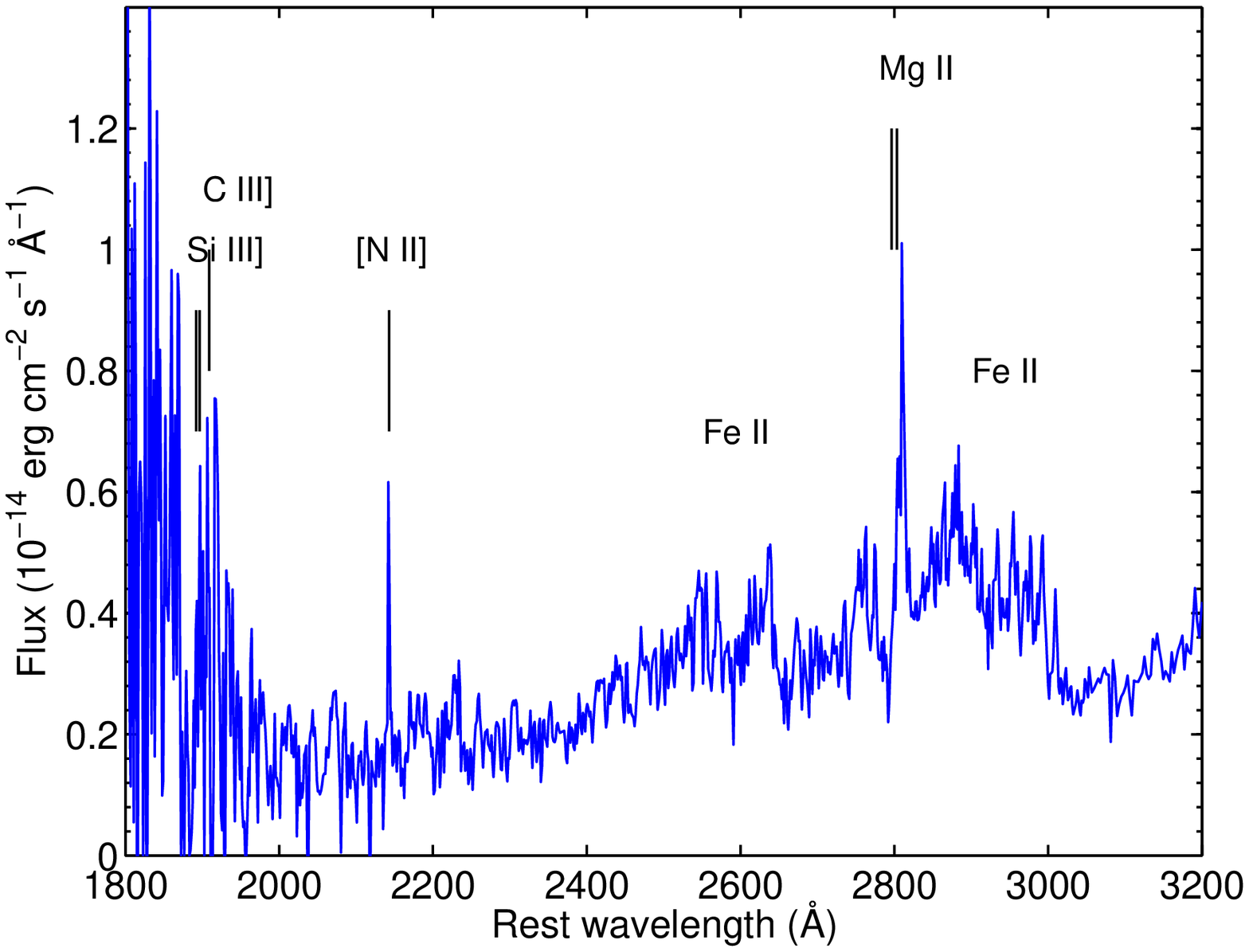}
\includegraphics[scale=0.44]{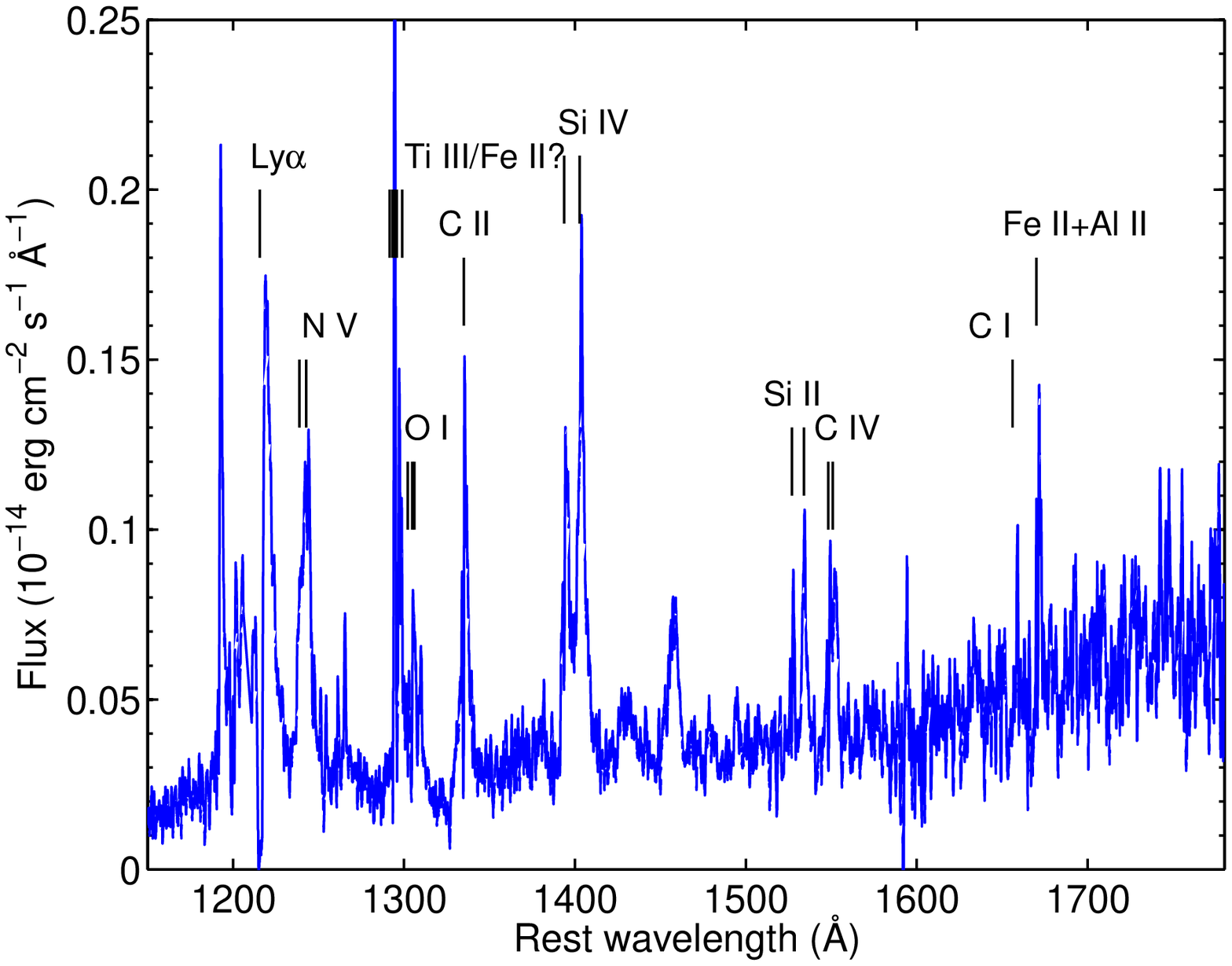}
\caption{\emph{Upper panel:} HST-STIS spectrum obtained on 2012 October 29 ($t_{\rm{pk}}+26$ days). The C III] and Si III] identifications are 
in a noisy part of the spectrum and are therefore uncertain.
\emph{Lower panel:} HST-COS spectrum obtained on 2012 November 6 ($t_{\rm{pk}}+34$ days).}
\label{Fig:HSTOct29UV}
\end{figure}

Motivated by the bright UV emission and very blue colors of SN\,2009ip we initiated
extensive UV spectral monitoring on 2012 September 27, $\sim6$ days before maximum UV light
(2012b explosion). Our campaign includes a total of  22 UVOT UV-grism low-resolution spectra 
and two epochs of  Hubble Space Telescope (HST) observations (PI R. Kirshner), 
covering the period $-6\,\rm{days}<t-t_{\rm{pk}}<+34\,\rm{days}$. 

Starting on 2012 September 27 ($t_{\rm{pk}}-6$ days), a series of spectra were taken with the \emph{Swift}
UVOT UV grism, with a cadence of one to two days, until 2012 October 28  ($t_{\rm{pk}}+25$ days),
with a final long observation on 2012 November 2  ($t_{\rm{pk}}+30$ days). 

Details are given in Table \ref{Tab:UVOTObsLog}. 
Over the course of the month the available roll angles changed.
The roll angle controls the position of the grism spectrum relative to the
strong zeroth orders from background stars in the grism image. The
best roll angle had the spectrum lying close to the first order of some other sources,
while some zeroth orders contaminated part of the spectrum in some observations.  Finally,
the second order overlap limits the usefulness of the red part of the UV grism
spectrum. To obtain the best possible uncontaminated spectra range, the spectra
were observed at a position on the detector where the second order lies next to
the first order, increasing the good, uncontaminated part of the first order from
about 1900 \AA\,  to 4500 \AA.

The spectra were extracted from the image using the UVOTPY package.
The wavelength anchor and flux calibration were 
the recent updates valid for locations other than the default position in the center
(Kuin et al. in prep., details can be found on-line\footnote{http://www.mssl.ucl.ac.uk/~npmk/Grism}). 
The spectra were extracted for a slit with the default 2.5$\,\sigma$ aperture and a 1$\,\sigma$
aperture. An aperture correction was made to the 1$\,\sigma$  aperture spectra which
were used. The 1$\,\sigma$ aperture does not suffer as much from contamination as the larger
aperture. Contamination from other sources and orders is readily seen when comparing the
extractions of the two apertures. The wavelength accuracy is 20\,\AA\, (1$\,\sigma$), 
the flux accuracy (systematic) is within 10\%, while the resolution is $R\sim75-110$
 depending on the wavelength range. 
The error in the flux was computed from  the Poisson noise in the data, as well as from the 
consistency of between the spectra extracted from the images on one day. 
The sequence of \emph{Swift}-UVOT spectra is shown in Fig. \ref{Fig:UVOTspec}.
Figure \ref{Fig:UVOTspecnormalised} shows the UV portion of the spectra, re-normalized 
using the black-body fits derived in Sec. \ref{Sec:SED}.

Starting from November 2012 SN\,2009ip is too faint for \emph{Swift}-UVOT spectroscopic observations.
We continued our UV spectroscopic campaign with HST  (Fig. \ref{Fig:HSTOct29UV}). 
Observations with the Space Telescope Imaging Spectrograph 
(STIS) were taken on 2012 October 29  ($t_{\rm{pk}}+26$ days) using aperture 52x0.2E1 with
gratings G230LB, G430L and G750L with exposures times of 1200 s, 400 s and 100 s, respectively.
The STIS 2-D images were cleaned of cosmic rays and dead pixel signatures before extraction.
The extracted spectra were then matched in flux to the STSDAS/STIS pipeline 1-D data product.
The spectrum is shown in Fig. \ref{Fig:HSTOct29UV}. 
Further HST-COS data were acquired on 2012 November 6 ($t_{\rm{pk}}+34$ days, Fig. \ref{Fig:HSTOct29UV}, lower panel).
Observations with the Cosmic Origin Spectrograph (COS) were acquired 
using MIRROR A + bright object aperture for 250 s.
The COS data were then reprocessed with the COS calibration pipeline, CALCOS v2.13.6, and combined
with the custom IDL co-addition procedure described by \cite{Danforth10} and \cite{Shull10}. The
co-addition routine interpolates all detector segments and grating settings onto a common wavelength grid, and makes
a correction for the detector quantum efficiency enhancement grid. No correction for the detector hex pattern is
performed. Data were obtained in four central wavelength settings in each far-UV grating mode (1291, 1300, 1309, and
1318 with G130M and 1577, 1589, 1600, and 1611 with G160M) at the default focal-plane split position.
The total exposure time for the far-UV observation was 3100 s and 3700 s for near-UV.
\subsection{Optical photometry}
\label{SubSec:Opticalphot}

Observations in the v, b and u filters were obtained with \emph{Swift}-UVOT
and reduced as explained in Sec. \ref{SubSec:UVphot}. The results from our observations
are listed in Table \ref{Tab:UVOTphot}. In Fig. \ref{Fig:OptPhot}
we apply a dynamical color term correction to the UVOT v, b and u filters to plot
the equivalent Johnson magnitudes as obtained following the prescriptions by 
\cite{Poole08}. This is a minor correction to the measured magnitudes and  it is
\emph{not} responsible for the observed light-curve bumps.

We complement our data set with R and I band photometry obtained  with 
the UIS Barber Observatory 20-inch telescope (Pleasant Plains, IL), a 0.40 m f/6.8 refracting
telescope operated by Josch Hambasch at the Remote Observatory Atacama Desert, a
Celestron C9.25 operated by TG Tan (Perth, Australia), and a C11 
Schmidt-Cassegrain telescope operated by Ivan Curtis (Adelaide, Australia).  
Exposure times ranged from 120 s to 600 s.
Images were reduced following standard procedure.
Each individual image in the series was measured and then averaged together 
over the course of the night.
The brightness was measured using circular apertures adjusted for seeing conditions 
and sky background from an annulus set around each aperture.  Twenty comparison stars
within 10$'$ of the target were selected from the AAVSO\footnote{http://www.aavso.org/apass.}  Photometric 
All-Sky Survey. 
Statistical errors in the photometry
for individual images were typically 0.05 magnitudes or less.  Photometry taken by 
different telescopes on the same night are comparable within the errors.  
Finally, the photometry was corrected to the photometric system of \cite{Pastorello12} 
using the corrections of dR = +0.046 and dI = +0.023 (Pastorello, personal communication).
R and I band photometry is reported in Table \ref{Tab:amateurphot}.

A single, late-time ($t_{\rm{pk}}+190$ days) V-band observation was obtained with the Inamori-Magellan Areal Camera 
and Spectrograph (IMACS, \citealt{Dressler06}) mounted on the Magellan/Baade $6.5$-m telescope on 2013 Apr 11.40.
Using standard tasks in IRAF to perform aperture photometry and 
calibrating to a standard star field at similar airmass, we measure $V=19.65 \pm 0.02$ mag (exposure time of 90 s).
\subsection{Optical spectroscopy}
\label{SubSec:Optspec}

\begin{figure*}
\vskip -0.0 true cm
\centering
\includegraphics[scale=0.9]{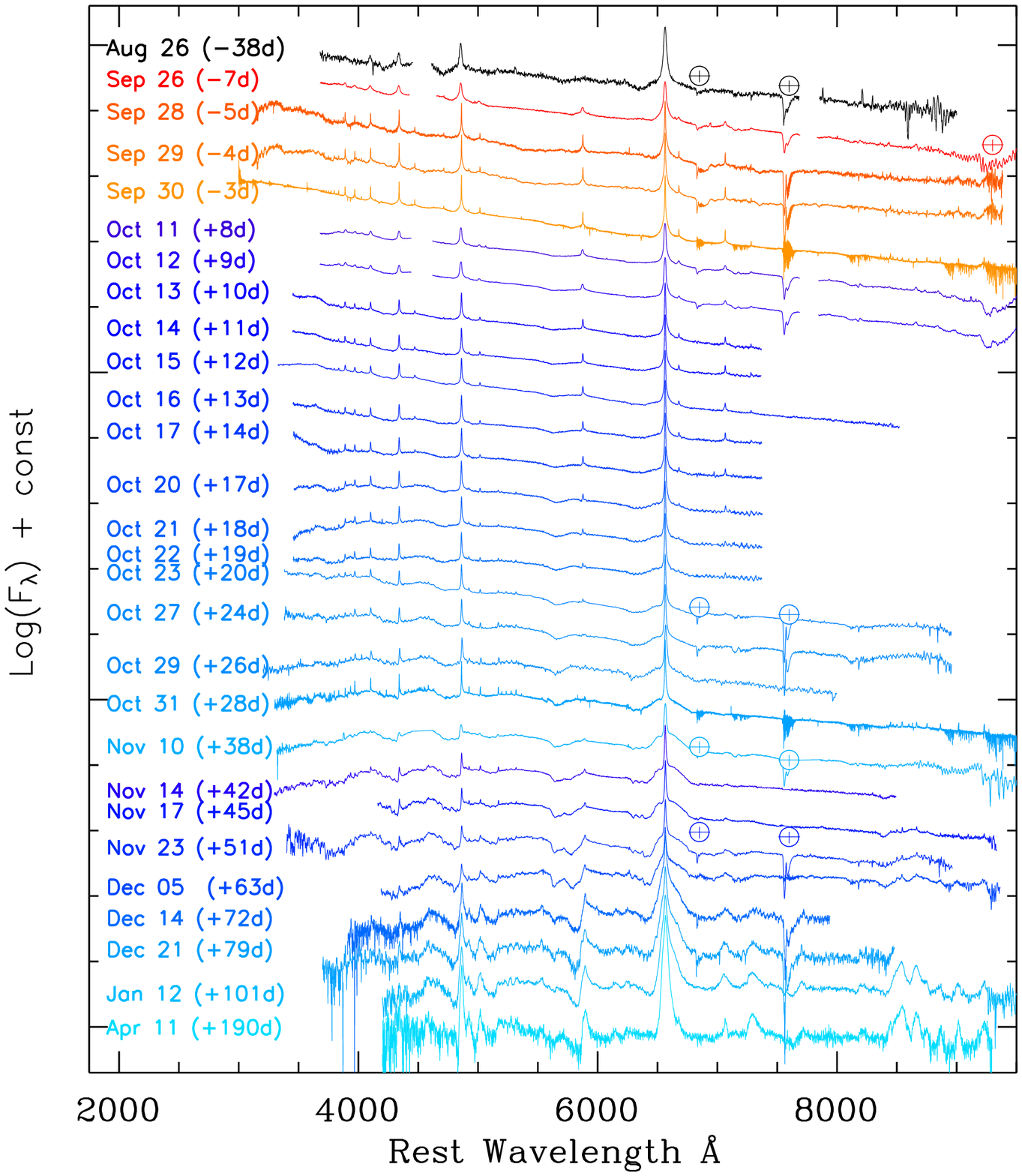}
\caption{Optical spectra of SN\,2009ip. Shades of
red (blue) are used for spectra obtained during the rise time (decay time) of the 2012b explosion.
Black is used for the 2012a outburst.} 
\label{Fig:OptSpec}
\end{figure*}

We obtained 28 epochs of optical spectroscopy of SN\,2009ip covering the time period 2012 August 26 to
2013 April 11  using a number of facilities
(see Table \ref{Tab:OptSpecObsLog}).

SN\,2009ip was observed with the MagE (Magellan Echellette)
Spectrograph mounted on the 6.5-m Magellan/Clay Telescope  at Las Campanas Observatory.
Data reduction was performed using a combination of Jack Baldwin's {\tt mtools} package 
and IRAF\footnote{IRAF is distributed by the National Optical Astronomy Observatories, 
which are operated by the Association of Universities for Research in Astronomy, Inc., 
under cooperative agreement with the National Science Foundation.}  
echelle tasks, as described in \cite{Massey12}.
Optical spectra were obtained at the F. ~L. \ Whipple Observatory 
(FLWO) 1.5-m Tillinghast telescope on several epochs 
using the FAST spectrograph \citep{Fabricant98}.   Data were reduced using 
a combination of standard IRAF and custom IDL procedures \citep{Matheson05}.
Low and medium resolution spectroscopy was obtained with the Robert Stobie Spectrograph mounted 
on the Southern African Large Telescope (SALT/RSS) at the South African Astronomical Observatory (SAAO)
in Sutherland, South Africa. 
Additional spectroscopy was acquired with the Goodman High Throughput Spectrograph (GHTS)
on the SOAR telescope. We also used IMACS
mounted on the Magellan telescope and the Low Dispersion Survey Spectrograph 3 (LDSS3)
on the Clay telescope (Magellan II).  The Multiple Mirror Telescope (MMT) equipped with the 
``Blue Channel'' spectrograph \citep{Schmidt89} was used to monitor the spectral evolution of SN\,2009ip over several
epochs.  Further optical spectroscopy was obtained with the R-C CCD Spectrograph (RCSpec)
mounted on the Mayall 4.0\,m telescope, a Kitt Peak National Observatory (KPNO) facility.
Spectra were extracted and calibrated following standard procedures using IRAF routines. 

We used the X-shooter echelle spectrograph \citep{DOdorico06} mounted at the 
Cassegrain focus of the {\it Kueyen} unit of the Very Large Telescope (VLT) 
at the European Southern Observatory (ESO) on Cerro Paranal (Chile) to obtain broad
band, high-resolution spectroscopy of SN\,2009ip on 2012 September 30 
($t_{\rm{pk}}-3$ days) and October 31 ($t_{\rm{pk}}+28$ days).  
The spectra were simultaneously observed 
in three different arms, covering the entire wavelength range 3000--25000 \AA: ultra-violet and blue (UVB), visual (VIS) and 
near-infrared (NIR) wavebands.
The main dispersion was achieved through a 180 grooves/mm echelle 
grating blazed at 41.77$^{\circ}$ (UVB), 99 grooves/mm echelle grating 
blazed at 41.77$^{\circ}$ (VIS) and 55 grooves/mm echelle grating 
blazed at 47.07$^{\circ}$ (NIR).
Observations were performed at parallactic angle under the following 
conditions: clear sky, the average seeing was $\sim$ 0.7$'' $ and  $\sim$ 1.0$'' $ 
and the airmass range was $\sim$ 1.1--1.23. 
We used the X-shooter pipeline \citep{Modigliani10} 
in physical mode to reduce both SN\,2009ip and the standard star spectra to 
two-dimensional bias-subtracted, flat-field corrected, order rectified and 
wavelength calibrated spectra in counts. To obtain 1-D spectra 
the 2-D spectra from the pipeline were optimally extracted
\citep[see][]{Horne86} using a custom IDL program. 
Furthermore, the spectra were slit-loss corrected, flux calibrated and corrected for
heliocentric velocities using a custom IDL program. 
The spectra were not (carefully) telluric corrected.
The sequence of optical spectra is shown in Fig. \ref{Fig:OptSpec}.

\subsection{NIR photometry}
\label{SubSec:NIRphot}

We obtained ZYJHK data using the Wide-field Infrared Camera
(WFCAM) on the United Kingdom Infrared Telescope (UKIRT).
The observation started on 2012 September 23 ($t_{\rm{pk}}-10$ days),
and continued on a nearly daily basis until 2012 December 31($t_{\rm{pk}}+89$ days)
when SN\,2009ip settled behind the Sun.
The data reduction was done through an automatic pipeline of
the Cambridge Astronomy Survey Unit. The flux of the object was
measured with AUTO-MAG of SExtractor \citep{Bertin96},
where the photometric calibration was done using 2MASS stars
within a radius of 8 arcmin from SN\,2009ip.
The 2MASS magnitudes of the stars were converted to
the UKIRT system following \cite{Hodgkin09}
and the stars with the magnitude errors smaller
than 0.10 mag were used for the photometry calibration,
which yields typically 20-30 stars. A more detailed description of the
NIR photometry can be found in Im et al. (in preparation).

Additional NIR photometry was obtained with PAIRITEL, the f/13.5 1.3-meter Peters 
Automated Infrared Imaging TELescope at the Fred Lawrence Whipple Observatory (FLWO) 
on Mount Hopkins, Arizona \citep{bloom06}. PAIRITEL data were processed with a single 
mosaicking pipeline that co-adds and registers PAIRITEL raw images into mosaics 
(see \citealt{woodvasey08}; \citealt{friedman12}).
Aperture photometry with a $3''$ aperture was performed at the SN position 
in the mosaicked images using the IDL routine aper.pro.  No aperture corrections or host galaxy 
subtraction were performed.  
Figure \ref{Fig:OptPhot} presents the complete SN\,2009ip NIR data set. 
The PAIRITEL photometry can be found in Table \ref{Tab:nirphot}. A table will
the UKIRT photometry will be published in Im et al. (in preparation). 
\subsection{NIR spectroscopy}
\label{SubSec:NIRspec}

\begin{figure*}
\vskip -0.0 true cm
\centering
\includegraphics[scale=0.9]{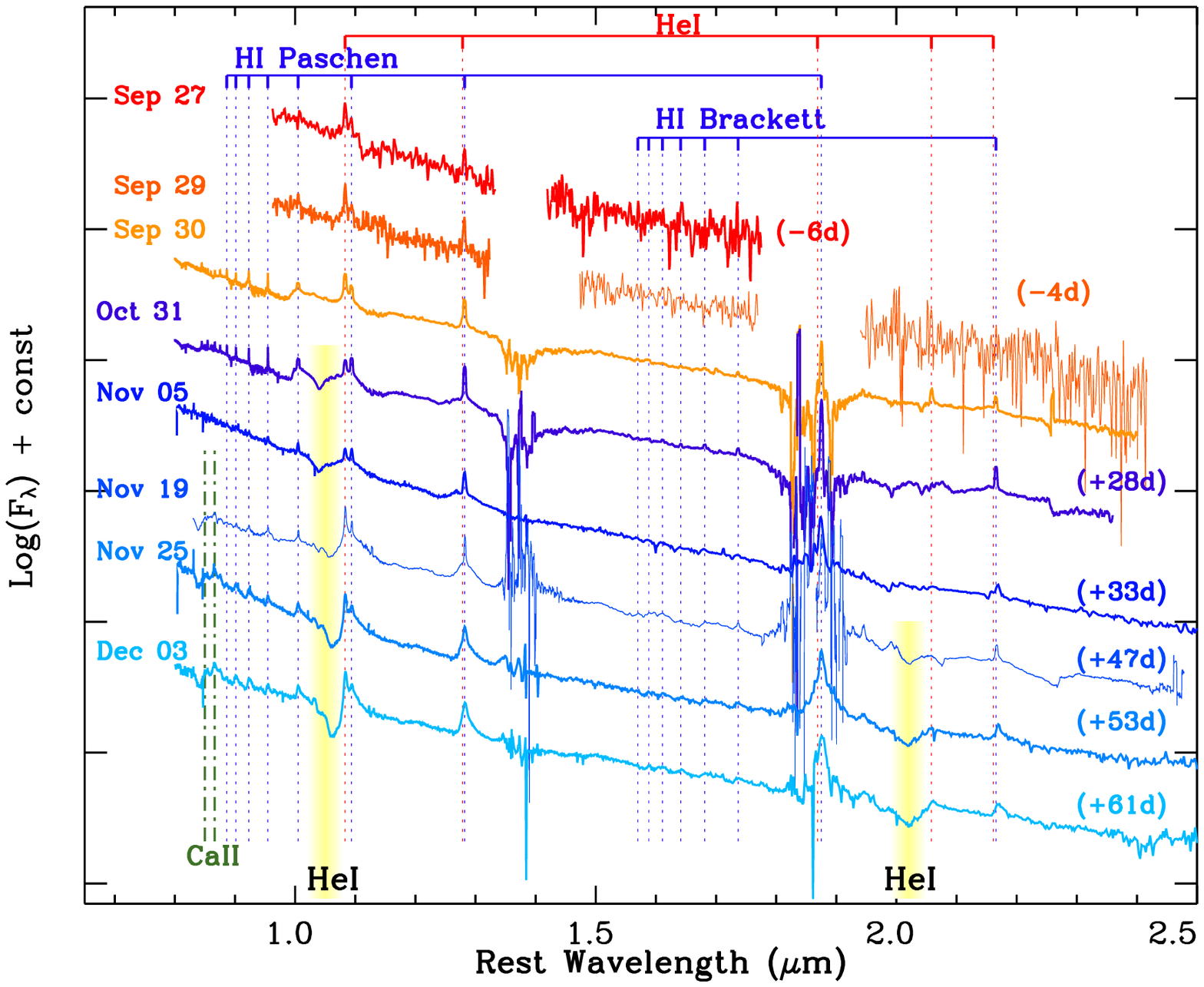}
\caption{NIR spectral sequence with line identifications overlaid. 
The high-resolution spectrum obtained on 2012 November 19 has been smoothed here for display purposes.
Shades of red (blue) have been used for spectra obtained during the rise (decay) time. A portion of the 
VLT/X-shooter spectra is also shown here.}
\label{Fig:NIRSpec}
\end{figure*}

In addition to the X-shooter spectra, early time, low-dispersion ($R \approx 700$) NIR
spectra  covering 0.9 to 2.4 $\mu \rm{m}$ were obtained with the
2.4m Hiltner telescope at MDM Observatory on 2012
September 27 ($t_{\rm{pk}}-6$ days) and September 29 
($t_{\rm{pk}}-4$ days). The data were collected using TIFKAM, a high-throughput
infrared imager and spectrograph with a $1024 \times 1024$ Rockwell
HgCdTe (HAWAII-1R) detector.
The target was dithered along the $0.6''$ slit in a ABBA
pattern to minimize the effect of detector defects and provide
first-order background subtractions. Data reduction followed standard
procedures using the IRAF software. Wavelength calibration of the
spectra was achieved by observing argon lamps at each position. The
spectra were corrected for telluric absorption by observing A0V stars
at similar airmasses, and stellar features were removed from the
spectra by dividing by an atmospheric model of Vega \citep{Kurucz93}.

Additional NIR low-resolution spectroscopy of SN\,2009ip was obtained
with  the Folded-Port Infrared Echellette (FIRE) spectrograph \citep{Simcoe13}
on the 6.5-m Magellan Baade Telescope,
with simultaneous coverage from 0.82 to 2.51 $\mu m$. Spectra were acquired
on 2012 November 5, 25 and December 3.
The object was nodded along the slit using the ABBA pattern.
The slit width was $0.6\arcsec$, yielding $R \approx 500$ in the J band.
Data were reduced following the standard procedures described in 
\cite{Vacca03}, \cite{Foley12} and \cite{Hsiao13}.
An A0V star was observed for telluric corrections. The resulting telluric 
correction spectrum was also used for the absolute flux calibration.

Moderate-resolution ($R\sim 6000$) NIR spectroscopy was obtained on 2012 November 19
with FIRE.
SN\,2009ip was observed in high-resolution echellette mode with the $0.6\arcsec$
slit. Eight frames were taken on source with 150 s exposures using ABBA nodding.  
Data were reduced using a custom-developed IDL pipeline 
(FIREHOSE), evolved from the MASE suite used for optical echelle reduction
\citep{Bochanski09}. Standard procedures were followed to apply
telluric corrections and relative flux calibrations as described above.
Finally, the corrected echelle orders were combined into single 1D spectrum for analysis.
The complete sequence of NIR spectra is shown in Fig. \ref{Fig:NIRSpec}.
The observing log can be found in Table \ref{Tab:NIRSpecObsLog}.

\subsection{Millimeter and Radio Observations: CARMA and EVLA}
\label{SubSec:RadioObs}

\begin{figure}
\vskip -0.0 true cm
\centering
\includegraphics[scale=0.45]{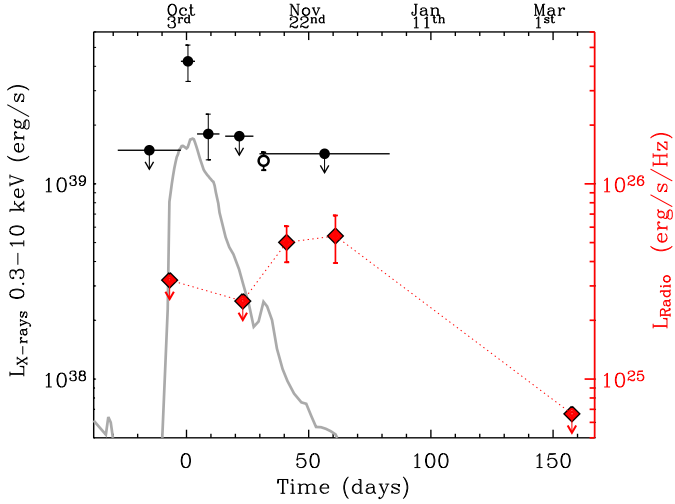}
\caption{X-ray (\emph{Swift}-XRT and XMM-Newton, filled and open circle, respectively) 
and  9 GHz radio light-curve (red squares, VLA)  of SN\,2009ip. X-rays are detected when the bolometric 
luminosity reaches its  peak. Radio emission  is detected at much later times. 
A re-scaled version of the bolometric light-curve is also shown for comparison. This plot does not
include the late time X-ray limit obtained on 2013 April 4.5 ($t_{\rm{pk}}+183$ days).}
\label{Fig:XrayRadioLC}
\end{figure}

\begin{deluxetable}{lrrrr}
\tablecaption{Radio and millimeter observations of SN\,2009ip}
\tablewidth{0pt}
\tablehead{
\colhead{Date} & & \colhead{$F_{\nu}$} & \colhead{$\nu$} & \colhead{Instrument} \\
\colhead{(UT start time)} & & \colhead{($\mu$Jy)} & \colhead{(GHz)} & \\}
\startdata
2012 Sep & 26.096& $<115.2$ & 21.25& VLA\\
2012 Sep & 26.096& $<46.5$  & 8.85 & VLA\\
2012 Sep\footnote{From \cite{Hancock12}.}  & 26.63& $<66$ &  18  &   ATCA\\
2012 Sep & 27.170& $<1000$ & 84.5 & CARMA\\
2012 Oct &16.049& $<70.5$  &  21.25 & VLA\\
2012 Oct &17.109 & $<104.1$  &  21.25 & VLA\\
2012 Oct & 17.120 & $<1500$ & 84.5 & CARMA\\
2012 Oct & 26.036& $<36.3$  &  8.85 &VLA\\
2012 Nov& 6.078 & $<59.1$  & 21.25 & VLA \\
2012 Nov& 12.966& $72.6\pm15.2$ & 8.85  & VLA \\
2012 Dec & 1.987 & $<70.5$ & 21.25 & VLA \\
2012 Dec & 2.932 & $78.3\pm21.4$ &8.99 & VLA \\
2013 Mar & 9.708 & $<9.6$ &9.00& VLA\\
\enddata
\tablecomments{Errors are $1\sigma$ and upper limits are $3\sigma$.}
\label{Tab:radio}
\end{deluxetable}

We obtained two sets of millimeter observations at mean frequency of $\sim$84.5 GHz ($\sim$7.5~GHz bandwidth) 
with the Combined Array for Research in Millimeter Astronomy (CARMA; \citealt{bock}) around
maximum light, beginning 2012 September 27.17 ($t_{\rm{pk}}-6$ days) and 2012 October 
17.12 ($t_{\rm{pk}}+14$ days).  
We utilized 2158-150 and 2258-279 for gain calibration,
2232+117 for bandpass calibration, and Neptune for flux calibration.  In $\sim$160 and $\sim$120 min integration
time at the position of SN\,2009ip, we obtain 3-$\sigma$ upper limits on the flux density of 1.5 and 1.0 mJy, respectively. 
The overall flux uncertainty with CARMA is $\sim$20\%.

We observed the position of SN\,2009ip with the Karl G. Jansky Very Large Array (VLA; \citealt{Perley+11})
on multiple epochs beginning 2012 September 26.10 ($t_{\rm{pk}}-7$ days), with the last epoch beginning 2012 Dec 2.93
($t_{\rm{pk}}+61$ days).  These observations were carried out at 21.25~GHz and 8.85~GHz with 2~GHz bandwidth in the VLA's most extended configuration 
(A; maximum baseline length $\sim$36.4 km) except for the first observations, which were obtained in the 
BnA configuration.  In most epochs our observations of flux calibrator, 3C48, were too contaminated with 
Radio Frequency Interference (RFI).  Therefore, upon
determining the flux of our gain calibrator J2213-2529 from the best observations of 3C48, we set the flux density of
J2213-2529 in every epoch to be 0.65 and 0.63 Jy, for 21.25 and 8.9 GHz, respectively. We note that this assumption 
might lead to slightly larger absolute flux uncertainties than usual ($\sim$15-20\%).  
In addition, the source-phase calibrator cycle time ($\sim$6 min) was a bit longer than standard for high frequency observations in 
an extended configuration, potentially increasing decoherence. We manually inspected the data and flagged edge channels
and RFI, effectively reducing the bandwidth by $\sim$15\%.  We reduced all data using standard procedures in the
Astronomical Image Processing System (AIPS; \citealt{gre03}).  
A summary of the observations is presented in Table~\ref{Tab:radio}. 

No source is detected at the position of SN\,2009ip at either frequencies during the first 50 days
since the onset of the major outburst in September 2012, enabling deep limits
on the radio emission around optical maximum. 
A source is detected at 8.85 GHz on 2012 November 13 ($t_{\rm{pk}}+41$ days), indicating a 
re-brightening of SN\,2009ip radio emission at the level of $F_{\nu}\sim 70~\mu$Jy. 
The source position is $\alpha$=22:23:08.29 $\pm0.01''$ 
and $\delta$= $-$28:56:52.4 $\pm0.1''$, consistent with the position determined from HST data.
We merged the two observations that yielded a detection to improve the signal
to noise and constrain the spectrum. Splitting the data into two 1 GHz slices
centered at 8.43 GHz and 9.43 GHz, we find integrated flux densities of 
$F_{\nu}=60.0\pm16.7  ~\mu$Jy (8.43 GHz) and $F_{\nu}=100.6\pm18.9~\mu$Jy (9.43 GHz), 
suggesting an optically thick spectrum. The upper
limit of $F_{\nu}<70.5~\mu$Jy at 21.25 GHz on 2012 December 2 indicates that the observed spectral peak 
frequency $\nu_{\rm{pk}}$ is between 9.43 GHz and 21.25 GHz. A late-time observation obtained
on March 9th shows that the radio source faded to $F_{\nu}<9.6~\mu$Jy at
9 GHz, pointing to a direct association with SN\,2009ip.
The radio light-curve at 9 GHz is shown in Figure \ref{Fig:XrayRadioLC}.
\subsection{X-ray observations: \emph{Swift}-XRT and XMM-Newton}
\label{SubSec:XRTObs}

\begin{figure}
\vskip -0.0 true cm
\centering
\includegraphics[scale=0.55]{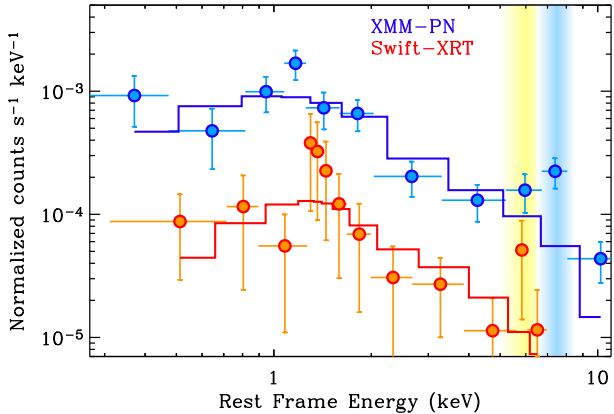}
\caption{X-ray spectra of SN\,2009ip. The \emph{Swift}-XRT spectrum collects observations
obtained around the optical peak ($t_{\rm{pk}}-2$ days until $t_{\rm{pk}}+13$ days, total exposure time
of 86 ks). The XMM EPIC-PN spectrum  was obtained on 2012 November 3 ($t_{\rm{pk}}+31$ days,
total exposure of 55 ks). The spectral model consists of absorbed bremsstrahlung
emission at $kT=60$ keV and intrinsic absorption ($\rm{NH_{int}}=0.10^{+0.06}_{-0.05}\times 
10^{22}\rm{cm^{-2}}$ and $\rm{NH_{int}}<3.1\times 10^{21}\rm{cm^{-2}}$ for the XMM and XRT spectra,
respectively). Contamination by a nearby source lying $\approx 6''$ from SN\,2009ip
is expected for the XMM spectrum. The color-coded shaded areas highlight the presence of possible 
emission excess with respect to the model. }
\label{Fig:XraySpec}
\end{figure}

\begin{figure}
\vskip -0.0 true cm
\centering
\includegraphics[scale=0.38]{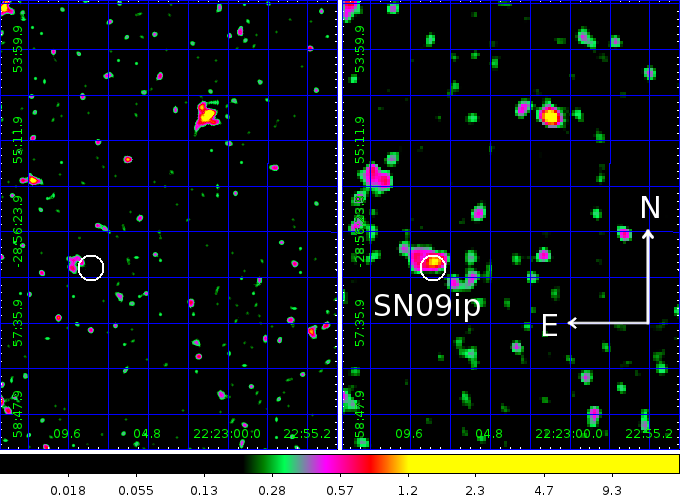}
\caption{Left panel: \emph{Swift}-XRT image of the field of SN\,2009ip collecting data
before and after the optical peak ($-32\,\rm{days}<t-t_{\rm{pk}}<-2\,\rm{days}$ and  $+29\,\rm{days}<t-t_{\rm{pk}}<+83\,\rm{days}$),
for a total exposure time of 110 ks. Right panel: same field imaged around the optical
peak ($-2 \,\rm{days}<t-t_{\rm{pk}}<+13\,\rm{days}$) for a total exposure of 86 ks. 
In both panels a white circle marks a $10''$ region around SN\,2009ip. 
An X-ray source is detected at a position consistent with SN\,2009ip around the
optical peak (right panel) with significance of  $6.1\sigma$. The contaminating source
discussed in the text is apparent in the left image.}
\label{Fig:Xrayimage}
\end{figure}

We observed SN\,2009ip with the \swift/X-Ray Telescope (XRT, \citealt{Burrows05}) 
from  2012 September 4 (20:36:23) until 2013 January 1 (13:43:55), 
for a total exposure of $260$ ks, covering the time period  $-29\,\rm{days}<t-t_{\rm{pk}}<+90\,\rm{days}$.
Data have been entirely acquired in Photon Counting (PC) mode\footnote{The \swift-XRT observing modes are
defined in \cite{Hill04}.} and analyzed using the latest HEASOFT (v6.12) release, 
with standard filtering and screening criteria.  No X-ray source is detected at the position
of SN\,2009ip during the decay of the 2012a outburst ($t<-11$ days), down to a 3$\sigma$ limit of 
$3\times10^{-3}\rm{cps}$ in the 0.3-10 keV energy band (total exposure of 12.2 ks). 
Observations sampling the rise time of the 2012b explosion ( $-11\,\rm{days}<t-t_{\rm{pk}}<-2\,\rm{days}$ ) also show no detection.
With  31.4 ks of total exposure the 0.3-10 keV count-rate limit at the SN position is
$<1.1\times10^{-3}\rm{cps}$. Correcting for PSF (Point Spread Function) losses and vignetting
and merging the two time intervals we find no evidence for X-ray emission originating
from SN\,2009ip in the time interval $-29\,\rm{days}<t-t_{\rm{pk}}< -2\,\rm{days}$ 
down to a limit of $<5.6 \times10^{-4}\rm{cps}$ (0.3-10 keV, exposure time of 43.6 ks).
X-ray emission is detected at a position consistent with SN\,2009ip starting from $t_{\rm{pk}}-2$ days,
when the 2012b explosion approached its peak luminosity in the UV/optical bands 
(Fig. \ref{Fig:XrayRadioLC}). The source is detected at the level of 5$\sigma$ and
$4\sigma$ in the time intervals $-2\,\rm{days}<t-t_{\rm{pk}}<+3\,\rm{days}$ and 
$+4\,\rm{days}<t-t_{\rm{pk}}<+13\,\rm{days}$, respectively, with PSF and vignetting corrected 
count-rates of $(1.6\pm0.3)\times 10^{-3}\rm{cps}$ and
$(6.8\pm1.8)\times 10^{-4}\rm{cps}$ (0.3-10 keV, exposure time of 42 and 44 ks).


Starting from $t_{\rm{pk}}+17$ days, the source is no longer detected by XRT. We therefore
activated our XMM-Newton program (PI P. Chandra) to follow the fading of 
the source. We carried out XMM-Newton observations starting 
from 2012 November 3 at 13:25:33 ($t_{\rm{pk}}+31$ days).  Observations
have been obtained with the EPIC-PN and EPIC-MOS cameras in full frame with thin filter
mode. The total exposures for the EPIC-MOS1 and EPIC-MOS2 are 62.62 ks and 62.64 ks,
respectively, and for the EPIC-PN, the exposure time is 54.82 ks.  A point-like
source is detected at the position of SN\,2009ip with significance of 4.5 $\sigma$ (for EPIC-PN),
and rate of  $(2.7\pm0.3)\times 10^{-3}$ cps in a region of $10''$ around the optical position of SN\,2009ip.

From January until April 2013 the source was Sun constrained for \emph{Swift}. 10 ks of  
\emph{Swift}-XRT data obtained on 2013 April 4.5 ($t_{\rm{pk}}+183$ days, when SN\,2009ip
became observable again) showed
no detectable X-ray emission at the position of the
transient down to a 3$\sigma$ limit of $4.1\times 10^{-3}$ cps (0.3-10 keV).

We use the EPIC-PN observation to constrain the
spectral parameters of the source. We extract photons from a region of $10''$ radius
to avoid contamination from a nearby source (Fig. \ref{Fig:Xrayimage}).
The XMM-Newton software SAS is used to 
extract the spectrum. Our spectrum contains a total of 132 photons. 
We model the spectrum with an absorption component (which combines the
contribution from the Galaxy and from SN\,2009ip local environment, $tbabs\times
ztbabs$ within \emph{Xspec}) and an
emission component. Both thermal bremsstrahlung and thermal emission
from an optically thin plasma in collisional equilibrium (\emph{Xspec} MEKAL model) 
can adequately fit the observed spectrum. In both cases
we find $kT> 10$ keV and intrinsic hydrogen absorption of $\rm{NH_{int}}\approx10^{21}
\,\rm{cm^{-2}}$. In the following we assume thermal bremsstrahlung emission with 
$kT=60 $ keV (this is the typical energy of
photons expected from shock break-out from a dense CSM shell, 
see Section \ref{SubSec:XraysBO}).\footnote{We consider a non-thermal
power-law emission model unlikely given the very hard best-fitting photon index 
of $\Gamma=0.87\pm0.15$ we obtain from this spectrum.} The Galactic
absorption in the direction of SN\,2009ip is $N_{\rm H,MW}=1.2\times 10^{20}$ cm$^{-2}$
 \citep{Kalberla05}. The best-fitting neutral hydrogen intrinsic absorption\footnote{This estimate assumes
 an absorbing medium with solar abundance and low level of ionization.} is 
 constrained to be $\rm{NH_{int}}=0.10^{+0.06}_{-0.05}\times 10^{22}\rm{cm^{-2}}$.
Using these parameters, the corresponding unabsorbed (absorbed) flux is
$(1.9 \pm 0.2)\times 10^{-14}\rm{erg\,s^{-1}cm^{-2}}$ ($(1.7 \pm 0.2)\times 10^{-14}\rm{erg\,s^{-1}cm^{-2}}$)  
in the $0.3-10$ keV band. The spectrum is displayed in Fig. \ref{Fig:XraySpec} and
shows some evidence for an excess of emission around $\sim7-8$ keV (rest-frame)
which might be linked to the presence of Ni or Fe emission lines 
(see e.g. SN2006jd and SN2010jl; \citealt{Chandra12a,Chandra12b}).

A  \emph{Swift}-XRT spectrum extracted around the peak ($-2\,\rm{days}<t-t_{\rm{pk}}<+13\,\rm{days}$, 
total exposure of 86 ks) can be fit by a thermal bremsstrahlung model, assuming
$kT=60$ keV and $\rm{NH_{int}}<3.1\times 10^{21}\rm{cm^{-2}}$
at the $3\sigma$ c.l. As for XMM, we use a $10''$ extraction region to avoid contamination
from a nearby source (Fig. \ref{Fig:Xrayimage}).
The count-to  flux conversion factor deduced from this spectrum is 
$3.8\times 10^{-11}\rm{erg\,s^{-1}cm^{-2}ct^{-1}}$ (0.3-10 keV, unabsorbed). We use this factor to calibrate 
our \emph{Swift}-XRT light-curve. The complete X-ray light-curve is shown in Fig. \ref{Fig:XrayRadioLC}. 

We note that at the resolution of XMM and \emph{Swift}-XRT 
we cannot exclude the presence of contaminating X-ray sources at a distance $\lesssim 10''$. 
We further investigate this issue constraining the level of the contaminating flux by 
merging the \emph{Swift}-XRT time intervals that yielded a
non-detection at the SN\,2009ip position. Using data collected between $t_{\rm{pk}}-29$ days and $t_{\rm{pk}}-2$
days, complemented by observations taken between $t_{\rm{pk}}+29$ days and $t_{\rm{pk}}+90$ days, we find 
evidence for an X-ray source located at  RA=\ra{22}{23}{09.19} and Dec=$-$\dec{28}{56}{48.7} (J2000), with an
uncertainty of $3.8''$ radius ($90\%$ containment), corresponding to $1''$ from SN\,2009ip. The source is detected
at the level of $3.4\sigma$ with a PSF, vignetting and exposure corrected count-rate 
of $(3.0\pm0.9)\times 10^{-4}$ cps (total exposure of 110 ks, 0.3-10 keV energy band). The field
is represented in Fig. \ref{Fig:Xrayimage}, left panel. This source
contaminates the reported SN\,2009ip flux at the level of $\sim 1.6 \times10^{-4}$ cps.
Adopting the count-to-flux conversion factor above, this translates into a contaminating
unabsorbed flux of $\sim6 \times 10^{-15}\rm{erg\,s^{-1}cm^{-2}}$ 
(luminosity of $\sim 5 \times 10^{38}\rm{erg\,s^{-1}}$ at the distance of SN\,2009ip), representing 
$\sim 10\%$ the X-ray luminosity of SN\,2009ip at peak. This
source does not dominate the X-ray energy release around the peak time.

Observations obtained with the \emph{Chandra} X-ray Observatory (PI D. Pooley) on $t_{\rm{pk}}+19$ days
reveal the presence of an additional X-ray source lying $\approx 6"$ from SN\,2009ip and
brighter than SN\,2009ip at that time. SN\,2009ip is also detected 
(Pooley, private communication). Our contemporaneous \emph{Swift}-XRT observations
constrain the luminosity of the contaminating source to be $\lesssim 1.5\times 10^{39}\rm{erg\,s^{-1}}$,
$\lesssim 30\%$ the X-ray luminosity of SN\,2009ip at peak.  We conclude that the contaminating source
is not dominating the X-ray emission of SN\,2009ip around peak, \emph{if} stable.
The temporal coincidence of the peaks of the X-ray and optical emission of SN\,2009ip is 
suggestive that the detected X-ray emission is physically associated with SN\,2009ip. 
However, given the uncertain contamination, in the following we conservatively 
assume $L_{\rm{x}}\lesssim 2.5\times 10^{39}\rm{erg\,s^{-1}}$
for  the peak X-ray luminosity of SN\,2009ip.

\subsection{Hard X-ray observations: Swift-BAT}
\label{SubSec:HardXrays}

Stellar explosions embedded in an optically thick medium have been shown to produce
a collisionless shock when the shock breaks out from the progenitor environment, 
generating photons with a typical energy $\gtrsim 60$ keV (\citealt{Murase11},
\citealt{Katz11}). We constrain the hard X-ray emission from SN\,2009ip
exploiting our \emph{Swift}-BAT  (Burst Alert Telescope, \citealt{Barthelmy05}) campaign 
with observations obtained between 2012 September 4 ($t_{\rm{pk}}-29$ days)  and 2013 January 1
($t_{\rm{pk}}+90$ days) in survey mode (15-150 keV energy range). 
We analyzed the {\em Swift}-BAT survey data following standard procedures:
sky images and source rates were obtained by processing the data with the {\sc batsurvey} tool
adopting standard screening and weighting based on the position of SN\,2009ip. Following the BAT survey
mode guidelines, fluxes were derived in the four standard energy channels, 14--24, 24--50,
50--100, and 100--195~keV. We converted the source rates to energy fluxes assuming a typical
conversion factor of $(5.9\pm1.0)\times10^{-7}$~erg~cm$^{-2}$~s$^{-1}$/count~s$^{-1}$, estimated
assuming a range of different photon indices of a power--law spectrum ($\Gamma=1-3$).
In particular, analyzing the data acquired around the optical peak we find evidence 
for a marginal detection at the level of $3.5\sigma$ in the time interval $-0.8\,\rm{days}<t-t_{\rm{pk}}<+0.2 \,\rm{days}$
(corresponding to 2012 October 2.2 -- 3.2). A spectrum extracted in this time interval can be fit
by a power--law spectrum with photon index $\Gamma=1.8\pm1.0$ (90\% c.l.) leading to a flux of
$(2.6\pm 1.4)\times 10^{-10} \rm{erg\,s^{-1}cm^{-2}}$ (90\% c.l., 15-150 keV, total exposure time 
of $7.0$ ks). The simultaneity of the hard X-ray emission with the optical peak is
intriguing. However, given the limited significance of the detection and the known presence of a
non-Gaussian tail in the BAT noise fluctuations (H. Krimm, private communication), we
conservatively use $F<7\times10^{-10}\rm{erg\,s^{-1}cm^{-2}}$ ($L<8\times10^{40}\rm{erg\,s^{-1}}$) 
as the $5\sigma$ upper limit to 
the hard X-ray emission from SN\,2009ip around maximum light, as derived from the spectrum 
above.

\subsection{GeV observations: Fermi-LAT}
\label{SubSec:GeVObs}

GeV photons are expected to arise when the SN shock collides with a dense
circumstellar shell of material, almost
simultaneous with the optical light-curve peak (\citealt{Murase11}, \citealt{Katz11}).
We searched for high-energy $\gamma$-ray emission from SN\,2009ip using all-sky survey 
data from the \textit{Fermi} Large Area Telescope \citep[LAT;][]{Atwood09}, starting from
2012 September 3 ($t_{\rm{pk}}-30$ days) until 2012 October 31 ($t_{\rm{pk}}+28$ days).
We use events between 100\,MeV and 10\,GeV from the \texttt{P7SOURCE\_V6} 
data class \citep{Ackermann12}, which is well suited for point-source analysis.
Contamination from $\gamma$-rays produced by cosmic-ray interactions with the 
Earth's atmosphere is reduced by selecting events arriving at LAT within 100\textdegree\,
of the zenith. Each interval is analyzed using a Region Of Interest (ROI) of 12\textdegree\, radius
centered on the position of the source.
In each time window, we performed a spectral analysis using the unbinned maximum likelihood algorithm \texttt{gtlike}.
The background is modeled with two templates for diffuse $\gamma$-ray background emission: a Galactic component produced by the interaction of cosmic rays with the gas and interstellar radiation fields of the Milky Way, and an isotropic component that includes both the contribution of the extragalactic diffuse emission and  the residual charged-particle backgrounds.\footnote{The models used for this analysis, \texttt{gal\_2yearp7v6\_v0.fits} and \texttt{iso\_p7v6source.txt}, are available from the {\it Fermi} Science Support Center, http://fermi.gsfc.nasa.gov/ssc/. This analysis
uses the \textit{Fermi}-LAT Science Tools, v. 09-28-00.}
We fix the normalization of the Galactic component but leave the normalization of the isotropic background as a free parameter.
We also include the bright source 2FGL J2158.8$-$3013, located at approximately  $5\fdg48$  from the location of SN\,2009ip, and we fixed its parameters according to the values reported in \citet{Nolan12}.

We find no significant emission at the position of SN\,2009ip. Assuming a simple power-law
spectrum with photon index $\Gamma=2$, the typical flux upper limits  in 1-day intervals are 
$\lesssim (1-3) \times 10^{-10}$ ergs cm$^{-2}$ s$^{-1}$ (100 MeV  -- 10 GeV  energy range,  95\% c.l.).
Integrating around the time of the optical peak ($-2\,\rm{days}<t-t_{\rm{pk}}<+4\,\rm{days}$ ) we find 
$F<[2.1, 1.9, 3.6]\times 10^{-11}$ ergs cm$^{-2}$ s$^{-1}$ for three energy bands  
(100 MeV--464 MeV, 464 MeV--2.1 GeV and 2.1 GeV--10 GeV).\footnote{We note the presence of a 
 data gap between $t_{\rm{pk}}-9\,\rm{days}$ and $t_{\rm{pk}}-2\,\rm{days}$ due to target-of-opportunity 
 observations by {\it Fermi} during that time.}
\section{Evolution of the continuum from the UV to the NIR}
\label{Sec:SED}

\begin{figure*}
\vskip -0.0 true cm
\centering
\includegraphics[scale=0.85]{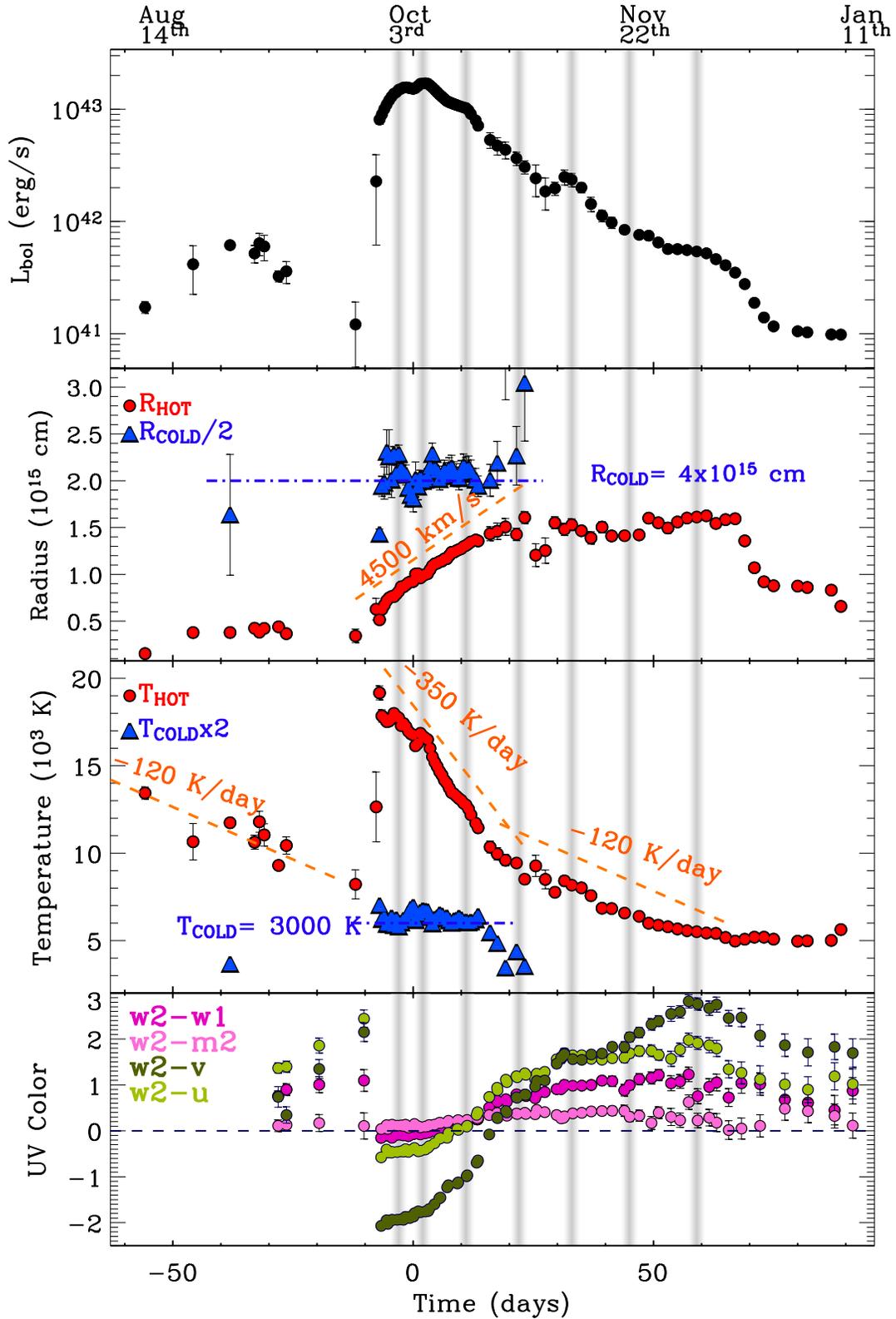}
\caption{\emph{Upper panel:} Bolometric light-curve of SN\,2009ip calculated from the best-fitting 
black-body temperatures and radii displayed in the intermediate panels. 
\emph{Lower panel:} UV color evolution with time. The onset of the 2012b explosion corresponds
to a sudden change in UV colors. After that, the  UV colors become progressively redder.
In this plot, v and u refers to the optical photometry in the UVOT system. Vertical shaded bands mark the
time of observed bumps in the photometry of Fig. \ref{Fig:OptPhot}: some are powerful enough to be
clearly visible  in the bolometric luminosity curve as well.}
\label{Fig:Lbol}
\end{figure*}

\begin{figure}
\vskip -0.0 true cm
\centering
\includegraphics[scale=0.6]{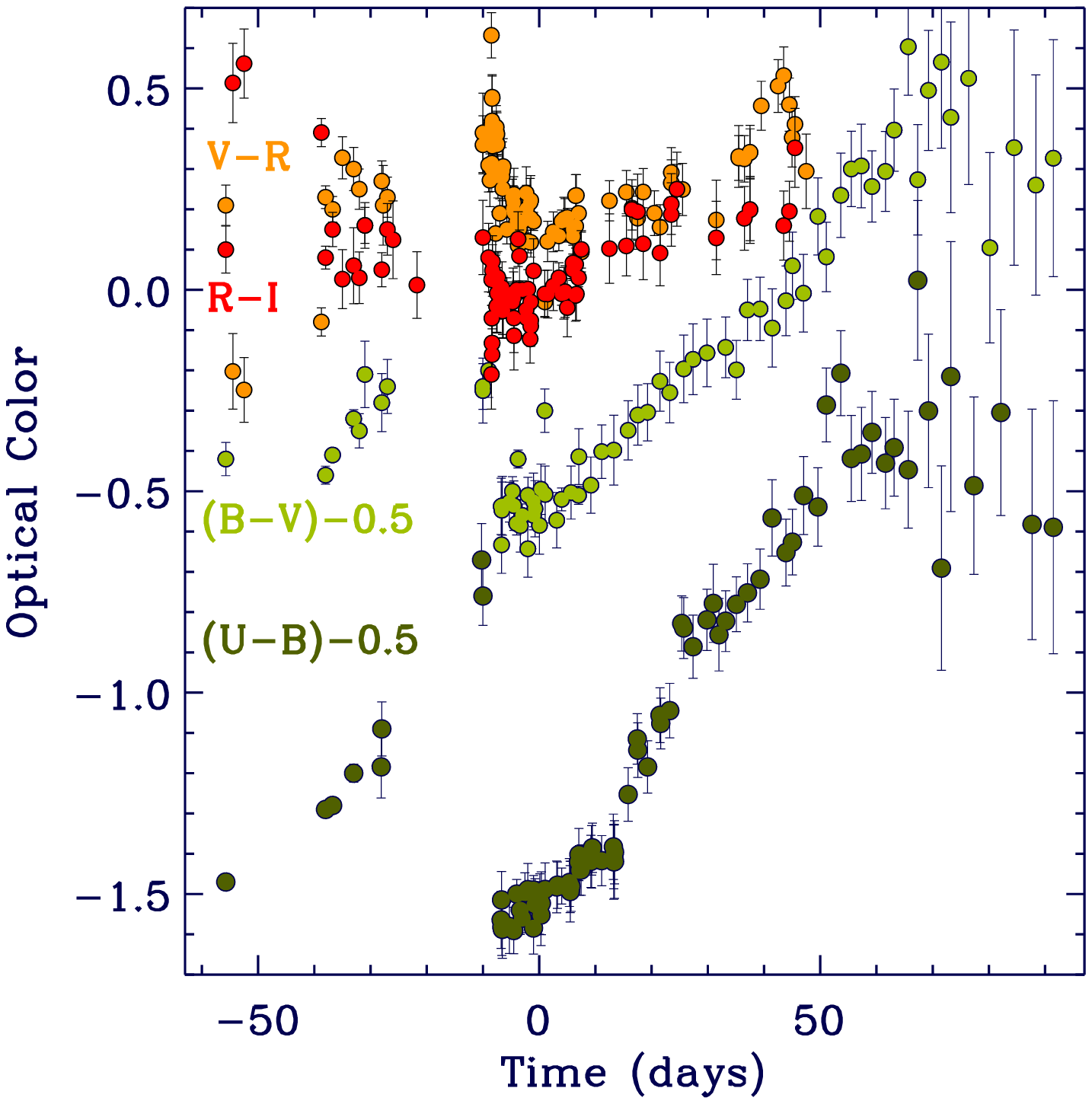}
\includegraphics[scale=0.6]{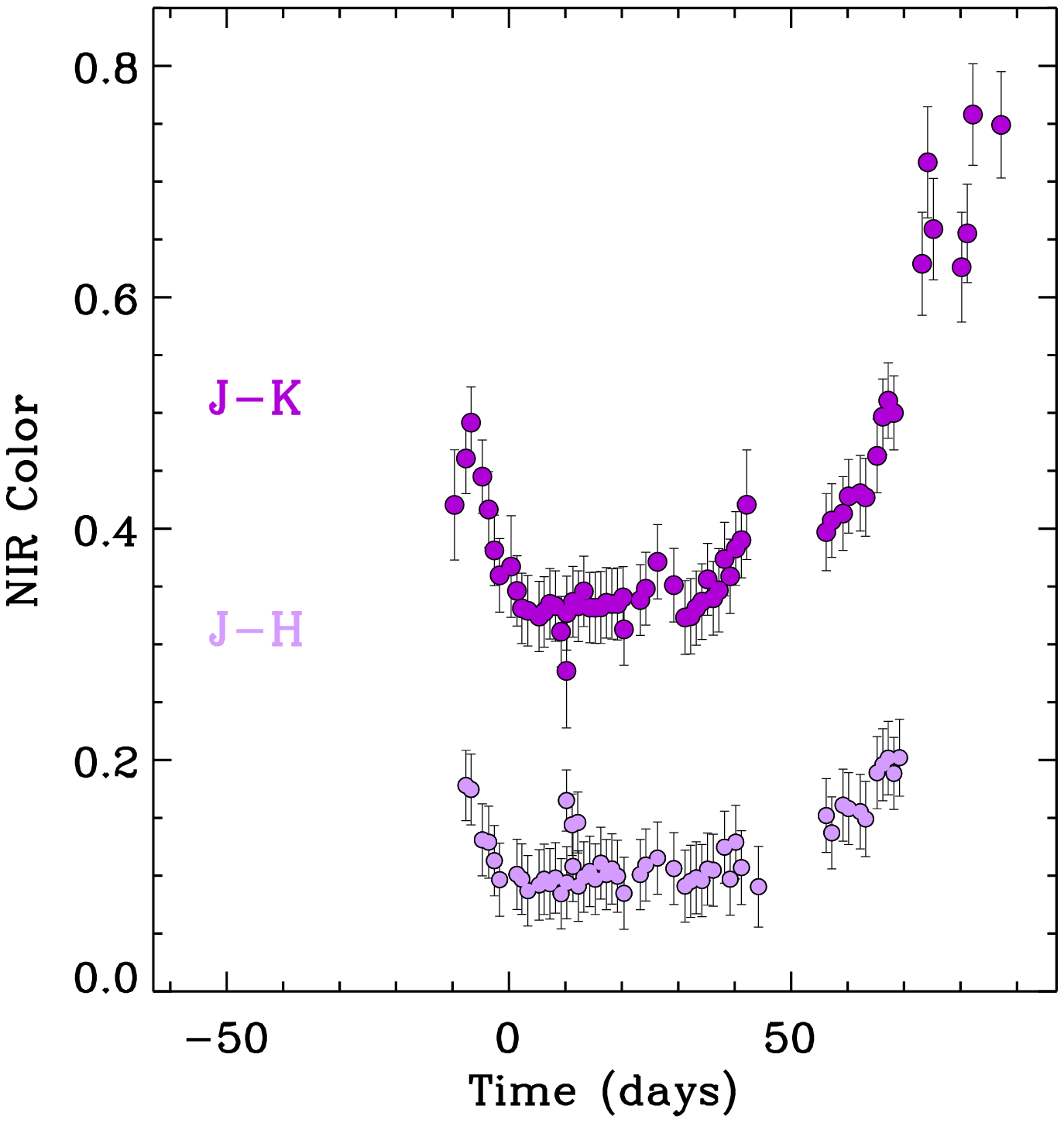}
\caption{\emph{Upper panel:} Optical colors. UVOT magnitudes have been converted into the Johnson
filters using a dynamical correction that accounts for the evolution of the color of the source.
\emph{Lower panel:} NIR colors. While SN\,2009ip clearly evolves towards
redder optical colors starting from $t_{\rm{pk}}-3$ days, no strong evolution is apparent 
in the NIR colors in the same time interval.}
\label{Fig:OptColor}
\end{figure}

\begin{figure}
\vskip -0.0 true cm
\centering
\includegraphics[scale=0.45]{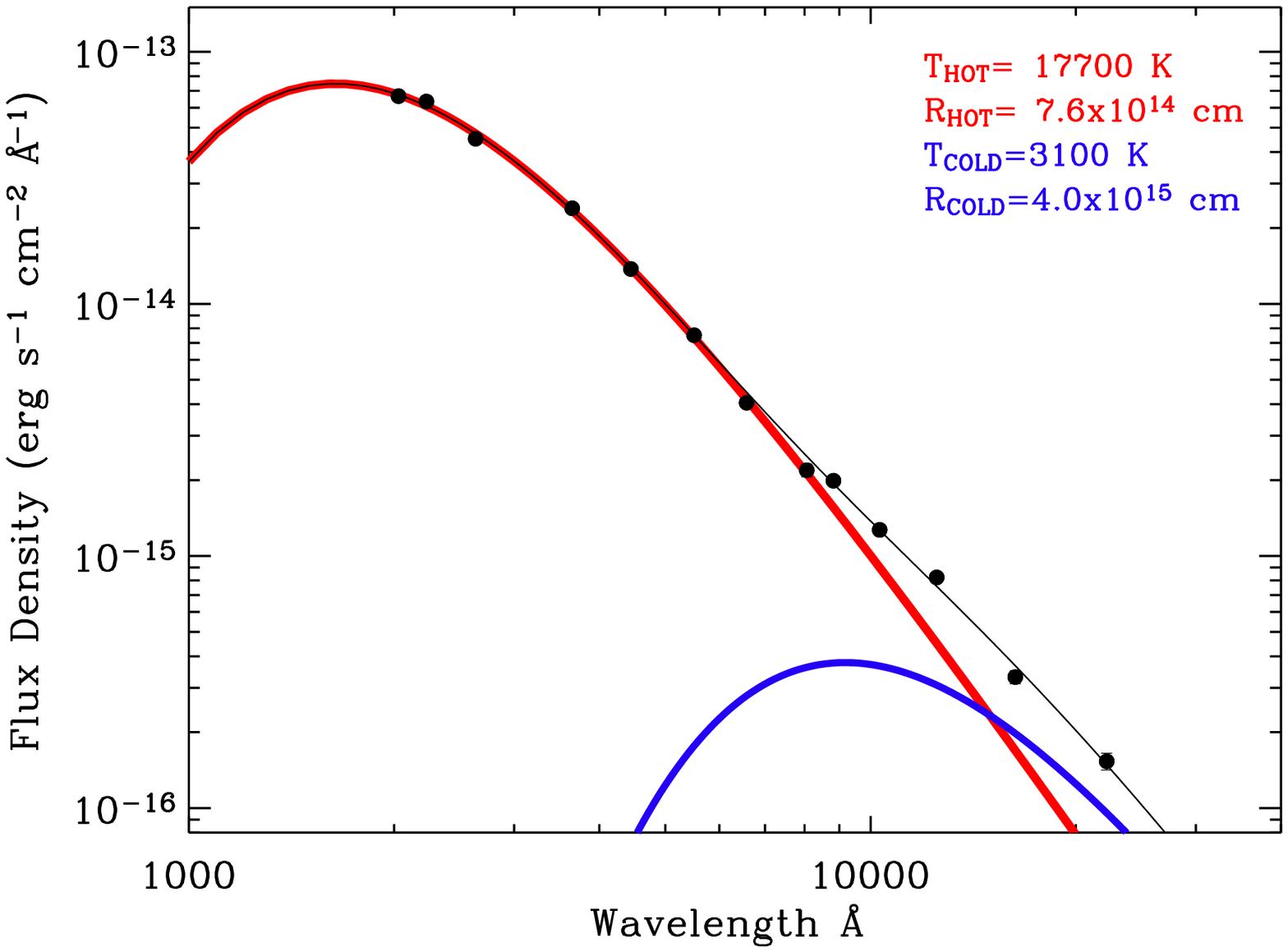}
\caption{Black solid line: best fitting SED model obtained at $t_{\rm{pk}}-4.5$ days
which clearly shows the presence of the "hot" (red line) and "cold" (blue line) 
components in the spectrum.}
\label{Fig:SED}
\end{figure}


Our 13-filter photometry 
allows us to constrain the evolution of the spectral energy distribution 
(SED) of SN\,2009ip with high accuracy. We fit a total of 84 SEDs, using data spanning from the UV to the NIR. 

The extremely blue colors and color evolution of SN\,2009ip (see Fig. \ref{Fig:Lbol}, lower panel, and
Fig. \ref{Fig:OptColor}) impose non-negligible deviations from the standard UVOT count-to-flux conversion 
factors. The filter passbands (e.g. the presence of the "red leak" in the
w2 and w1 filters) also affects the energy distribution of the detected photons for different incoming spectra.
Because of the rapidly changing spectral shape in the UV, even the ratio of intrinsic flux to observed counts 
through the m2 filter, which has no significant red leak, is strongly dependent on the spectral shape.  
We account for these effects as follows:
first, for each filter, we determine a grid of count-to-flux conversion factors at the effective UVOT filters
Vega wavelengths listed in \citealt{Poole08}, following the prescriptions by \cite{Brown10}. We assume a black-body spectrum
as indicated by our analysis of the SED of SN\,2009ip at optical wavelengths. Our grid spans the temperature
range between 2000 K and 38000 K with intervals of 200 K. We observe a variation in the conversion factor of
$90\%$, $17\%$, $7\%$  and $5\%$  in the w2, m2 , w1  and u filters as the temperature goes from 6000 K 
to 20000 K. For the v and b filters the variation is below $1\%$. As a second step we iteratively fit each
SED consisting of UVOT plus ground-based observations until the input black-body temperature assumed
to calibrate the UVOT filters matches the best-fitting temperature within uncertainties. 

For $t>t_{\rm{pk}}-7$ days the UV+BVRI  SED is well fitted by a black-body spectrum with 
a progressively larger radius ("hot" black-body component in Fig. \ref{Fig:Lbol}). 
The temperature evolution tracks the bolometric luminosity, with the photosphere becoming appreciably hotter
in correspondence with light-curve bumps and then cooling down after the peak occurred. Around
$t_{\rm{pk}}+70$ days the temperature settles to a floor around $5000$ K and remains nearly 
constant in the following 20 days. The temperature has been observed to plateau
at similar times in some SNe IIn (e.g. SN\,2005gj and SN\,1998S where the black-body temperature 
reached a floor at $\sim6000-6500$ K; \citealt{Prieto07}, \citealt{Fassia00}) and in SNe IIP as well
(e.g. SN\,1999em, with a plateau at $\sim 5000$ K; \citealt{Leonard02}).

The black body radius increases from $\sim5.1\times 10^{14}\,\rm{cm}$ to $\sim6.3\times 10^{14}\rm{cm}$
in $\sim0.3$ days (from $t=t_{\rm{pk}}-6.8$ days to $t=t_{\rm{pk}}-6.5$ days), then makes a transition to a linear evolution with 
average velocity of $\sim4000-4500\,\rm{km\,s^{-1}}$  until $t_{\rm{pk}}+20$ days, followed by a plateau around 
$R_{\rm{HOT}}=1.6\times 10^{15}\rm{cm}$. In the context of the interaction scenario 
of  Section \ref{Sec:Discussion} this change in the black-body radius evolution with time
likely marks the transition to when the interaction shell starts to become optically thin (the black-body
radius is a measure of the effective radius of emission: the shock radius obviously keeps increasing with time).
A rapid decrease in radius is observed around  $t_{\rm{pk}}+70$ days. After this time the $R_{\rm{HOT}}$ mimics 
the temporal evolution of the bolometric light-curve (see Fig. \ref{Fig:Lbol}). In SNe dominated by interaction 
with pre-existing material, the black-body radius typically increases steadily with time, reaches a peak and 
then smoothly transitions to a decrease (see e.g. SN\,1998S, \citealt{Fassia00}; SN\,2005gj, \citealt{Prieto07}).
The more complex behavior we observe for SN\,2009ip likely results from 
a more complex structure of the immediate progenitor environment (Section \ref{Sec:Discussion}).

Starting from $t_{\rm{pk}}+16$ days, the best-fitting black body model tends to over-predict the 
observed flux in the UV, an effect likely due to increasing line-blanketing. As the temperature
goes below $\sim10^{4}$ K, the recombination of the ejecta 
induces a progressive strengthening of metal-line blanketing which is 
responsible for partially blocking the UV light.
We account for line-blanketing by restricting our fits to the UBVRI flux densities for $t>t_{\rm{pk}}+16$ days. 
Our fits still indicate a rapidly decreasing temperature with time.
We conclude that the rapid drop in UV light observed starting from $t_{\rm{pk}}+12$ days mainly results from
the cooling of the  photosphere. 

Starting around $t_{\rm{pk}}+59$ days the UV emission fades more slowly and we observe a change 
in the evolution of the UV colors: from red to blue (Fig. \ref{Fig:Lbol}, lower panel).
The same evolution is observed in the (U-B) color of Fig. \ref{Fig:OptColor}.
This can also be seen from Fig. \ref{Fig:OptPhot}, where the NIR emission displays a more 
rapid decay than the UV.  This manifests as an excess of UV emission with respect to the 
black-body fit.\footnote{This is especially true in the case of the UVOT m2 filter, which does
\emph{not} suffer from the "red leak".} 
After $t_{\rm{pk}}+67$ days a pure black-body spectral shape 
provides a poor representation of the UV to NIR SED.

We furthermore find clear evidence for excess of 
NIR emission with respect to the hot black body (see Fig. \ref{Fig:SED}) as we first  reported in 
\cite{Gall12}, based on the analysis of the VLT/X-shooter spectra (Fig. \ref{Fig:Xshooter1} and \ref{Fig:Xshooter2}). 
Modeling the NIR excess with an 
additional black-body component, we obtain the radius and temperature evolution displayed 
in Fig. \ref{Fig:Lbol} ("cold" black body). The cold black-body radius is consistent with no
evolution after $t_{\rm{pk}}-4$ days, with $R_{\rm{COLD}}\sim4\times10^{15}\rm{cm}$. $T_{\rm{COLD}}$
is also found to be $T_{\rm{COLD}}\sim 3000$ K until $t_{\rm{pk}}+14$ days, which implies
$L_{\rm{COLD}}\approx const$ for $-4\,\rm{days}<t-t_{\rm{pk}}<+14\,\rm{days}$ (together with the almost unchanged NIR colors
of Fig. \ref{Fig:OptColor}).\footnote{This is also consistent with the almost flat K-band photometry. In this
time interval the K-band photometry is dominated by the cold component. For $t\gtrsim t_{\rm{pk}}+12$ days
$L_{\rm{\lambda,HOT}}>L_{\rm{\lambda,COLD}}$ at $\lambda= \lambda_{\rm{K}}$, so that the K band flux
starts to more closely follow the temporal evolution seen at bluer wavelengths.}
Starting from $t_{\rm{pk}}+16$ days $T_{\rm{COLD}}$ cools down to reach $T_{\rm{COLD}}\sim 2000$ K
on $t_{\rm{pk}}+23$ days. At this stage the hot black body with $T_{\rm{HOT}}\sim 8500$ K
completely dominates the emission at NIR wavelengths and the fit is no longer able to 
constrain the parameters of the cold component. Our NIR spectra of Fig.  \ref{Fig:NIRSpec}
clearly rule out line-emission as a source of the NIR excess.

Applying the same analysis to the 2012a outburst we find that the temperature of the
photosphere evolved from $\sim 13400$ K  (at $t_{\rm{pk}}-56$ days) to $8000$ K 
($t_{\rm{pk}}-12$ days), with an average decay of $\sim 120\, \rm{K/day}$.
Our modeling shows a slightly suppressed UV flux which we 
interpret as originating from metal line-blanketing.
Notably, the SED at $t_{\rm{pk}}-38$ days (when we have almost 
contemporaneous coverage in the UBVRI and JHK bands) shows evidence for a NIR excess
corresponding to $T_{\rm{COLD}}\sim 2000$ K at the radius consistent with 
$R_{\rm{COLD}}\sim4\times10^{15}\rm{cm}$ (as found for the NIR excess during the 2012b explosion).

Finally, we use the SED best-fitting models above to compute the bolometric luminosity of SN\,2009ip.
Displayed in Fig. \ref{Fig:Lbol} is the contribution of the "hot" black body. The "cold" black-body
contribution is marginal, being  always $(2-4)\%$ the luminosity of the "hot" component.
\section{Spectral changes at UV/optical/NIR frequencies}
\label{Sec:SpecLinesOpt}

\begin{figure*}
\vskip -0.0 true cm
\centering
\includegraphics[scale=0.55]{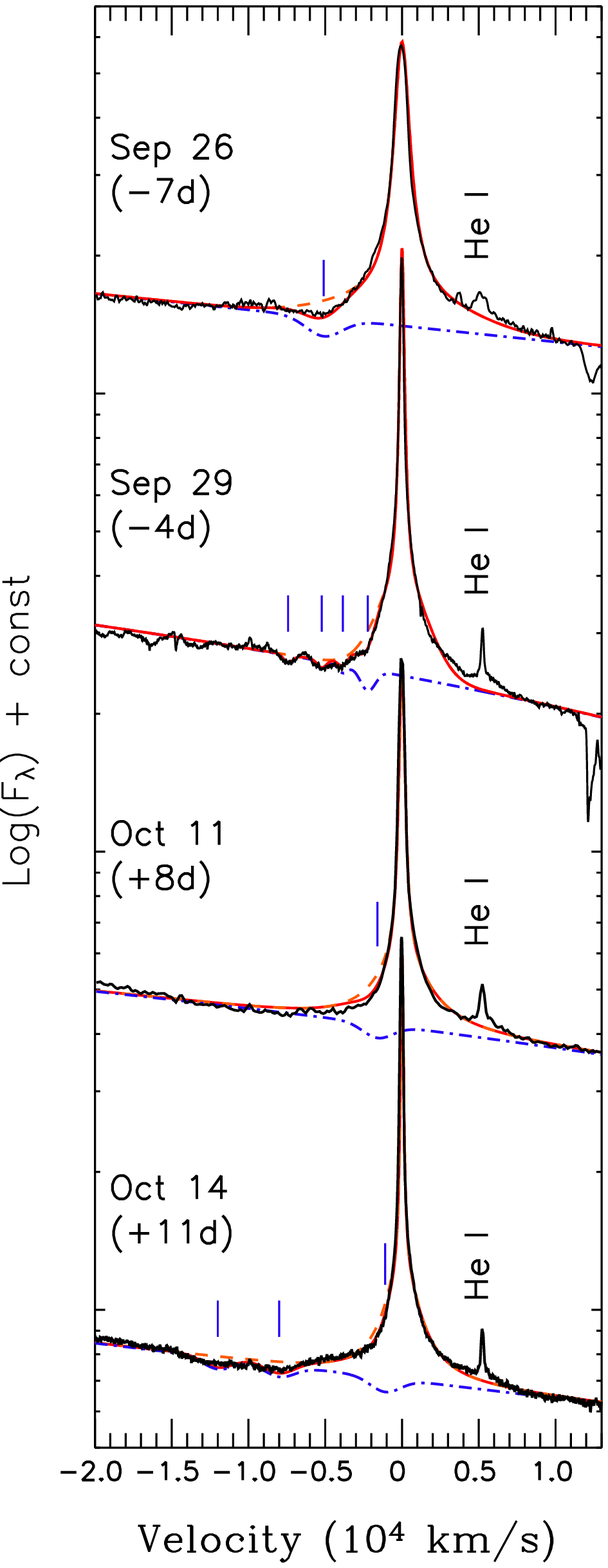}
\includegraphics[scale=0.55]{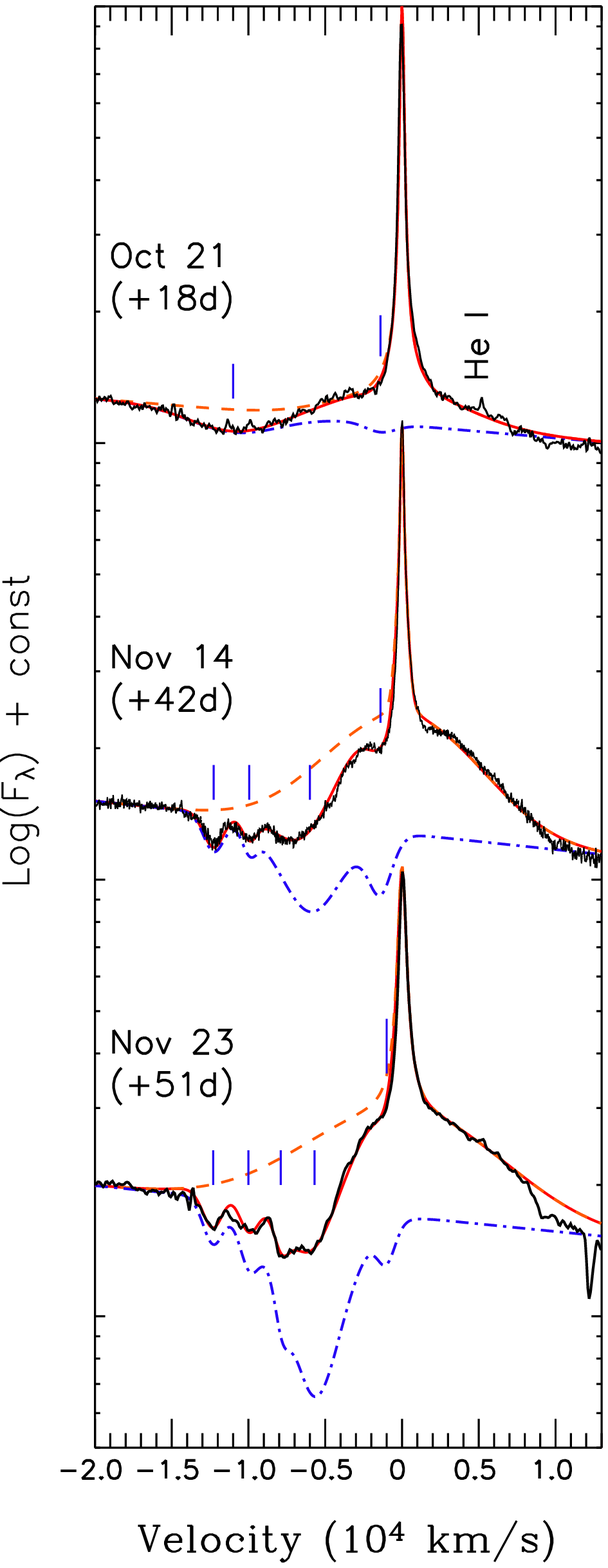}
\includegraphics[scale=0.55]{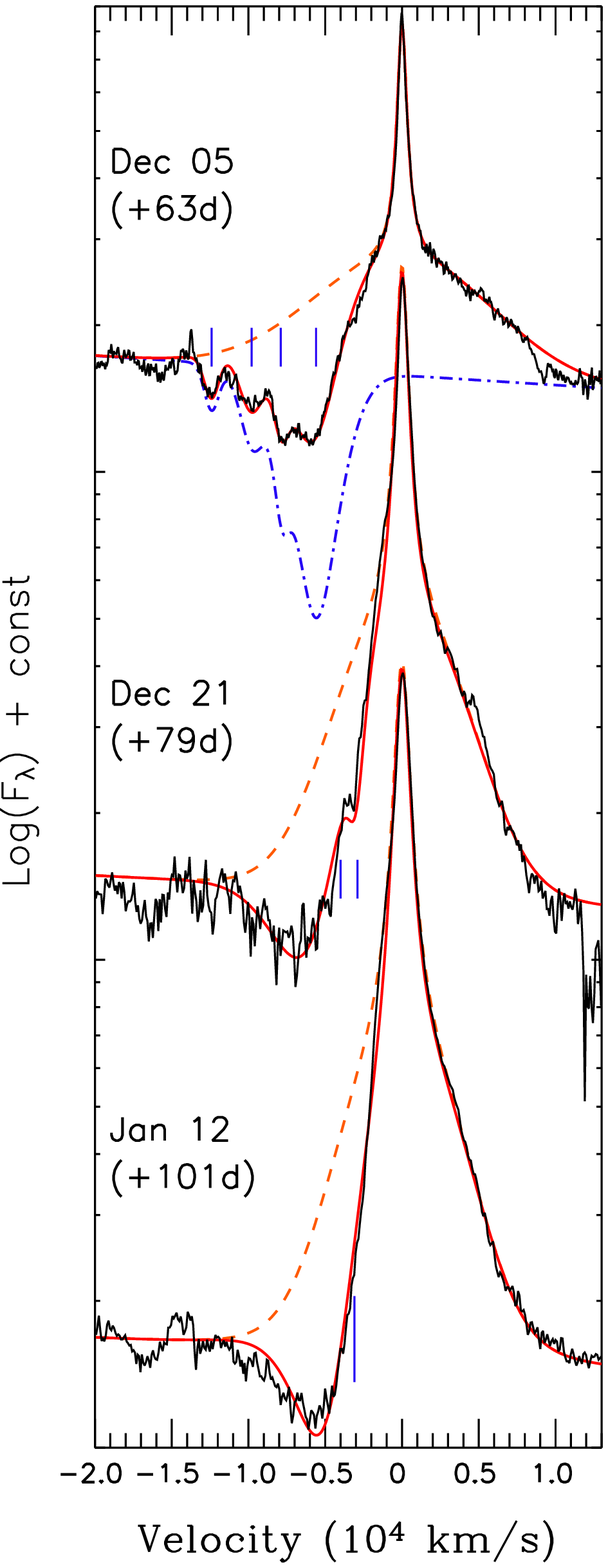}
\caption{Evolution of the H$\alpha$ line profile with time. 
Orange dashed line: emission components.  Blue dot-dashed line: absorption components.
Red thick line: composite line profile. The vertical blue lines mark the velocity of the absorption
components.}
\label{Fig:Halpha}
\end{figure*}

\begin{figure*}
\vskip -0.0 true cm
\centering
\includegraphics[scale=0.45]{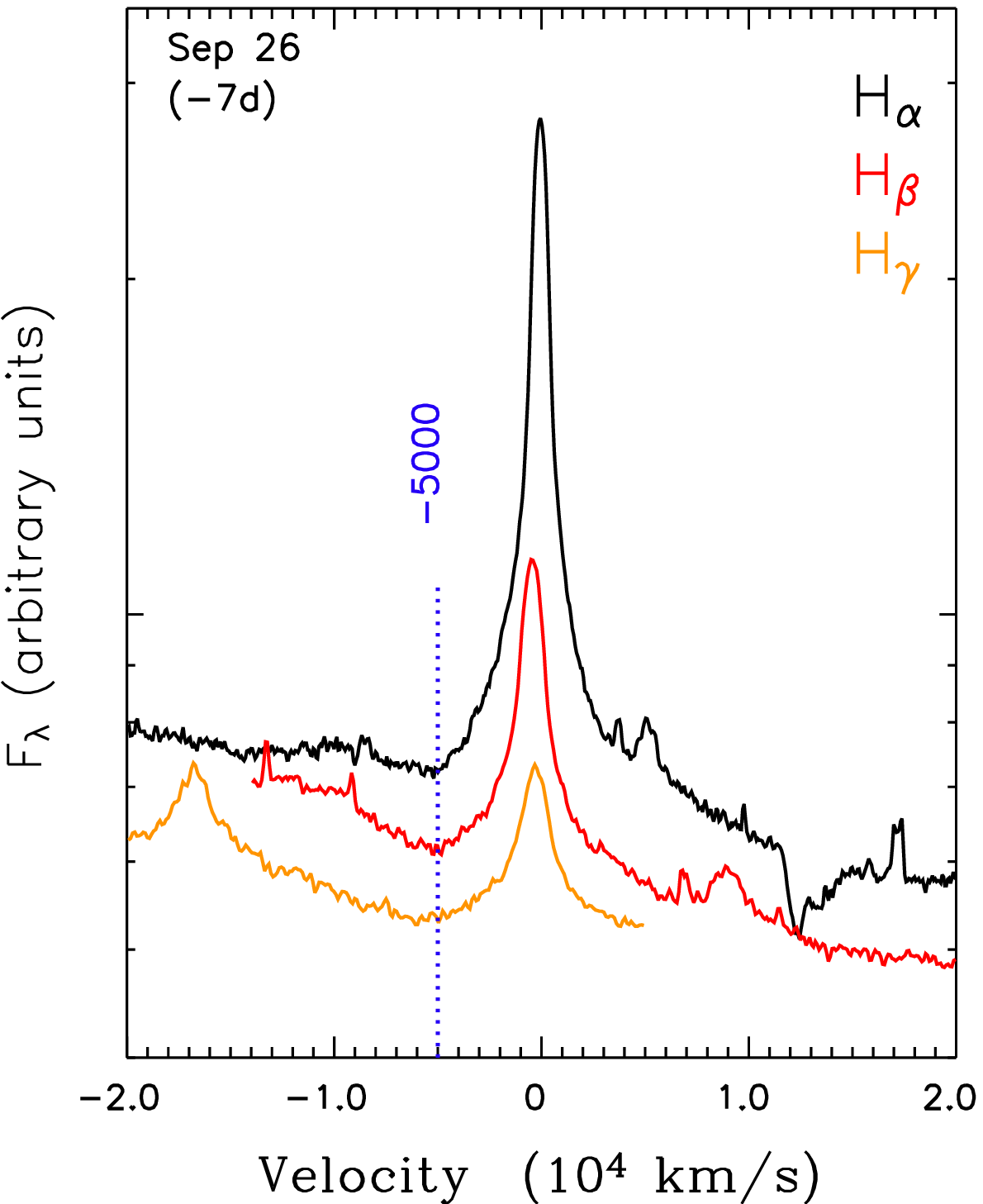}
\includegraphics[scale=0.45]{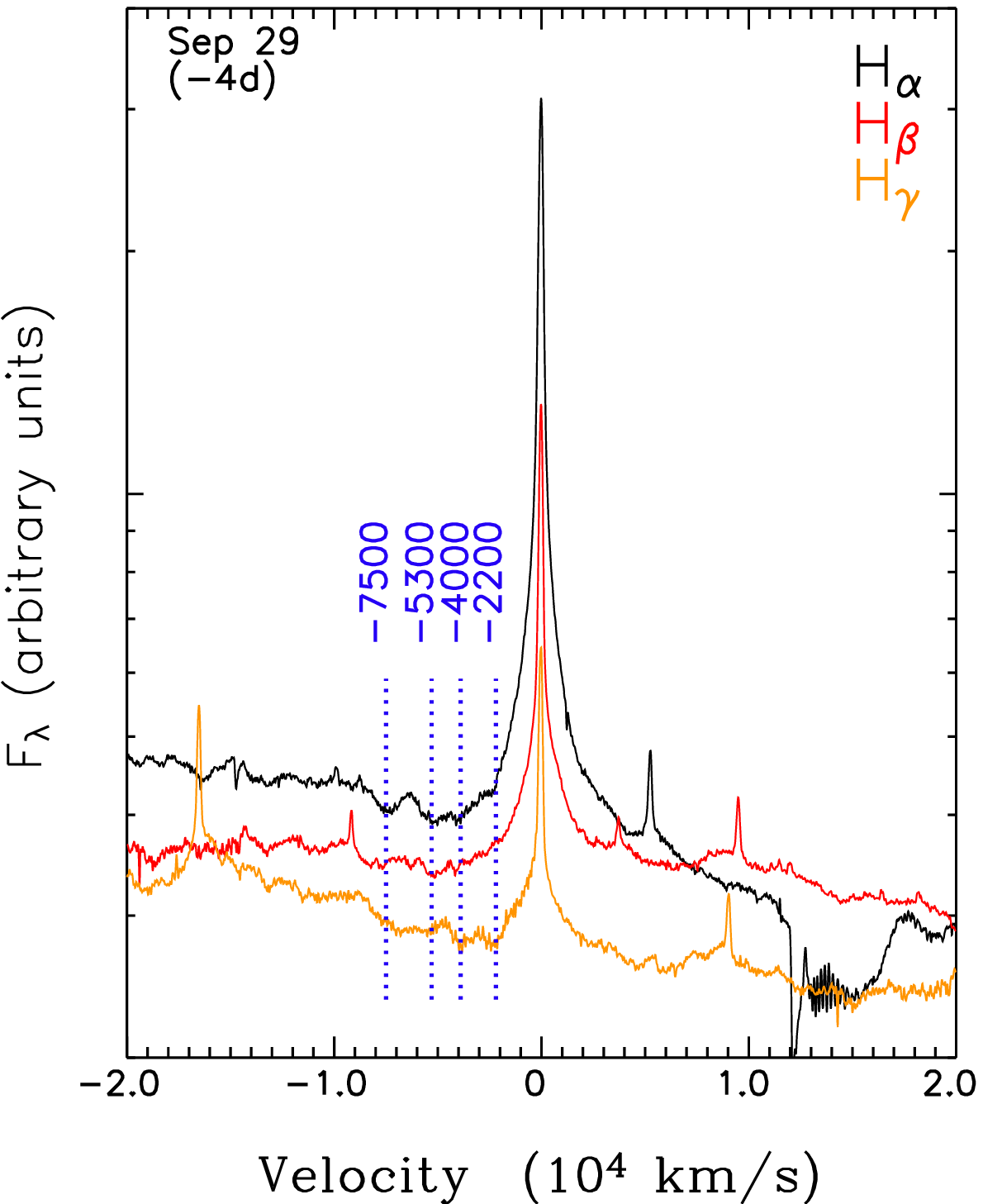}
\includegraphics[scale=0.45]{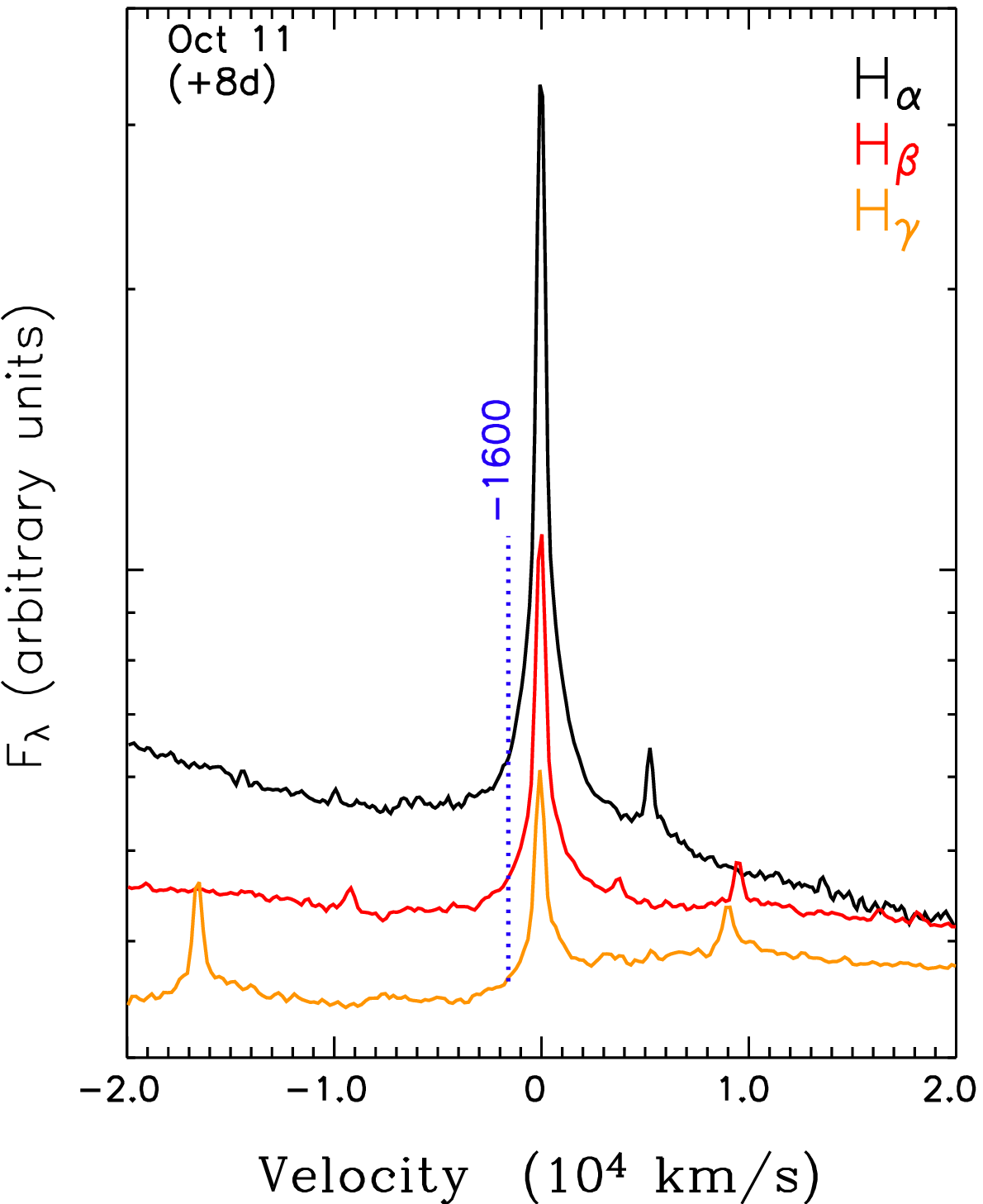}\\
\includegraphics[scale=0.45]{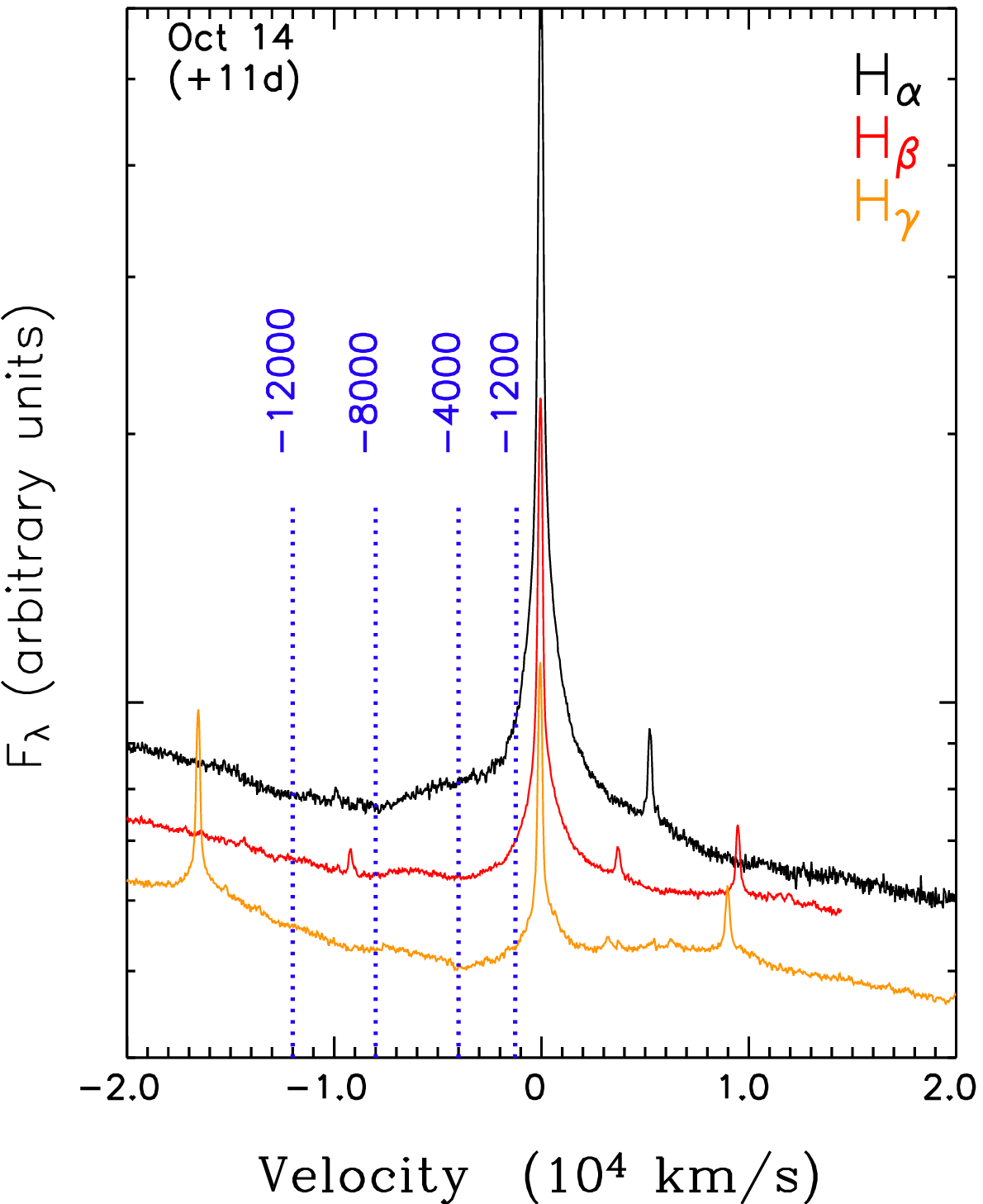} 
\includegraphics[scale=0.45]{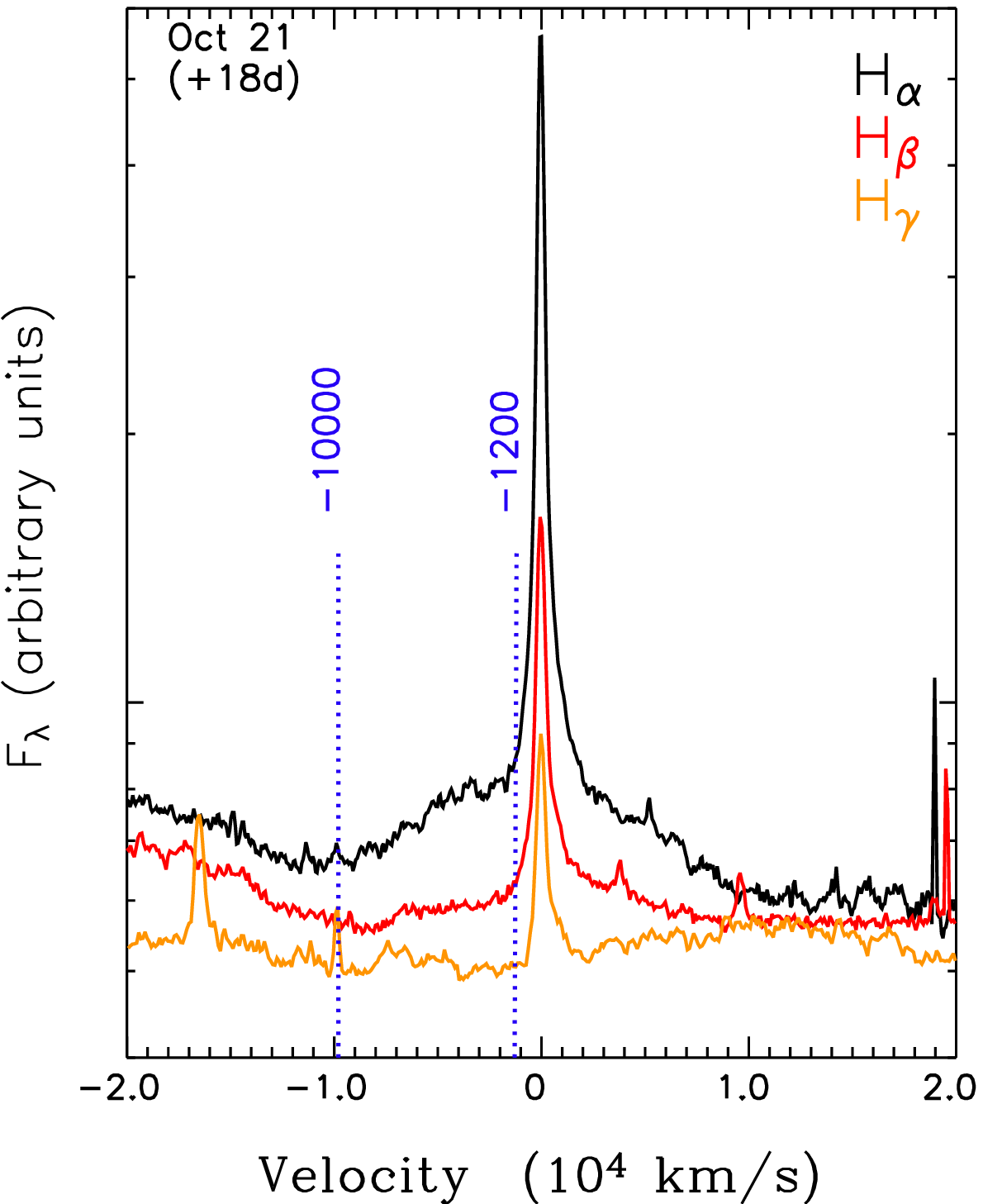}
\includegraphics[scale=0.45]{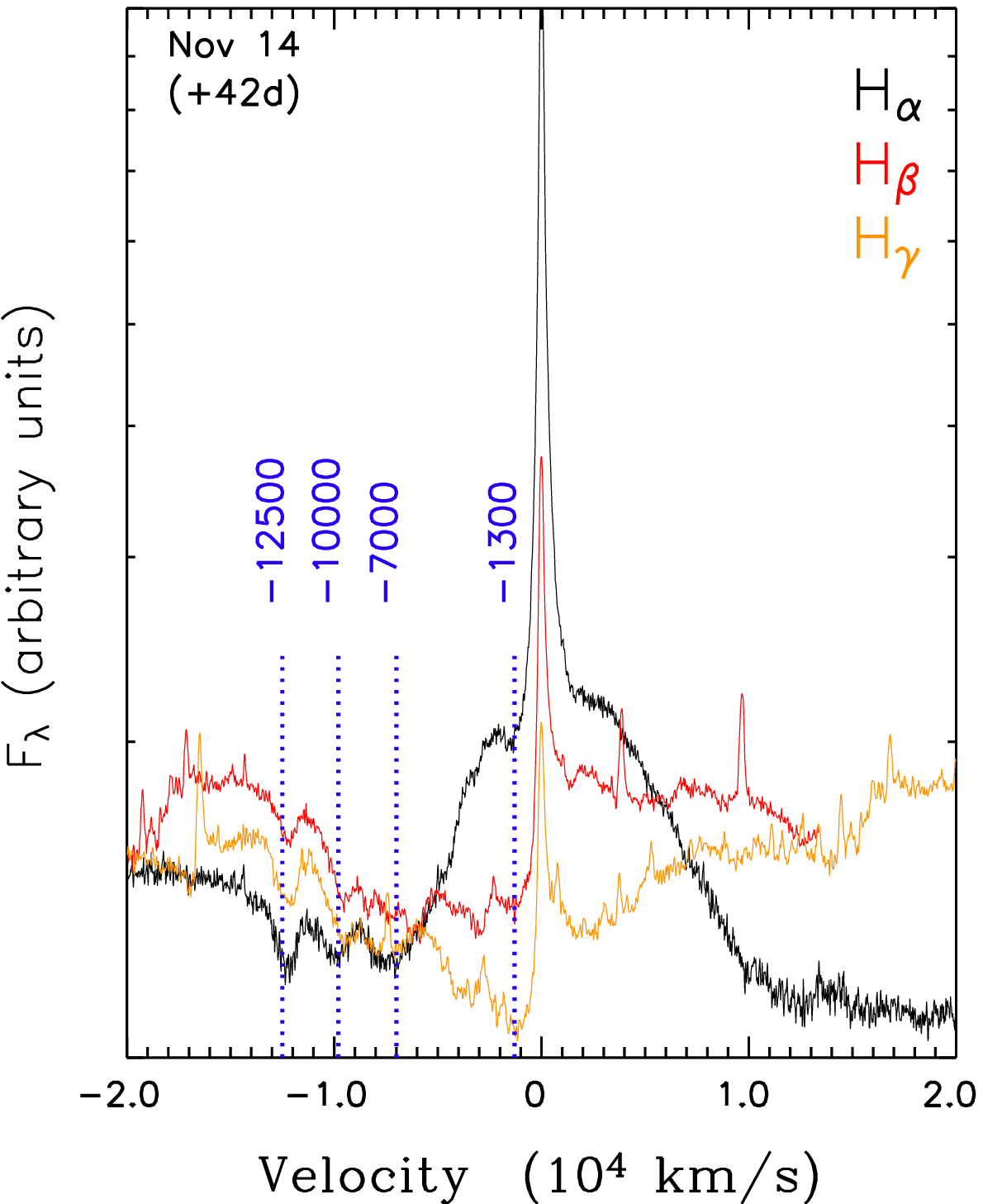}\\
\includegraphics[scale=0.45]{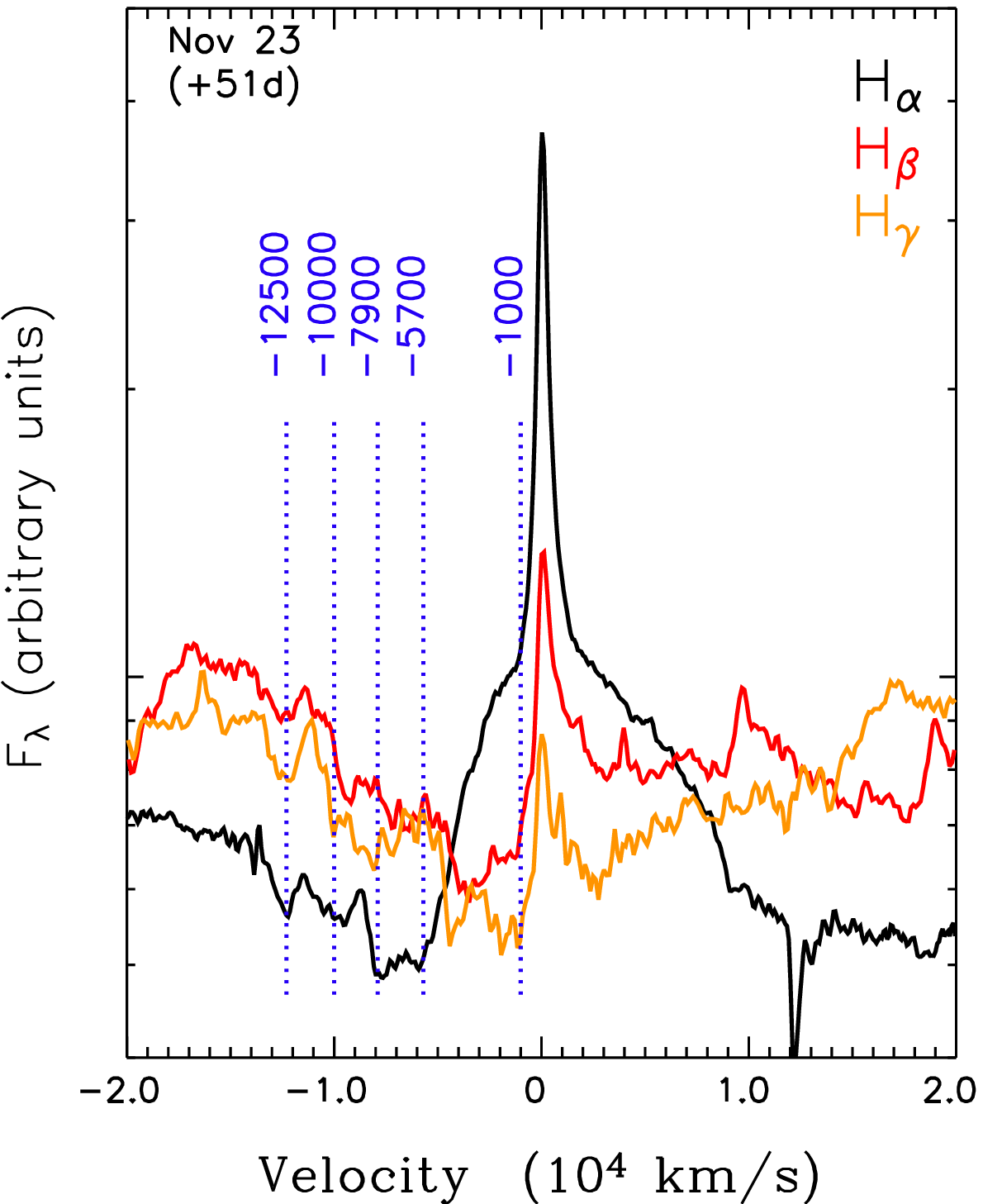}
\includegraphics[scale=0.45]{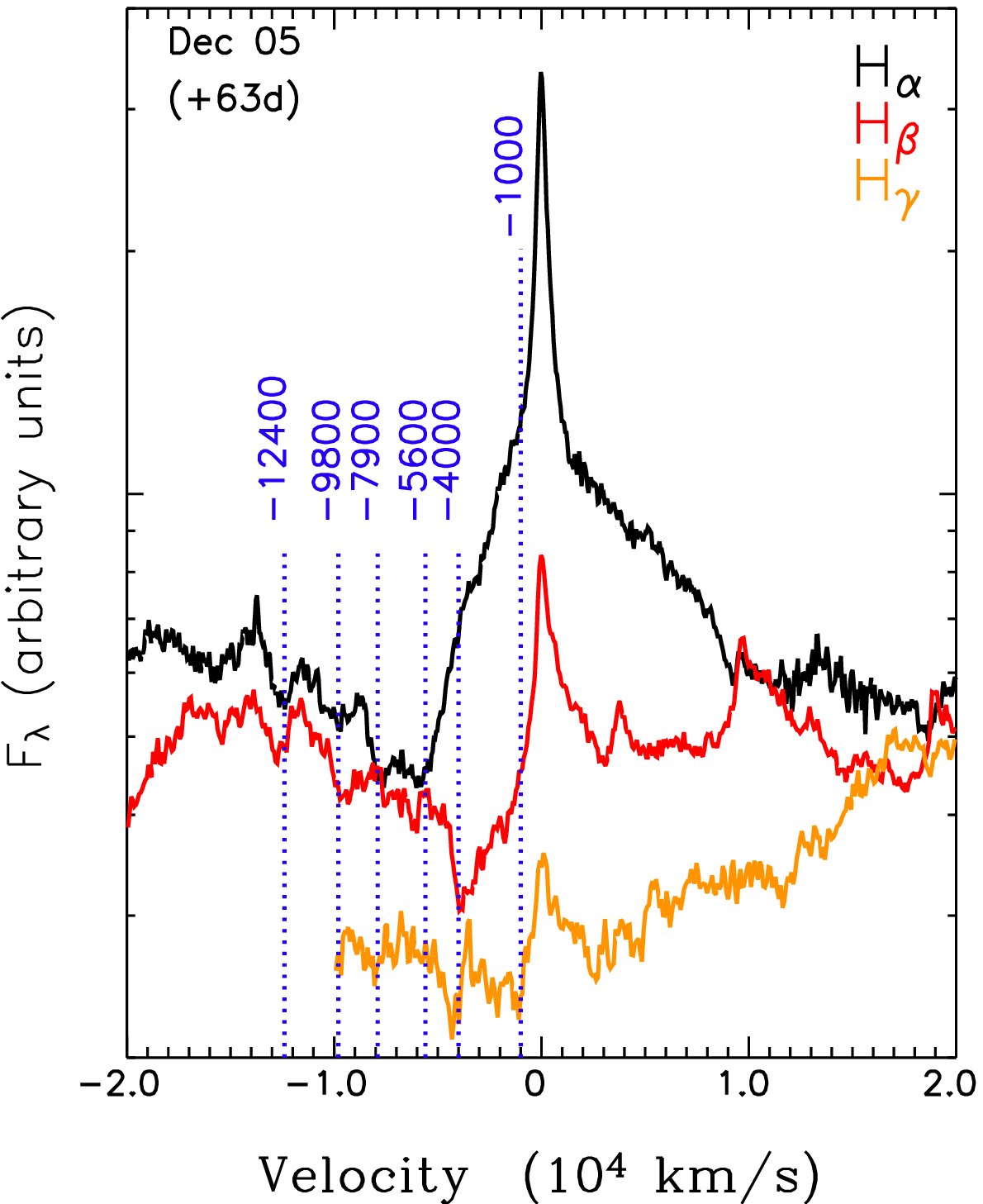}
\includegraphics[scale=0.45]{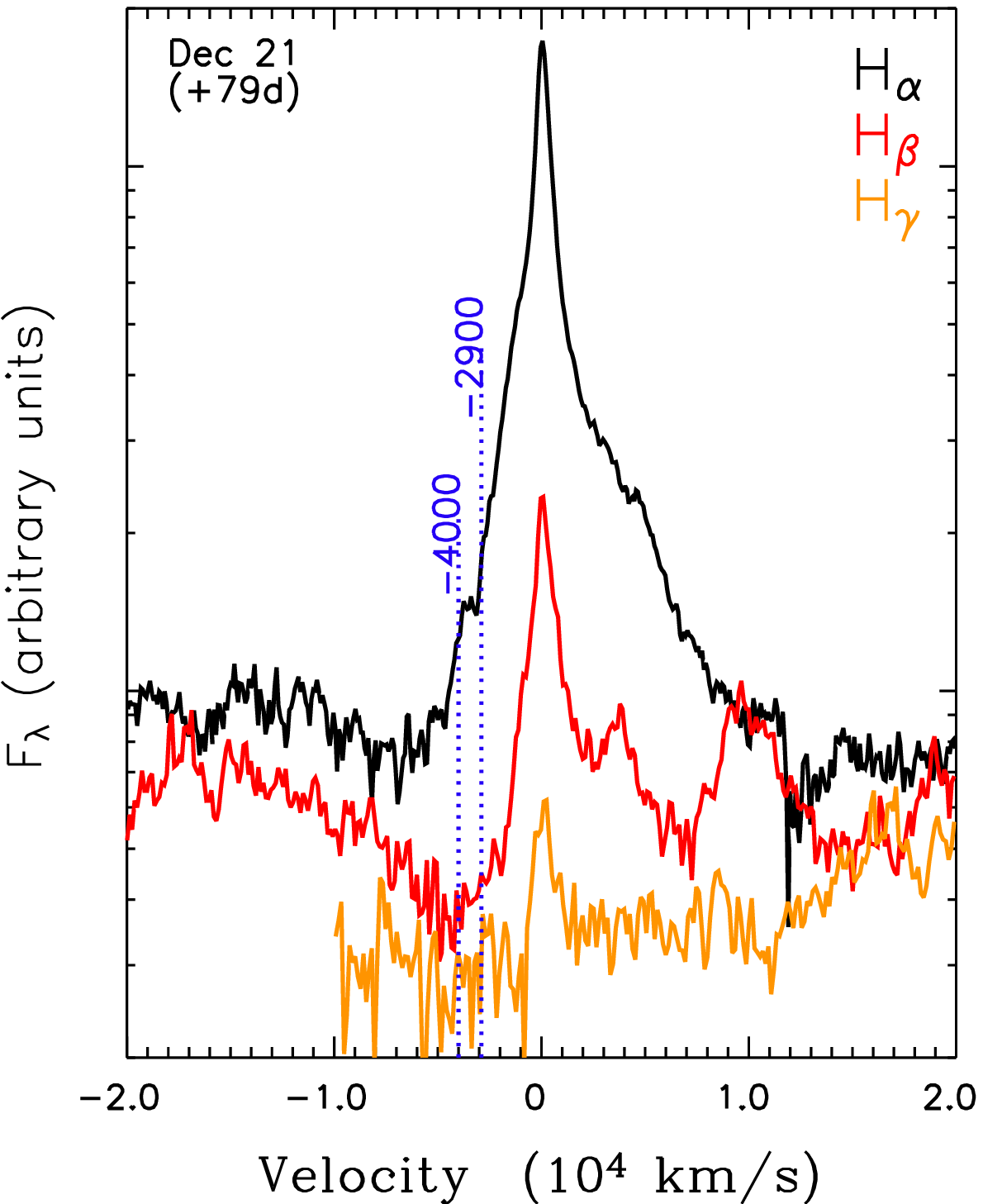}
\caption{H$\alpha$, H$\beta$ and $H\gamma$ line profiles of SN\,2009ip
at representative epochs. The blue dotted lines mark the velocity of the
major absorption components identified by our fits of the H$\alpha$ line 
profile. For 2012 December 5 we also  added two absorption components at
$ -1000\,\rm{km\,s^{-1}}$ and  $-4000\,\rm{km\,s^{-1}}$ identified
in the H$\beta$ and H$\gamma$ lines. The late-time spectrum acquired on 
$t_{\rm{pk}}+101$ days (2013 January 12) shows limited evolution in the 
H$\alpha$ profile with respect to the previous epoch and it is not shown here. }
\label{Fig:HBalmer}
\end{figure*}

\begin{figure}
\vskip -0.0 true cm
\centering
\includegraphics[scale=0.75]{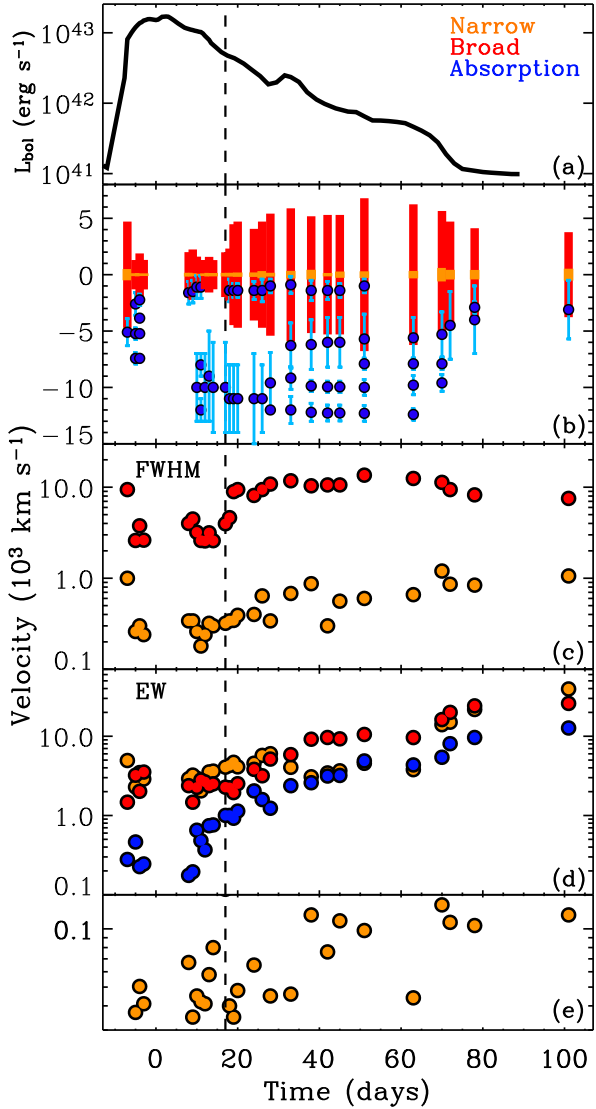}
\caption{Evolution of the H$\alpha$ line with time. We model the H$\alpha$ line
with a combination of Lorentzian and Gaussian profiles. Orange markers
represent the narrow
Lorentzian profile. Red is used for the broad component, while blue is  
associated with the blue-shifted absorption components. Panel (a):
bolometric light-curve for reference. Panel (b): the red (orange) bars span the
FWHM of the broad (narrow) component. These values are also reported in panel (c).
Blue dots: absorption minima as obtained by modeling the absorption with a combination of
Gaussians. Negative values indicate blue-shifted components. We use
light-blue bars to mark the $1\sigma$ width as obtained from the fit. Panel (d) shows the
evolution of the equivalent width of the narrow (orange), broad (red) and absorption
(blue) components. The peak of the narrow component progressively shifts to
 larger redshifted velocities as illustrated in Panel (e) and independently found by 
 \cite{Fraser13}.
 The vertical dashed line marks an important time
 in the evolution of SN\,2009ip from different perspectives: from this plot it is
 clear that around this time the width of broad component undergoes 
 a remarkable transition from FWHM$\sim3000\,\rm{km\,s^{-1}}$ to  
 FWHM$\sim10000\,\rm{km\,s^{-1}}$.}
\label{Fig:Halpha_evol}
\end{figure}

\begin{figure}
\vskip -0.0 true cm
\centering
\includegraphics[scale=0.72]{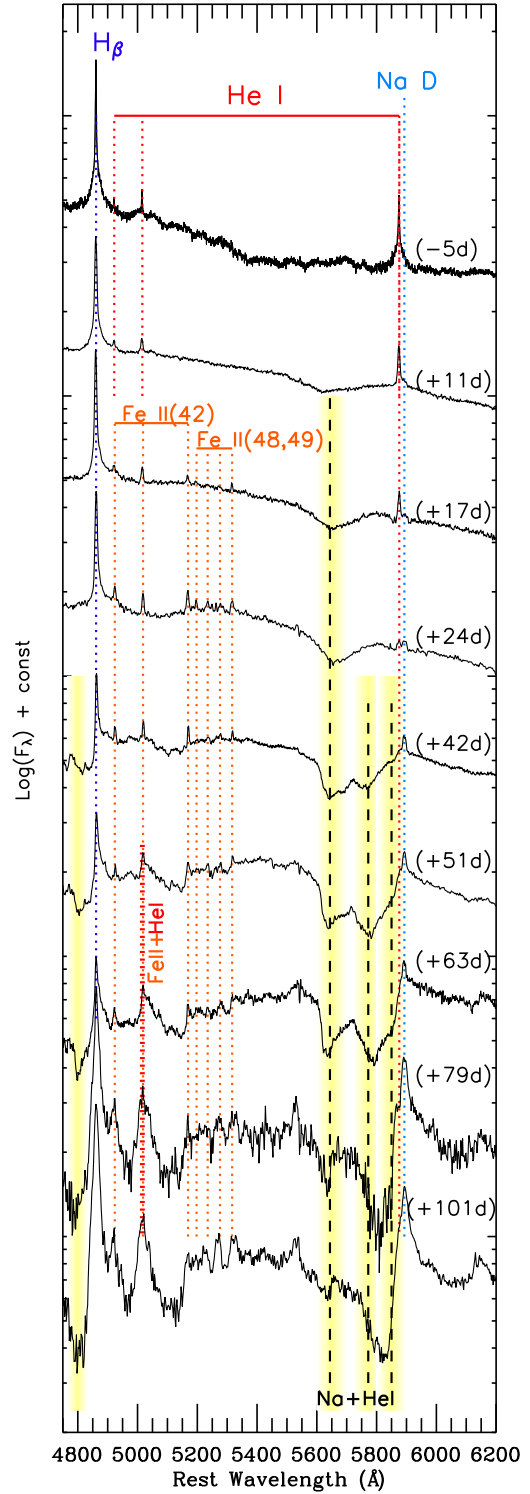}
\caption{Key spectral changes in SN\,2009ip between 4500$-$6200 \AA:
narrow \ion{He}{1} emission lines subside while  \ion{Na}{1} D emission
grows in strength. \ion{He}{1} later re-appears with the broad/intermediate component. 
\ion{Fe}{2} emission lines emerge, while
broad absorption dips develop red-wards the \ion{Na}{1} D (and \ion{He}{1}) lines, around 5650 \AA.
Starting from $\sim t_{\rm{pk}}+30$ days, additional broad absorption features
around  5770 \AA\, and  5850 \AA\, appear, associated with the  \ion{He}{1} and 
the \ion{Na}{1} D lines;  \ion{H}{1} lines
develop strong absorption on their blue wing. 
}
\label{Fig:FeNaHe}
\end{figure}

\begin{figure}
\vskip -0.0 true cm
\centering
\includegraphics[scale=0.57]{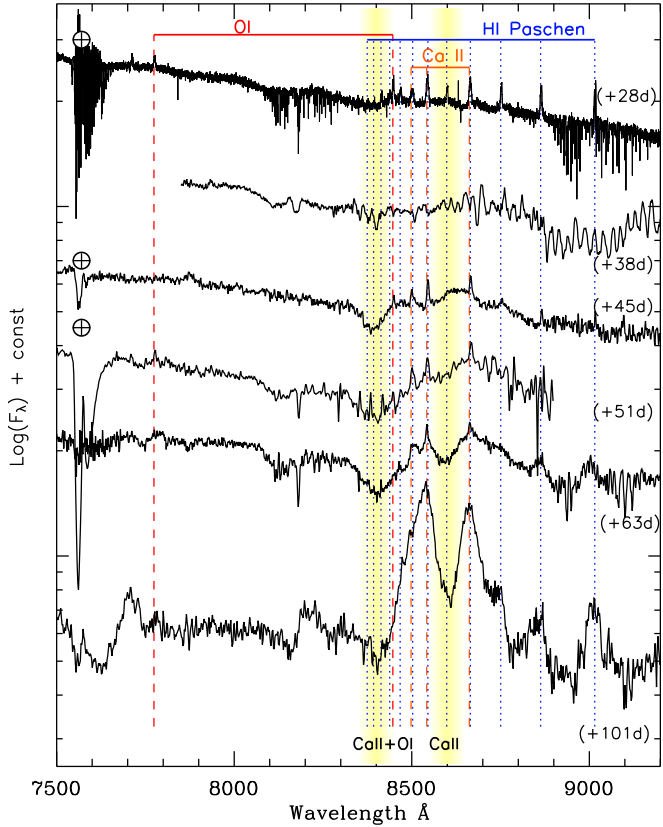}
\caption{Beginning at $t_{\rm{pk}}+30$ days SN\,2009ip develops broad emission
and absorption components between  8300 \AA\, and 9000 \AA\, we attribute to
Ca II. NIR emission from the CaII triplet is typical of IIP SNe (e.g. \citealt{Pastorello06}).}
\label{Fig:CaII}
\end{figure}

\begin{figure*}
\vskip -0.0 true cm
\centering
\includegraphics[scale=0.95]{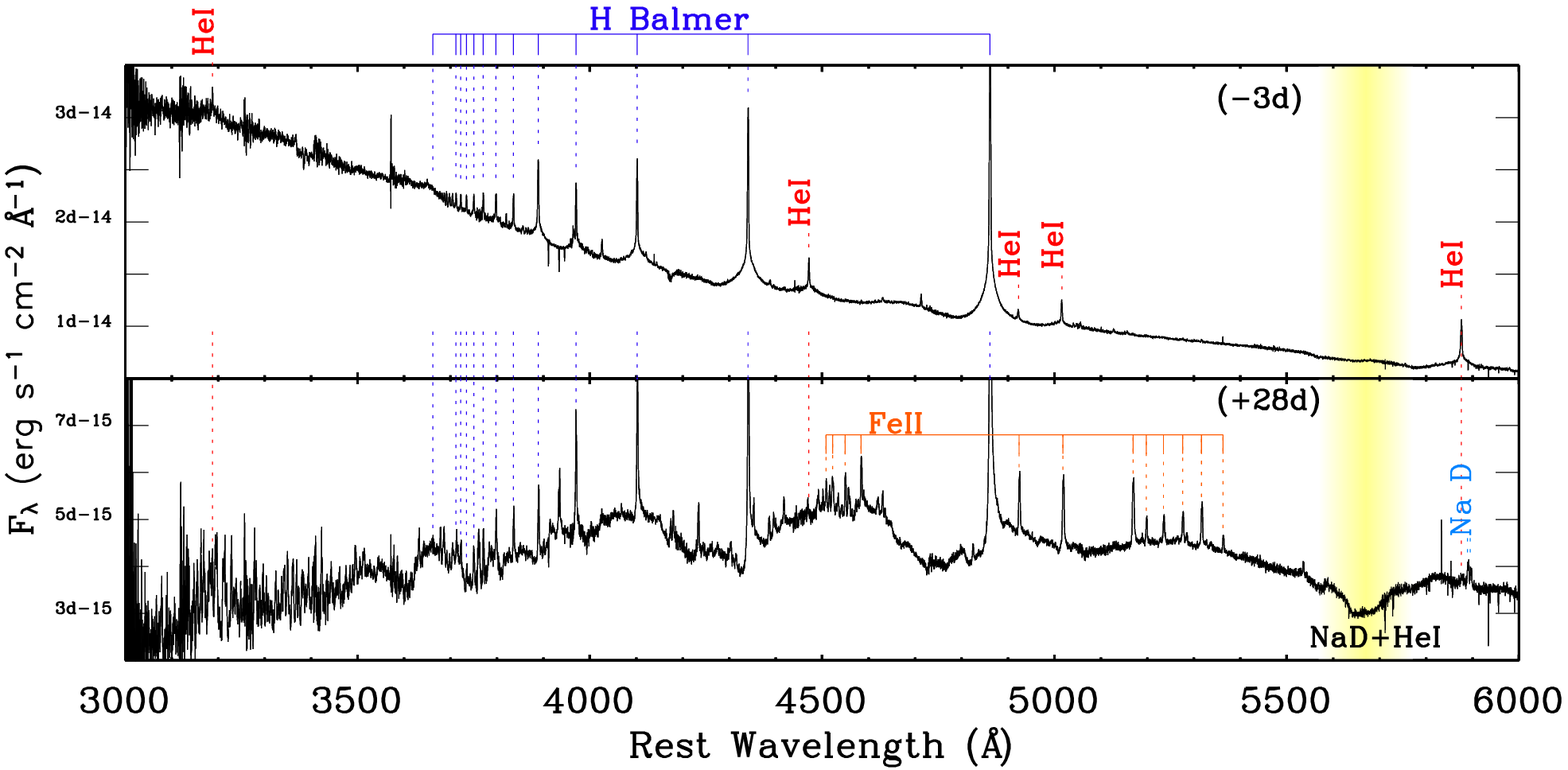}\\
\includegraphics[scale=0.95]{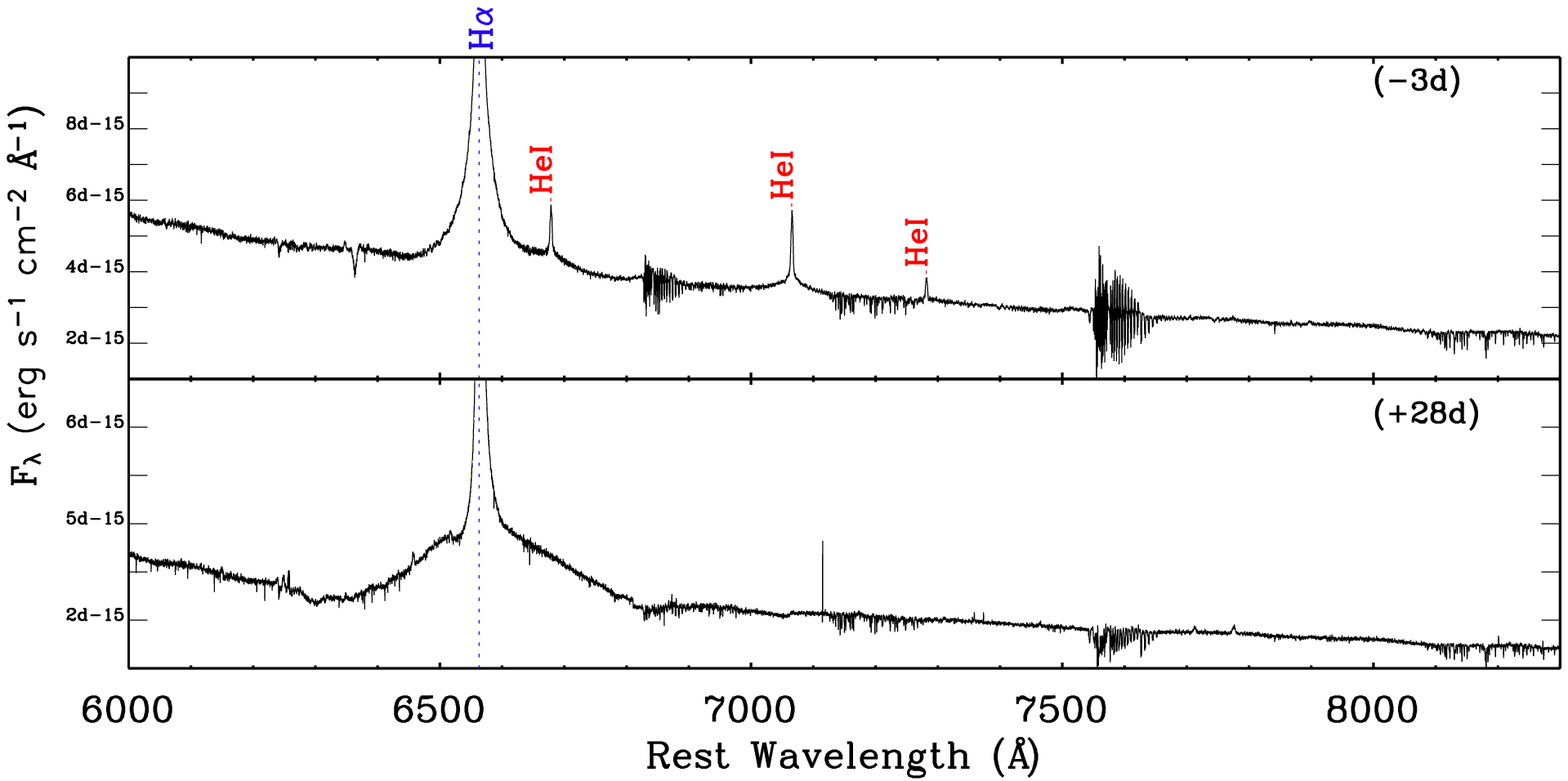}\\
\caption{High-resolution VLT/X-shooter spectra captured the evolution
of SN\,2009ip in fine detail. \emph{Upper panel:} around the optical peak,
on 2012 September 30 ($t_{\rm{pk}}-3$ days), SN\,2009ip shows a narrow-line dominated spectrum 
typical of SNe (and LBVs) interacting with a medium.  
One month later (\emph{lower panel} of each plot) SN\,2009ip started to develop
broad emission components (see in particular the H$\alpha$ line) and deep absorption features (e.g.
the yellow-shaded band around $5650\,$\AA\,) more typical of SNe IIP. The 
complementary $10000-24500$ \AA\, wavelength range is shown in Fig. \ref{Fig:Xshooter2}.
Data have been corrected for Galactic extinction.}
\label{Fig:Xshooter1}
\end{figure*}

\begin{figure*}
\vskip -0.0 true cm
\centering
\includegraphics[scale=0.95]{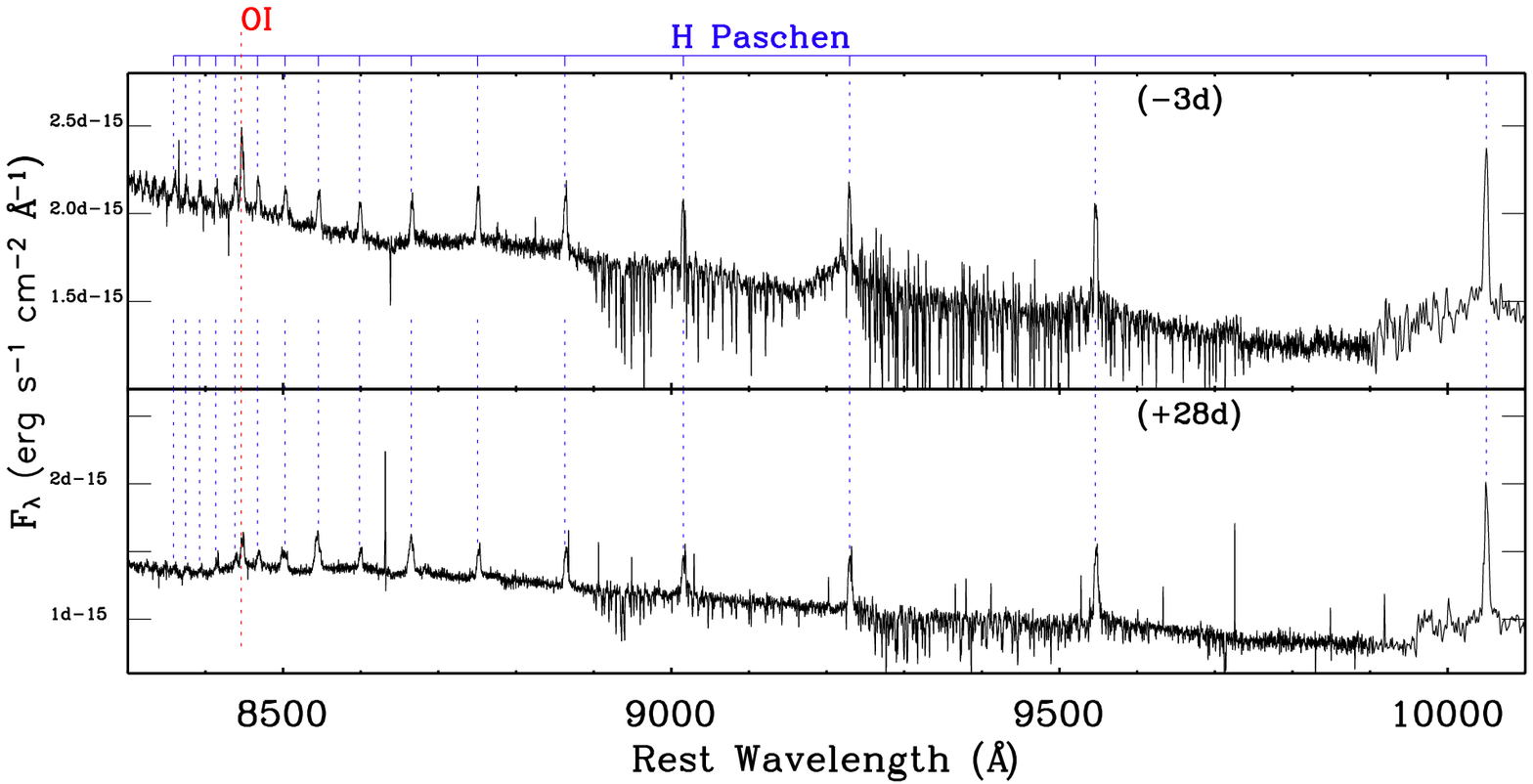}\\
\includegraphics[scale=0.95]{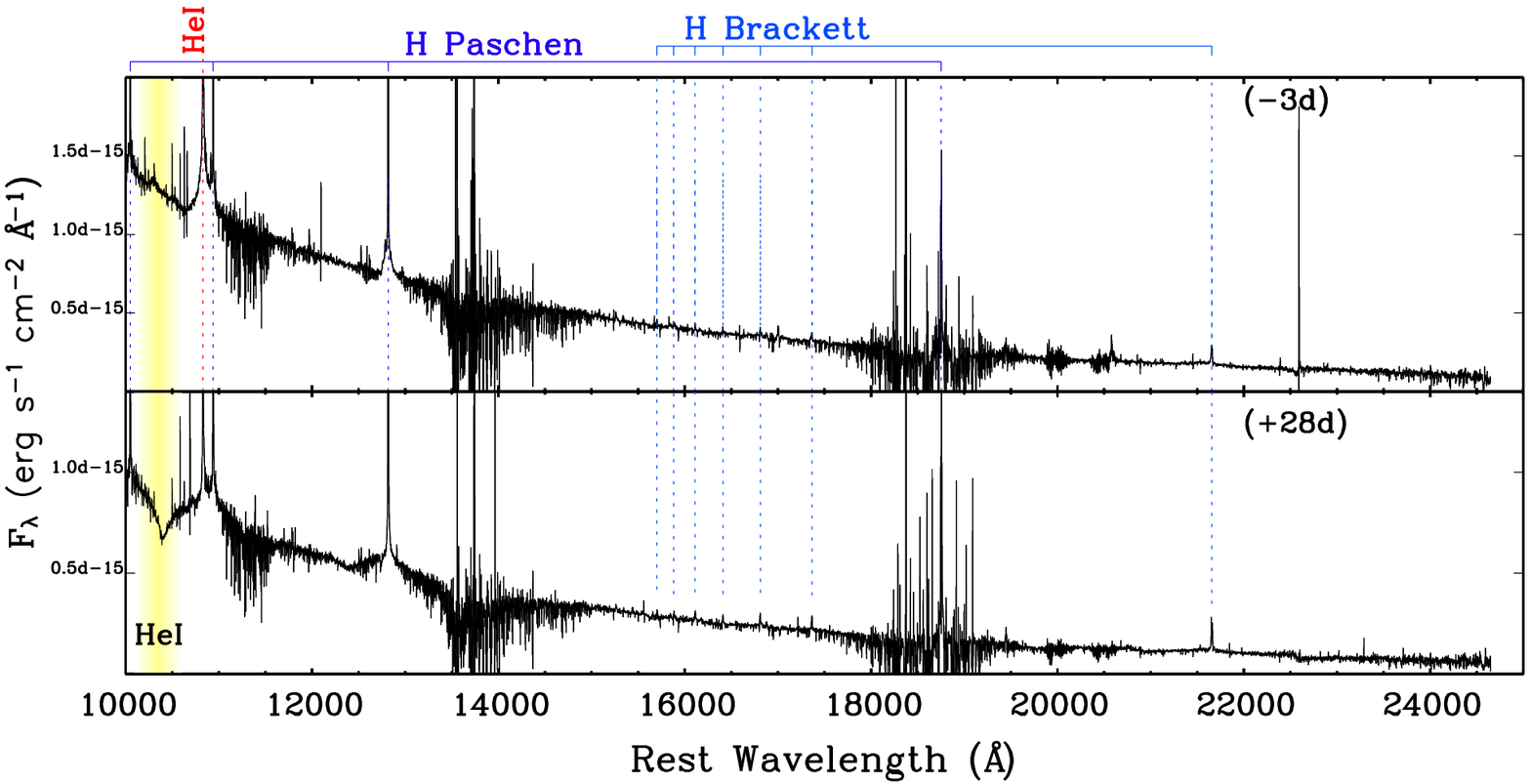}\\
\caption{High-resolution VLT/X-shooter spectra from $10000$ to $24500$ \AA\,.
Continued from Fig. \ref{Fig:Xshooter1}. }
\label{Fig:Xshooter2}
\end{figure*}

\begin{figure*}
\vskip -0.0 true cm
\centering
\includegraphics[scale=0.6]{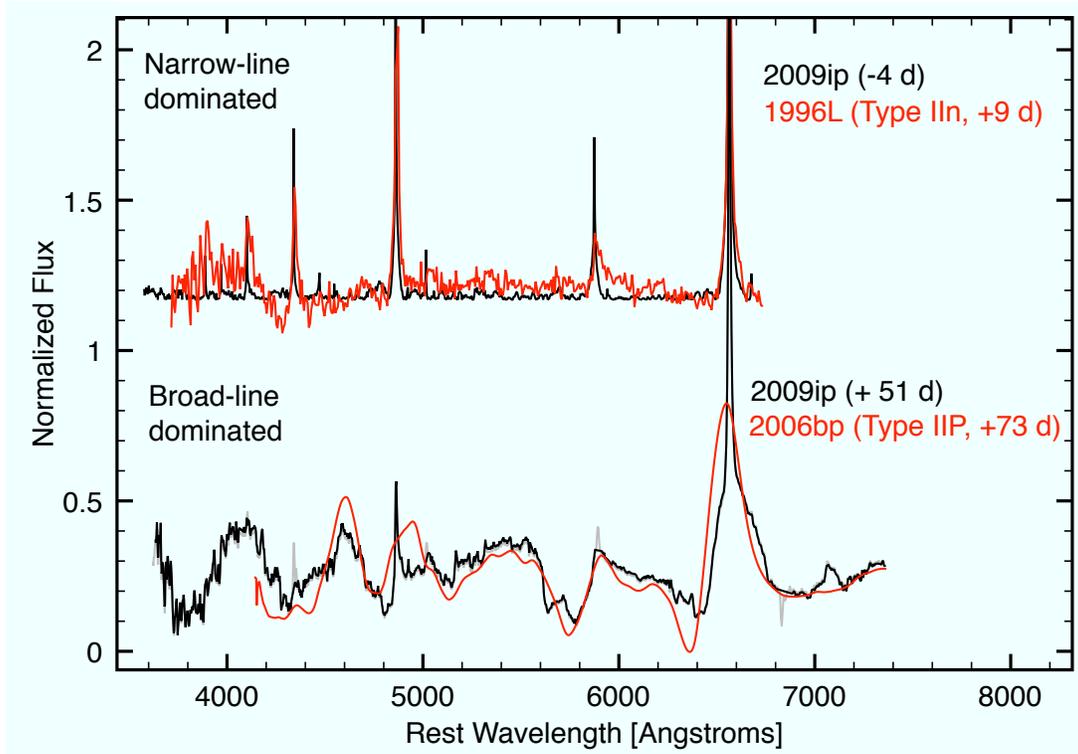}
\caption{SN\,2009ip evolved from a narrow-line dominated spectrum typical of Type IIn SN explosions
to a spectrum that clearly shows broad absorption features more typical of Type IIP SNe. Here we show the
spectrum of Type IIn SN\,1996L \citep{Benetti99} and Type IIP SN\,2006bp \citep{Quimby07}.}
\label{Fig:09ipV06bp}
\end{figure*}

\begin{figure}
\vskip -0.0 true cm
\centering
\includegraphics[scale=0.8]{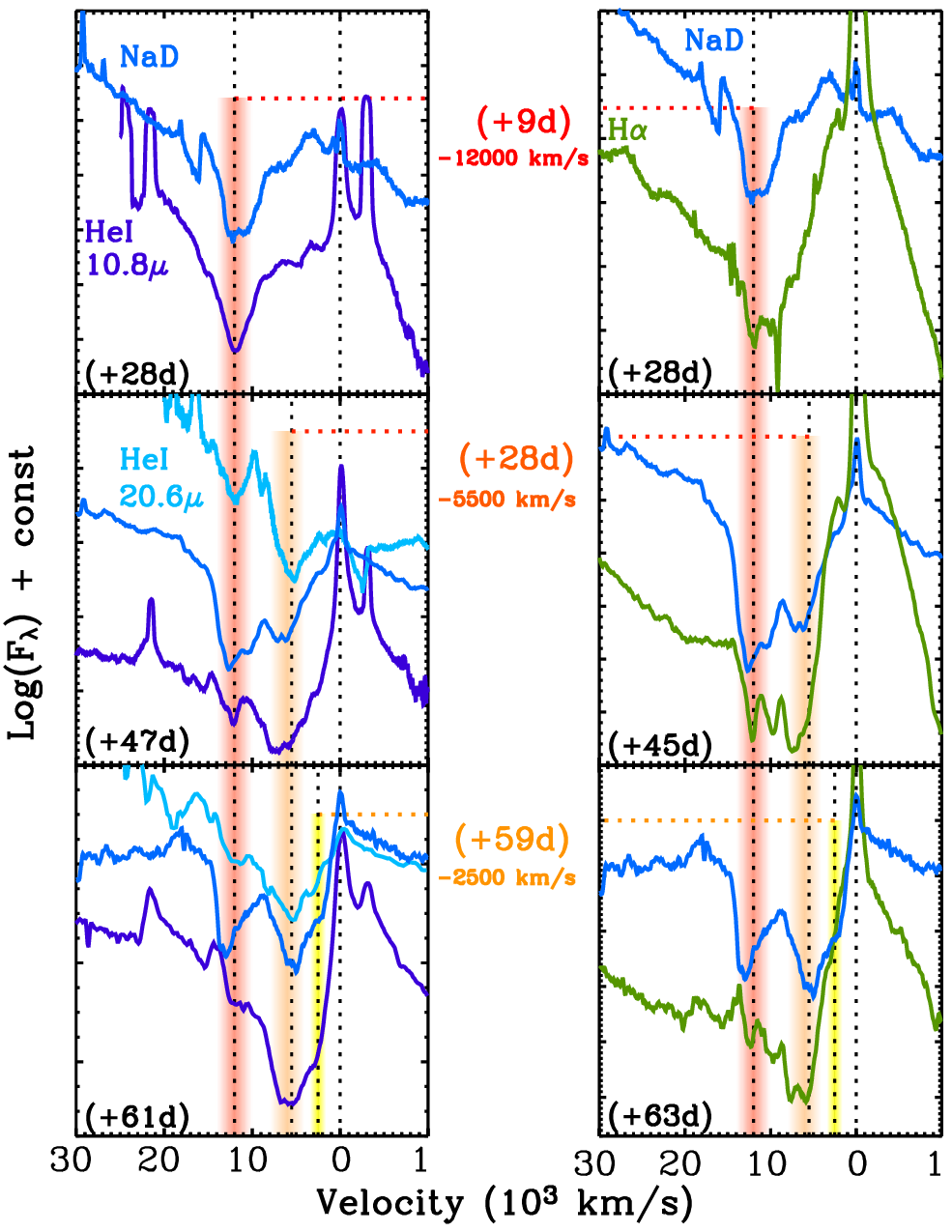}
\caption{Evolution of the broad absorption features associated with H$\alpha$,
\ion{He}{1} and \ion{Na}{1} D lines.
As the photosphere recedes into the ejecta broad absorption features appear in the
spectra with 3 typical velocities: $v\sim -12000\,\rm{km\,s^{-1}}$ (red band); 
$v\sim -5500\,\rm{km\,s^{-1}}$ (orange band) and $v\sim -2500\,\rm{km\,s^{-1}}$ (yellow band). 
Absorption features at higher velocity are revealed at earlier times: we clearly detect
material with $v\sim -12000\,\rm{km\,s^{-1}}$ starting around $t_{\rm{pk}}+9$ days; material 
with $v\sim -5500\,\rm{km\,s^{-1}}$ starts to be detected around $t_{\rm{pk}}+28$ days, while slowly
moving ejecta with  $v\sim -2500\,\rm{km\,s^{-1}}$ is only detected $\sim60$ days after peak.
\emph{Left panel}: velocity profile of the absorption arising from \ion{Na}{1} D plus 
\ion{He}{1} ($\lambda = 5876$ \AA), 
compared with HeI $1.08\,\mu\rm{m}$ and HeI $2.06\,\mu\rm{m}$ velocity profiles. \emph{Right panel}:
\ion{Na}{1} D plus \ion{He}{1} ($\lambda = 5876$ \AA) vs. H$\alpha$ velocity profile.}
\label{Fig:absorptions}
\end{figure}

\cite{Pastorello12} find the spectrum of SN\,2009ip during the 2012a outburst
to be dominated by prominent Balmer lines. In particular, spectra collected in 
August and September 2012 show clear evidence for narrow emission 
components (FWHM$\approx 800\,\rm{km\,s^{-1}}$ for $\rm{H}\alpha$) accompanied by 
absorption features, indicating the presence of high velocity material with 
velocities extending to $v\approx -14000\,\rm{km\,s^{-1}}$  (\citealt{Mauerhan12}).
Our 2012 August 26 spectrum confirms these findings. SN\,2009ip experienced a sudden re-brightening
around 2012 September 23 ($t_{\rm{pk}}-10$ days, \citealt{Brimacombe12}; \citealt{Margutti12}), signaling
the beginning of the 2012b explosion.
By this time the $\rm{H}\alpha$ line  developed a prominent broad emission component with 
FWHM$\approx 8000\,\rm{km\,s^{-1}}$ \cite[their Fig. 5]{Mauerhan12}.
The broad component disappeared 3 days later: 
our spectrum obtained on 2012 September 26  ($t_{\rm{pk}}-7$ days) indicates that the
$\rm{H}\alpha$ line evolved back to the narrow profile (Fig. \ref{Fig:Halpha}), 
yet still retained evidence for absorption with a core velocity 
$v\approx -5000\,\rm{km\,s^{-1}}$, possibly extending to $v\approx -7500\,\rm{km\,s^{-1}}$.
By 2012 September 30 ($t_{\rm{pk}}-3$ days, Fig. \ref{Fig:Xshooter1} and \ref{Fig:Xshooter2}) 
the spectrum no longer shows evidence for the high velocity
components in absorption and  is instead dominated by \ion{He}{1} and \ion{H}{1} lines 
with narrow profiles. 

In the following months SN\,2009ip progressively evolves from a typical 
SN IIn (or LBV-like) spectrum with clear signs of interaction with the medium, to a 
spectrum dominated by broad absorption features, more typical of  SNe IIP (Fig. \ref{Fig:09ipV06bp}).
Our two Xshooter spectra (Fig. \ref{Fig:Xshooter1} and \ref{Fig:Xshooter2}) sample two 
key points in this metamorphosis, providing a broad band 
view of these spectral changes at high resolution.  Broad features
completely disappear by the time of our observations in April 2013 ($t_{\rm{pk}}+190$ days, Fig. \ref{Fig:latetimeSpec}).
At no epoch we find evidence for very narrow, low velocity blue shifted absorption 
at $v\sim -100\,\rm{km\,s^{-1}}$, differently from what typically observed in Type IIn SNe and LBVs (see e.g.
SN\,2010jl, \citealt{Smith12b}). The major spectral changes during the 2012b explosion can be summarized as follows:
\begin{itemize}
\item Broad/intermediate  absorption/emission features progressively re-appear
in the H Balmer lines, with evidence of multiple velocity components (Fig. \ref{Fig:Halpha}, \ref{Fig:HBalmer}
and  \ref{Fig:Halpha_evol}).
\item Narrow \ion{He}{1} lines weaken with time (Fig. \ref{Fig:FeNaHe}); \ion{He}{1}
later re-emerges with the intermediate component only.
\item Fe II features re-emerge and later develop P Cygni profiles.
\item Emission originating from \ion{Na}{1} D is detected (Fig. \ref{Fig:FeNaHe}).
\item A broad near-infrared Ca II triplet feature typical of Type IIP SNe
develops starting around 2012 November 15 (Fig. \ref{Fig:CaII}).
\item More importantly, SN\,2009ip progressively develops broad absorption dips which have never
been observed in LBV-like eruptions, while being typical of a variety of SN explosions
(Fig. \ref{Fig:FeNaHe}). Broad absorption dips disappear $\sim 200$ days after peak.
\end{itemize}

Around $100$ days after peak,
emission from forbidden transitions (see e.g. [CaII]  $\lambda\lambda$ 7291, 7324
 in Fig. \ref{Fig:OptSpec}) starts to emerge. At this time SN\,2009ip settles
 behind the Sun. Despite limited spectral evolution between $t_{\rm{pk}}+100$ days 
 and  $t_{\rm{pk}}+200$ days (when SN\,2009ip re-emerges from the Sun constraint)
 we do observe the absorption features to migrate
 to lower velocities. We discuss each of the items below. 
 
 Additional optical/NIR spectroscopy of SN\,2009ip during the 2012b explosion
 has been published by \cite{Mauerhan12},  \cite{Pastorello12}, \cite{Levesque12}, \cite{Smith13}
 and \cite{Fraser13}: we refer to these works for a complementary description 
 of the spectral changes underwent by SN\,2009ip.
\subsection{Evolution of the \ion{H}{1} line profiles}
\label{SubSec:Halpha}

The H$\alpha$ line profile experienced a dramatic change in morphology after the 
source suddenly re-brightened on 2012 September 23. Figure \ref{Fig:Halpha} shows the 
H$\alpha$ line at representative epochs: at any epoch the
H$\alpha$ line has a complex profile resulting from the combination of 
a narrow (Lorentzian) component (FWHM$< 1000\,\rm{km\,s^{-1}}$), intermediate/broad width 
components (FWHM$> 1000\,\rm{km\,s^{-1}}$) and blue absorption features with evidence 
for clearly distinguished velocity components. Emission and absorption 
components with similar velocity are also found in the H$\beta$ and H$\gamma$
line profiles (Fig. \ref{Fig:HBalmer}).

The evolution of the line profile results from changes in the relative strengths of the different
components in addition to the appearance (or disappearance) of high-velocity blue
absorption edges. The evolution of the width and relative strength of the
different components is schematically represented in Fig. \ref{Fig:Halpha_evol}. The broad component
dominates over the narrow emission starting from $t_{\rm{pk}}+33$ days and reaches its maximum width 
at $t_{\rm{pk}}+51$ days. After this time, the width of the broad component decreases.
There is evidence for an increasing width of the narrow component with time,
accompanied by a progressive shift of the peak to higher velocities.
Finally, high-velocity ($v>1000\,\rm{km\,s^{-1}}$) absorption features get stronger
as the light-curve makes the transition from the rise to the decay phase. 
The spectral changes are detailed below. 
 
By $t_{\rm{pk}}-7$ days the broad components dominating the line profile 10 days before
(\citealt{Mauerhan12}) have weakened to the level that most of the emission originates
from a much narrower component which is well described by a Lorentzian
profile with FWHM$\approx 1000\,\rm{km\,s^{-1}}$. Absorption from high velocity  material  
($v\approx -5000\,\rm{km\,s^{-1}}$, measured at the minimum of the 
absorption feature) is still detected when the 2012b explosion luminosity is still rising.
The high-resolution spectra collected on $t_{\rm{pk}}-5$ and $t_{\rm{pk}}-4$ days allow us to resolve different 
blue absorption components: modeling these absorption features with Gaussians, the central 
velocities are found to be $v\approx-2200\,\rm{km\,s^{-1}}$, $\approx-4000\,\rm{km\,s^{-1}}$, 
$\approx-5300\,\rm{km\,s^{-1}}$, $\approx-7500\,\rm{km\,s^{-1}}$ with $\sigma\approx 300-500\,\rm{km\,s^{-1}}$.
These absorption features are detected in the $H\beta$ and $H\gamma$ lines as well (Fig. \ref{Fig:HBalmer}).
The width of the  narrow component of emission decreases to FWHM$\approx 280\,\rm{km\,s^{-1}}$.

On 2012 September 30 ($t_{\rm{pk}}-3$ days) SN\,2009ip approaches its maximum luminosity (Fig. \ref{Fig:Lbol}).
From our high-resolution spectrum the H$\alpha$ line is well modeled by the 
combination of two Lorenztian profiles with FWHM$\approx 240\,\rm{km\,s^{-1}}$ and
FWHM$\approx 2600\,\rm{km\,s^{-1}}$. We find no clear evidence for absorption components.
Interpreting the broad wings as a result of multiple Thomson scattering in the circumstellar shell
of the narrow-line radiation
\citep{Chugai01} suggests that the optical depth of the unaccelerated circumstellar
shell envelope to Thomson scattering is  $\tau\sim3$.


High-velocity absorption features in the blue wing of the H$\alpha$ line progressively re-appear as the 
luminosity of the explosion enters its declining phase. Eight days after peak the H$\alpha$ line exhibits 
a combination of narrow (FWHM$\approx 340\,\rm{km\,s^{-1}}$) and broad 
(FWHM$\approx 2000-3000\,\rm{km\,s^{-1}}$)
Lorentzian profiles and a weak P Cygni profile with an absorption minimum around
$-1600\,\rm{km\,s^{-1}}$. Three days later ($t_{\rm{pk}}+11$ days) the broad component 
(FWHM$\approx 2600\,\rm{km\,s^{-1}}$) of emission becomes more prominent while the 
width of the narrow Lorentzian profile decreases again to 
FWHM$\approx 220\,\rm{km\,s^{-1}}$. At this time the bolometric light-curve exhibits
a third bump (Fig. \ref{Fig:Lbol}). High-velocity absorption features re-appear in the blue
wing of the H$\alpha$ line with absorption minima at $v\approx -12000\,\rm{km\,s^{-1}}$
and $v\approx -8000\,\rm{km\,s^{-1}}$ ($\sigma\sim1000\,\rm{km\,s^{-1}}$). 
The low velocity P Cygni absorption is also detected at $v\approx-1200\,\rm{km\,s^{-1}}$.
The $H\beta$ and $H\gamma$ lines possibly show evidence for an additional absorption 
edge at $v\approx -4000\,\rm{km\,s^{-1}}$  (Fig. \ref{Fig:HBalmer}).

A lower resolution spectrum obtained on $t_{\rm{pk}}+18$ days shows the development of an even stronger
broad emission component with FWHM$\approx 9400\,\rm{km\,s^{-1}}$.  While
we cannot resolve the different components of velocity responsible for the blue absorption, 
we find clear evidence for a deep minimum at $v\approx -10000\,\rm{km\,s^{-1}}$ with 
edges extending to $v\approx -14000-15000\,\rm{km\,s^{-1}}$. The broad emission component 
keeps growing with time: at $t_{\rm{pk}}+42$ days it clearly dominates the H$\alpha$ profile. At this
epoch the H$\alpha$ line consists of a narrow component with FWHM$\approx240\,\rm{km\,s^{-1}}$,
a broad emission component (FWHM$\approx 10600\,\rm{km\,s^{-1}}$) and
a series of absorption features on the blue wing (both at high and low velocity). 
Our high-resolution spectrum resolve the absorption minima at $v\approx -12500\,\rm{km\,s^{-1}}$, 
$\approx -10000\,\rm{km\,s^{-1}}$, $\approx -7000\,\rm{km\,s^{-1}}$ and $\approx -1300\,\rm{km\,s^{-1}}$
(Fig. \ref{Fig:Halpha}). The $H\beta$ and $H\gamma$ lines exhibit an additional blue absorption
at $v\approx 3000\,\rm{km\,s^{-1}}$ (Fig. \ref{Fig:HBalmer}). 

By $t_{\rm{pk}}+51$ days the broad component which dominates the H$\alpha$ line reaches 
FWHM$\approx 13600\,\rm{km\,s^{-1}}$. High velocity absorption features are still detected 
at $v\approx -12500\,\rm{km\,s^{-1}}$ and $\approx -10000\,\rm{km\,s^{-1}}$. The absorption feature
at $v\approx -7000\,\rm{km\,s^{-1}}$ becomes considerably more pronounced and shows
clear evidence for two velocity components with minima at $v\approx -7900\,\rm{km\,s^{-1}}$
and $v\approx -5700\,\rm{km\,s^{-1}}$.  The low-velocity absorbing component is also 
detected with a minimum at $v\approx -1000\,\rm{km\,s^{-1}}$.  A spectrum obtained  
63 days after maximum shows little evolution in the H$\alpha$ profile, the only difference being a more 
pronounced absorption at $v\approx -5700\,\rm{km\,s^{-1}}$. At $t_{\rm{pk}}+79$ days we find 
a less prominent broad component: by this time its width decreased from 
FWHM$\approx12500\,\rm{km\,s^{-1}}$  to FWHM$\approx8200\,\rm{km\,s^{-1}}$.
A spectrum obtained at  $t_{\rm{pk}}+101$ days confirms this trend (FWHM of the broad component  
$\approx7500\,\rm{km\,s^{-1}}$): the bulk of the absorption is now at lower velocities 
$v\approx -3100\,\rm{km\,s^{-1}}$ (with a tail possibly extending to $v\approx -8000\,\rm{km\,s^{-1}}$).
At $t_{\rm{pk}}+190$ days the blue-shifted absorption is found peaking at even lower
velocities of  $v\lesssim -2400\,\rm{km\,s^{-1}}$, and the "broad" (now intermediate) component
has  FWHM of only $\approx2000\,\rm{km\,s^{-1}}$. 

Finally, comparing the H Paschen and Brackett emission lines using our two highest
resolution spectra collected around the peak (narrow-line emission dominated spectrum
at $t_{\rm{pk}}-3$ days)
and 28 days after peak (when broad components start to emerge, see Fig. \ref{Fig:Xshooter1}
and \ref{Fig:Xshooter2}), we find that for both epochs
the line profiles are dominated by the narrow component 
(FWHM$\approx170\,\rm{km\,s^{-1}}$) with limited evolution between the two. The Paschen $\beta$ line
clearly develops an intermediate-broad component starting from $t_{\rm{pk}}+33$ days
(see Fig. \ref{Fig:NIRSpec}). Spectra obtained 
by \cite{Pastorello12} before the sudden re-brightening of 
2012 September 23 ($t_{\rm{pk}}-10$ days) show a similar
narrow plus broad component structure, with the broad emission dominating
the narrow lines between 2012 August 26 and 2012 September 23. As for the H Balmer lines, the 
broad component completely disappeared as the light-curve approached its
maximum.  

We conclude by noting that, observationally, $t_{\rm{pk}}+17$ days 
(i.e. 2012 October 20) marks an important transition in the evolution of SN\,2009ip: 
around this time the broad H$\alpha$ component evolves from FWHM$\sim 3000\,\rm{km\,s^{-1}}$ to 
FWHM$\sim 10000\,\rm{km\,s^{-1}}$ (Fig. \ref{Fig:Halpha_evol}); the photospheric
radius $R_{\rm{HOT}}$ flattens to $R_{\rm{HOT}}\sim 1.6 \times 10^{15}\rm{cm}$
while the hot black-body temperature transitions to a milder decay in time (Section
\ref{Sec:SED}, Fig. \ref{Fig:Lbol}). It is intriguing to note that our modeling described in Section
\ref{Sec:source} independently suggests that this is roughly the time when the explosion 
shock reaches the edge of the dense shell of material previously ejected by the progenitor.
\subsection{The evolution of \ion{He}{1} lines}
\label{SubSec:HeI}

Conspicuous \ion{He}{1} lines are not
unambiguously detected in our spectrum obtained on 2012 August 26. They are, however,  
detected in our spectrum acquired one month later,  $\sim 3$
days after SN\,2009ip re-brightened\footnote{Note that \ion{He}{1} was clearly detected during the
LBV-like eruption episodes in 2011 \citep{Pastorello12}}.
At this epoch the light curve of SN\,2009ip is still rising. 
Similarly to H Balmer lines, HeI features (the brightest being at 5876 \AA\,,
and $7065$ \AA\footnote{We also detect  \ion{He}{1} $\lambda4713$ (weak),
\ion{He}{1} $\lambda5016$  (later blended with Fe II $\lambda5019$), \ion{He}{1} $\lambda6678$,  
on the red wing of H$\alpha$, 
\ion{He}{1} $\lambda7281$ (weak) and \ion{He}{1} $\lambda10830$ (blended with Pa$\gamma$). 
\ion{He}{1} $\lambda5876$
is also blended with \ion{Na}{1} D emission.}) exhibit  
a combination of a narrow-intermediate profile (FWHM$\approx1000\,\rm{km\,s^{-1}}$), 
a weak broad component (FWHM$\approx5000\,\rm{km\,s^{-1}}$)
together with evidence for a P Cygni absorption at velocity $v\approx -5000\,\rm{km\,s^{-1}}$.

As for the H Balmer lines, high-resolution spectroscopy obtained at  
$t_{\rm{pk}}-5$ and $t_{\rm{pk}}-4$ days
shows the appearance of multiple absorption components on the blue wing of the 
\ion{He}{1} $\lambda5876$  and $\lambda7065$  lines, with velocities $v\approx-2000\,\rm{km\,s^{-1}}$,
$\approx-4800\,\rm{km\,s^{-1}}$ and  $\approx-7000\,\rm{km\,s^{-1}}$ measured at the absorption
minima (to be compared with Fig. \ref{Fig:HBalmer}).  High velocity absorption features disappear 
by $t_{\rm{pk}}-3$ days: \ion{He}{1} $\lambda5876$  and $\lambda7065$  show the combination of a narrow plus 
broader intermediate Lorentzian profiles with  FWHM$\approx2000\,\rm{km\,s^{-1}}$ and 
FWHM$\approx240\,\rm{km\,s^{-1}}$, respectively.

Starting from $t_{\rm{pk}}-3$ days,  \ion{He}{1} features become weaker
until \ion{He}{1} $\lambda7065$ is not detected in our high-resolution spectrum 
acquired at $t_{\rm{pk}}+28$ days (Fig. \ref{Fig:Xshooter1} and Fig. \ref{Fig:Xshooter1}).
\ion{He}{1} later re-appears in our spectra taken in the second half of November
($t>t_{\rm{pk}}+43$ days) showing the broad/intermediate component only (FWHM$\approx 2500\,\rm{km\,s^{-1}}$ as 
measured at $t_{\rm{pk}}+63$ days). At  $t_{\rm{pk}}+79$ days 
\ion{He}{1} $\lambda7065$ shows an intermediate-broad emission profile with 
FWHM$\approx 3000\,\rm{km\,s^{-1}}$. A similar value is obtained at $t_{\rm{pk}}+101$ days. Roughly $100$ days
later, on 2013 April 11 \ion{He}{1} 7065 \AA\,  is clearly detected with considerably narrower emission
(FWHM$\approx 1000\,\rm{km\,s^{-1}}$). \ion{He}{1} $\lambda6678$ also re-emerges on the red wing 
of the H$\alpha$ profile (Fig. \ref{Fig:latetimeSpec}).
\subsection{The evolution of \ion{Fe}{2} lines}
\label{SubSec:FeII}

A number of \ion{Fe}{2} lines from different multiplets have been observed during  previous 
SN\,2009ip outbursts (both in 2009, 2011 and the 2012a outburst, see \citealt{Pastorello12}, their Fig. 5 and 6). 
The Fe responsible for this emission is therefore pre-existent the 2012 explosion.
\ion{Fe}{2} is instead not detected in our spectra until
$t_{\rm{pk}}+17$ days (Fig. \ref{Fig:FeNaHe}). From the 
Xshooter spectrum acquired at $t_{\rm{pk}}+28$ days we measure the FWHM of the
narrow \ion{Fe}{2} lines $\lambda 5018$ and $\lambda 5169$ 
(multiplet 42): FWHM$\approx 240 \,\rm{km\,s^{-1}}$. A similar value has been
measured by \cite{Pastorello12} from their 2012 August 18 and September 5 spectra. As a comparison,
the FWHM of the narrow (Lorentzian) component of the H$\alpha$ line measured
from the same spectrum is $\approx 170 \,\rm{km\,s^{-1}}$.
By $t_{\rm{pk}}+63$ days the \ion{Fe}{2} emission lines develop a 
P Cygni profile (Fig. \ref{Fig:FeNaHe}), with absorption minimum velocity of 
$v\approx -1000 \,\rm{km\,s^{-1}}$, possibly 
extending to $v\approx -4000 \,\rm{km\,s^{-1}}$.
\subsection{The NIR \ion{Ca}{2} feature}
\label{SubSec:CaII}

Starting from $\sim30$ days after peak, our spectra (Fig. \ref{Fig:CaII}) show the progressive emergence
of broad NIR emission originating from the \ion{Ca}{2} triplet $\lambda\lambda8498$, $8542$, $8662$
(see also \citealt{Fraser13}, their Fig. 4).
The appearance of this feature is typically observed during the evolution of
Type II SN explosions (see e.g. \citealt{Pastorello06}).
Interestingly, no previous outburst of SN\,2009ip
showed this feature (2012a outburst included, see \citealt{Pastorello12}). No broad \ion{Ca}{2} triplet
feature has ever been observed in an LBV-like eruption.

Figure \ref{Fig:CaII} also sjows the emergence of broad absorption dips
around 8400 \AA\, and 8600 \AA. If \ion{Ca}{2} $\lambda8662$ is causing the absorption
around 8600 \AA, the corresponding velocity at the absorption minimum is
$v\approx -2400\,\rm{km\,s^{-1}}$. This absorption developed between 51 days and
63 days after peak. The absorption at  $\lambda \approx 8400$ \AA\, is instead clearly detected
in our spectra starting from $t_{\rm{pk}}+45$ days and  likely results from the combination of 
OI and CaII. If OI (8447 \AA) is dominating the absorption at
minimum, the corresponding velocity is $v\approx -1500 \,\rm{km\,s^{-1}}$.

\subsection{The development of broad absorption features}
\label{SubSec:absorption}

High-velocity, broad absorption features appear in our spectra
starting 9 days after peak  (see yellow bands in Fig. \ref{Fig:NIRSpec},  
Fig. \ref{Fig:FeNaHe}, Fig. \ref{Fig:Xshooter1}, Fig. \ref{Fig:Xshooter2}).
Absorption features of similar strength and velocity
have never been associated with an LBV-like eruption to date,
and are more typical of SNe 
(Fig. \ref{Fig:09ipV06bp}). These absorption features are unique to the 2012b explosion
and have not been observed during the previous outbursts of SN\,2009ip 
(see \citealt{Smith10},  \citealt{Foley11}, \citealt{Pastorello12}).

As the photosphere recedes into the ejecta it illuminates 
material moving towards the observer with different velocities. Our observations identify 
\ion{He}{1}, \ion{Na}{1} D and \ion{H}{1} absorbing at 3 typical velocities (Fig. \ref{Fig:absorptions}).
The blue absorption edge of  \ion{He}{1} plus \ion{Na}{1} D extends to $v\approx 18000\,\rm{km\,s^{-1}}$,
as noted by \cite{Mauerhan12}.
High-velocity $v\sim -12000\,\rm{km\,s^{-1}}$ absorption appears first, around $t_{\rm{pk}}+9$ days
followed by the  $v\sim -5500\,\rm{km\,s^{-1}}$ absorption around $t_{\rm{pk}}+28$ days,
which in turn is followed by slower material with $v\sim -2500\,\rm{km\,s^{-1}}$, seen in absorption
only starting from $\sim t_{\rm{pk}}+60$ days. This happens since material with lower velocity 
naturally overtakes the photosphere at later times. 
Material moving at three \emph{distinct} velocities argues against a
continuous distribution in velocity of the ejecta and suggests instead the presence
of distinct shells of ejecta expanding with typical velocity 
$v\sim -12000\,\rm{km\,s^{-1}}$, $v\sim -5500\,\rm{km\,s^{-1}}$ and $v\sim -2500\,\rm{km\,s^{-1}}$.
\subsection{UV spectral properties}

Our \emph{Swift}-UVOT low-resolution spectroscopic monitoring 
campaign maps the evolution of SN\,2009ip during the first  month after
its major peak in 2012  (Fig. \ref{Fig:UVOTspec}). We do not find evidence for strong spectral 
evolution at UV wavelengths (Fig. \ref{Fig:UVOTspecnormalised}): as time proceeds the \ion{Fe}{3} absorption
features become
weaker while \ion{Fe}{2} develops stronger absorption features, consistent with the progressive
 decrease of the black-body temperature with time (Fig. \ref{Fig:Lbol}).
UVOT spectra show the progressive emergence
of an emission feature around $2500-3000$ \AA\, that is later well
resolved by HST/STIS as emission from \ion{Mg}{2} $\lambda \lambda 2796, 2803$  lines
as well as \ion{Fe}{2} multiplets at $\sim 2550, 2630, 2880$ \AA\, (Fig. \ref{Fig:HSTOct29UV}).
The \ion{Mg}{2} line profiles are similar to the \ion{H}{1} line profiles, with 
a narrow component and broad, blue-shifted absorption features. As for the
 \ion{H}{1} lines, the narrow component originates from the interaction with 
 slowly moving CSM.
We further identify strong, narrow emission from N II] at $\lambda \lambda 2140, 2143$.
Emission from C III] ($\lambda 1909$) and  Si III] ($\lambda \lambda 1892, 1896$)
might also be present, but the noise level does not allow a firm identification.

At shorter wavelengths, the HST/COS spectrum taken 34 days after peak shows 
a mixture of high and low ionization lines (Fig. \ref{Fig:HSTOct29UV}, lower panel). We identify
strong lines of C II  ($\lambda \lambda 1334.5, 1335.7$), O I 
($\lambda \lambda$ 1302.2-1306.0), Si II ($\lambda \lambda$ 1526.7, 1533.5).
Of the higher ionization lines one notes C IV ($\lambda \lambda 1548.2, 1550.8$) and 
N V  ($\lambda \lambda 1238.8, 1242.8$). Interestingly, N IV] $\lambda 1486.5$
is either very weak or absent which indicates a medium with density 
$n\gtrsim 10^9 \,\rm{ cm^{-3}}$. Fe II is also present, although the identification
of the individual lines is not straightforward (e.g. the Fe II feature at $\sim 1294$ \AA \ 
may also be consistent with Ti III). Ly$\alpha$ emission is also very well detected. 

Around  this time, both the optical, NIR and UV spectra are dominated by permitted transitions:
in particular, despite the presence of high ionization lines  there are no forbidden lines of,
e.g., [\ion{O}{3}] $\lambda \lambda$ 4959, 5007, \ion{N}{4}]  $\lambda\lambda$1486 or \ion{O}{3}] 
$\lambda\lambda $ 1664, consistent with the picture of high density in the line forming region.
(The [\ion{Ca}{2}] $\lambda$ 7300 lines will clearly emerge only after $t_{\rm{pk}}+79$ days). The main exceptions 
are the  [\ion{N}{2}]   $\lambda \lambda$ 2140, 2143 lines (Fig. \ref{Fig:HSTOct29UV}). The
explanation could be a comparatively high critical density, $\sim 3 \times 10^9\, \rm{cm^{-3}}$
in combination with a high N abundance.

A comparison of high (\ion{C}{4} $\lambda \lambda 
1548.2, 1550.8$ and N V  $\lambda \lambda 1238.8, 1242.8$) and low 
(\ion{C}{2} $\lambda$ 1335) ionization emission line profiles in velocity space reveals no 
significant difference:  the three lines extend to $\sim850\, \rm{km\,s^{-1}}$ on the red side, 
while there is an indication of a somewhat smaller extent on the blue wing, 
$\sim500\, \rm{km\,s^{-1}}$. This is however complicated by the P Cygni absorption features
and the doublet nature of the \ion{C}{4} and \ion{N}{5} lines. 
The mixture of low and high ionization lines indicates that there are several components
present in the line emitting region. This may either be in the form of different density 
components, or different ionization zones. The similar line profiles argue for a similar
location of the ionization zones, supporting the idea of a complex emission region
with different density components. The observed X-ray emission can in principle be
responsible for the ionization. 

\section{Metallicity at the explosion site and host environment}
\label{Sec:Metallicity}

\begin{figure}
\includegraphics[scale=0.8]{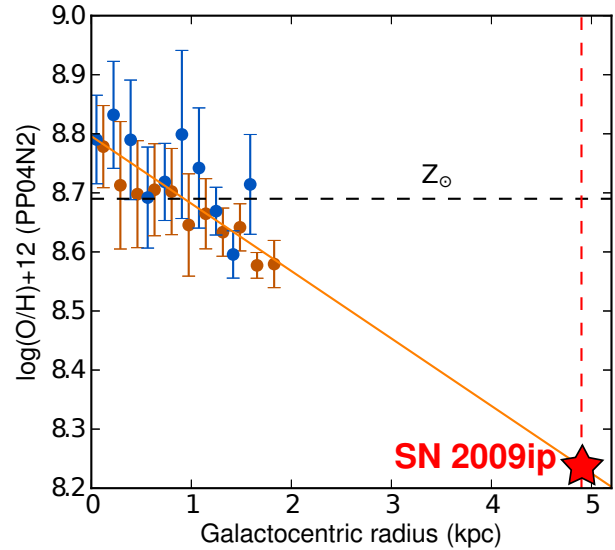}
\caption{Metallicity profile of NGC~7259, the host galaxy of SN\,2009ip, as derived 
from our long slit spectroscopy.  The solid line shows the best fit metallicity gradient.  
The dashed horizontal line marks the solar metallicity and the vertical dashed line 
marks the SN galactocentric radius.  The colors distinguish measurements from 
opposite sides of the galaxy center.  The error bars reflect propagation of the emission 
line flux uncertainties only. This analysis constrains the metallicity at the explosion site of SN\,2009ip
to be  $8.2<\log(\rm{O/H})+12<8.6$ ($0.4\,Z_{\sun}<Z<0.9\,Z_{\sun}$).}
\label{fig:Z}
\end{figure}

The final fate of a massive star is controlled by the mass of its
helium core (e.g. \citealt{Woosley07}), which is strongly dependent on the initial stellar mass, 
rotation and composition. Metallicity has a key role in determining the mass-loss
history of the progenitor, with low metallicity generally leading to a suppression
of mass loss, therefore allowing lower-mass stars to end their lives with 
massive cores. SN\,2009ip is positioned in the outskirts of NGC~7259 (Fig. \ref{Fig:SN2009ip}).
The remote location of SN\,2009ip has been discussed by \cite{Fraser13}.
Our data reveal no evidence for an \ion{H}{2} region in the
vicinity of SN\,2009ip that would allow us to directly measure the metallicity 
of the immediate environment.  Thus, we inferred 
the explosion metallicity by measuring the host galaxy metallicity gradient.
The longslit was placed along the galaxy center at parallactic angle.
We extracted spectra of the galaxy at positions in a sequence across our slit, producing a set of integrated light 
spectra from $\sim0-2$~kpc from either side of the galaxy center.

We use the ``PP04~N2'' diagnostic of \cite{PP04} to estimate gas phase metallicity using 
the H$\alpha$ and [\ion{N}{2}]~$\lambda6584$ emission lines.  We estimate the uncertainty 
in the metallicity measurements by Monte Carlo propagation of the uncertainty in the 
individual line fluxes.  The median uncertainty is $0.09$~dex, which is similar to the systematic 
uncertainty in the calibration of the strong line diagnostic \citep{KE08}.    Robust metallicity 
profiles can not be recovered in other diagnostics due to the faintness of the [\ion{O}{3}] lines in 
our spectroscopy.

Figure~\ref{fig:Z} shows the resulting metallicity profile of NGC~7259.  The metallicity at the 
galaxy center is $\log(\rm{O/H})+12=\sim8.8$, $\sim1.3~\rm{Z}_\odot$ on the PP04~N2 scale, 
but declines sharply with radius.  The metallicity profiles on each side of the galaxy center in our 
longslit spectrum are consistent.  We therefore assume that the metallicity profile is azimuthally 
symmetric. We estimate the metallicity gradient by fitting a linear profile. 
The best fit gradient intercept and slope are $8.8\pm0.02$~dex and $-0.11\pm0.02~\rm{dex~kpc^{-1}}$, 
respectively.

SN\,2009ip is located $\sim43.4"$ from the center of the galaxy NGC~7259 (equal to 
$\sim5.0$~kpc at $d_{\rm{L}}=24\,\rm{Mpc}$). This is more than twice the distance to which our metallicity profile observations 
extend.  Extrapolating directly from this gradient would imply an explosion site metallicity of 
$\log(\rm{O/H})+12=\sim8.2$, or $\sim0.4~\rm{Z}_\odot$.  This metallicity would place SN~2009ip at the 
extreme low metallicity end of the distribution of observed host environments of Type~II SN  \citep{Stoll12}, 
and nearer to the low metallicity regime of broad-lined Type~Ic supernovae \citep{Kelly12,blindsample}.  
However, the metallicity properties of galaxies at distances well beyond a scale radius have not been 
well studied.  It is likely that a simple extrapolation is not appropriate, and the metallicity profile in the 
outskirts of the galaxy may flatten \citep{Werk11} or drop significantly \citep{Moran12}.  In either case, 
it is unlikely that the explosion site metallicity is significantly enriched relative to the gas we observe at 
$R\sim2$~kpc, with $\log(\rm{O/H})+12\sim8.6$ ($\sim0.9~\rm{Z}_\odot$).  If we adopt this value as the 
explosion site metallicity, it is fully consistent with the observed distribution of SNe II, Ib, and Ic 
\citep{Kelly12,blindsample,Stoll12}. 

Our best constraint on the explosion site metallicity is therefore $0.4\,Z_{\sun}<Z<0.9\,Z_{\sun}$,
pointing to a (mildly) sub-solar environment. 
\section{Energetics of the explosion}
\label{Sec:Energetics}

\begin{figure}
\vskip -0.0 true cm
\centering
\includegraphics[scale=0.7]{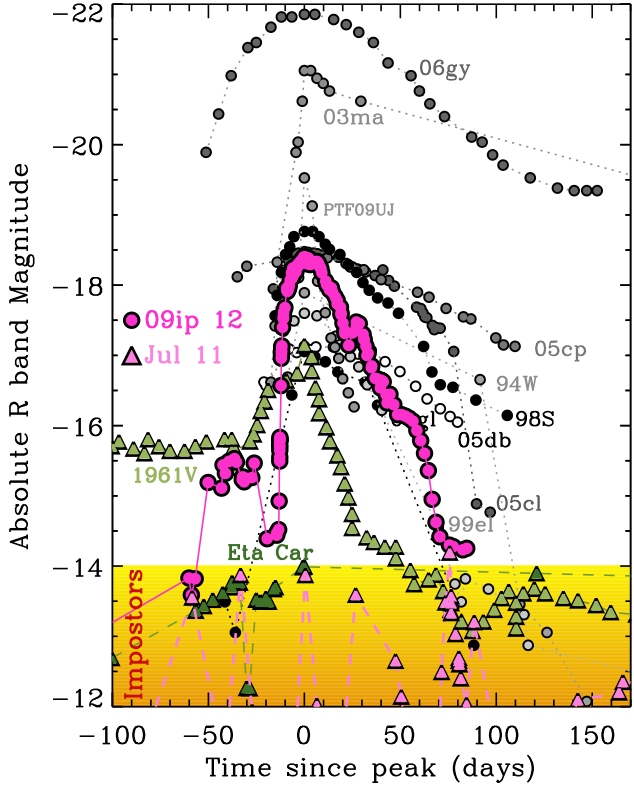}
\caption{Absolute R-band magnitude of SN\,2009ip (pink dots) compared to the sample of Type IIn SNe
with R-band photometric coverage around the peak (from the literature). 
For SNe 1994W, 1998S, 1999el, 2003ma, 2005cp, 2005cl, 2005db, 2005gl and 2006gy
we refer the reader to  \cite{Kiewe12} and references therein. The photometry of PTF09UJ has been presented by  
\cite{Ofek10}. Colored area: typical absolute magnitude of LBV-like eruption episodes. Pink triangles
mark the luminosity of SN\,2009ip during the 2011 outburst ($t=0$ corresponds here to 2011 July 8 for convenience).
The exceptional SN (impostor?) SN1961V (photographic plate magnitudes, from \citealt{Pastorello12})  and $\eta$ Carinae
during the 19th century Great Eruption ($M_{vis}$ as compiled by \citealt{Frew04}, see also \citealt{Smith11a}) 
are shown with green triangles. The comparison with SN\,2010mc is shown in Fig.
\ref{Fig:09ipvs10mc}. }
\label{Fig:AbsRband}
\end{figure}

\begin{figure}
\vskip -0.0 true cm
\centering
\includegraphics[scale=0.6]{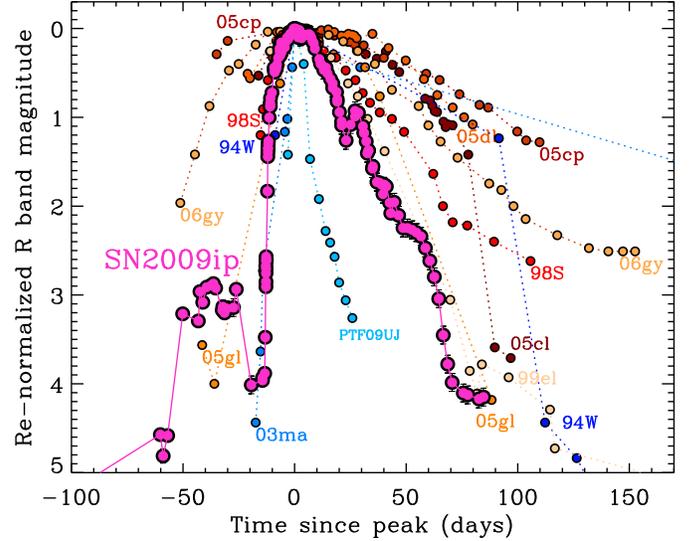}
\caption{Re-normalized R-band magnitude of SN\,2009ip compared with the sample of SNe IIn
of Fig. \ref{Fig:AbsRband}. Shades of orange (blue) are used for SNe with a slower (faster) rise time.
SN\,2009ip is characterized by  a fast  rise and fast decay.}
\label{Fig:AbsRbandRenorm}
\end{figure}

The extensive photometric coverage (both in wavelength and in time) gives us the opportunity
to accurately constrain the bolometric luminosity and total energy radiated by SN\,2009ip.
SN\,2009ip reaches a peak luminosity of 
$L_{\rm{pk}}=(1.7\pm0.1)\times 10^{43}\rm{erg\,s^{-1}}$ (Fig. \ref{Fig:Lbol}). The total 
energy radiated during the 2012a outburst (from 2012 August 1 to September 23) is $(1.5\pm0.4)\times
10^{48}\rm{erg}$ while for the 2012b explosion we measure 
$E_{\rm{rad2}}=(3.2\pm 0.3)\times 10^{49}$ erg.  As much as
$\sim35$\% of this energy was released before the peak, while
$50$\% of $E_{\rm{rad2}}$ was radiated during the first $\sim15$ days.
Subsequent re-brightenings (which constitute a peculiarity of SN\,2009ip) 
only contributed to small fractions of the total energy.

The peak luminosity of SN\,2009ip is not uncommon among the heterogeneous
class of SNe IIn, corresponding to $M_{\rm{R}}\approx -18$ mag (Fig.  \ref{Fig:AbsRband}).
Its radiated energy of $(3.2\pm 0.3)\times 10^{49}$ erg falls instead into the low energy tail of the 
distribution mainly because of the very rapid rise and decay times of the bolometric luminosity 
(Fig. \ref{Fig:AbsRbandRenorm}). The limited energy radiated by SN\,2009ip brings
into question the final fate of the progenitor star: was the total energy released 
sufficient to fully unbind the star (i.e., terminal explosion) or does SN\,2009ip
results from a lower-energy ejection of only the outer stellar envelope (i.e., non-terminal explosion)?
This topic is explored in Section  \ref{Sec:Discussion}.
Indeed, stars might be able to survive eruptive/explosive
events that reach a visual absolute magnitude of $M_{\rm{vis}}\approx -17$ mag (e.g.
 SN\,1961V in Fig. \ref{Fig:AbsRbandRenorm}, \citealt{VanDyk12}, \citealt{Kochanek11}), so that the
peak luminosity is \emph{not} a reliable indicator of a terminal vs. non-terminal
explosion.\footnote{The same line of reasoning applies to the velocity  of the fastest moving 
material measured from optical spectroscopy as pointed out by \cite{Pastorello12}:  very fast
material with $v\sim 12500\,\rm{km\,s^{-1}}$ was observed on the LBV-like outburst 
of SN\,2009ip of September 2011, proving that high-velocity ejecta can be observed even
\emph{without} a terminal explosion.} With an estimated radiated energy of
$3.2\times 10^{49}$ erg \citep{Davidson97} the "Great Eruption" of $\eta$-Carinae 
(see e.g. \citealt{Smith13b} and references therein) demonstrated 
that it is also possible to survive the release of comparable amount of energy, even if
on time scales much longer than those observed for SN\,2009ip (the "Great Eruption" 
lasted about 20 yrs).

SN\,2010mc shows instead striking similarities to SN\,2009ip both in terms
of timescales and of released energy (Fig.
\ref{Fig:09ipvs10mc}).  As in SN\,2009ip, a precursor bump was detected $\sim 40$ days
before the major outburst. \cite{Ofek13b} calculate $E_{\rm{rad}}> 6\times 10^{47}\rm{erg}$
(precursor-bump) and $E_{\rm{rad}}\sim 3\times 10^{49}\rm{erg}$ (major outburst) for SN\,2010mc,
compared with $E_{\rm{rad1}}=(1.5\pm0.4)\times10^{48}\rm{erg}$ and
$E_{\rm{rad2}}=(3.2\pm 0.3)\times 10^{49}\rm{erg}$ we calculated above for SN\,2009ip.
The very close similarity of SN\,2010mc and SN\,2009ip originally noted by \cite{Smith13}
 has important implications on the nature of both explosions (see Section \ref{Sec:Discussion}).
\section{Source of energy and properties of the immediate environment }
\label{Sec:source}

In the previous sections we concentrated on the properties of the
explosion (e.g. energetics, evolution of the emission/absorption features) 
and of the environment (i.e. the metallicity) that can be
directly \emph{measured}; here we focus on properties that can be \emph{inferred}
from the data.

The light-curve of SN\,2009ip shows two major episodes of emission:
the precursor bump (2012a outburst) and the major re-brightening  
(2012b explosion). Is this phenomenology due to \emph{two} distinct explosions or 
is the double-peaked light-curve the result of a single explosion? The  main
argument against a single explosion producing the two peaks is the observed
evolution of the photospheric radius in Fig. \ref{Fig:Lbol}.  In the single-explosion
scenario material can only decelerate with time: at  $t_{\rm{pk}}+7$ days
the photospheric radius is $R_{\rm{HOT}}\sim1.2\times 10^{15}\rm{cm}$ and the velocity is
$v=4500\,\rm{km\,s^{-1}}$. Extrapolating back in time, this implies that 
the zero-time of the 2012b explosion is later than $t_{\rm{pk}}-24$ days. This is much
later than the observed onset of the 2012a outburst that occurred at  $t_{\rm{pk}}-56$ days
and favors against a single-explosion scenario. Models where the first bump in the light-curve
is a SN explosion while the second peak is due to the interaction of the
SN ejecta with the CSM (\citealt{Mauerhan12}) belong to this category. 
In the following we proceed instead with a two-explosion 
hypothesis and argue that we witnessed
two separate but causally connected explosions from the progenitor of SN\,2009ip.
\subsection{Limit on the Nickel mass synthesized by the 2012b explosion}
\label{subsec:LimitNi}

Narrow emission lines in the optical spectra of SN\,2009ip require that 
interaction with previously ejected material (either in the form of a stable wind or
from erratic mass-loss episodes) is occurring at some level. The multiple outbursts of
SN\,2009ip detected in the 2009, 2011 and August 2012 (from 3 years to $\sim1$
month before the major 2012 explosion) are likely to have ejected conspicuous 
material in the immediate progenitor surroundings so that \emph{interaction} 
of the 2012b explosion shock with this material qualifies as an efficient 
way to convert kinetic energy into radiation. 

The radioactive decay of $^{56}\rm{Ni}$ represents another obvious source of
energy. We employ the nebular phase formalism developed by \cite{Valenti08} 
(expanding on the original work by  \citealt{Sutherland84} and \citealt{Cappellaro97}) 
to constrain the amount of Nickel synthesized by the 2012b explosion using late time
observations. If the observed light-curve
were to be entirely powered by the energy deposition of the $^{56}\rm{Ni}$ radioactive decay
chain, our latest photometry would imply $M_{\rm{Ni}}\sim0.03\,\rm{M_{\sun}}$
for a standard explosion kinetic energy of $E_{k}=10^{51}\,\rm{erg}$.\footnote{
This limit is also sensitive to the ejecta mass $M_{\rm{ej}}$. 
We solve for the degeneracy between $M_{\rm{ej}}$ and $E_{k}$
using the observed photospheric velocity at maximum light $v_{\rm{phot}}\sim
4500\,\rm{km\,s^{-1}}$, which implies $(M_{\rm{ej}}/M_{\odot})\sim 3.0 (E_{k}/10^{51}\,\rm{erg})$.} 
For a low energy explosion with $E_{k}=10^{50}\,\rm{erg}$,  $M_{\rm{Ni}}\sim0.08\,\rm{M_{\sun}}$.
Allowing for other possible
sources of energy contributing to the observed luminosity (like interaction), we conclude
$M_{\rm{Ni}}<0.08\,\rm{M_{\sun}}$. Using this value (and the photospheric
formalism by \citealt{Valenti08}, based on \citealt{Arnett82}) we largely 
underpredict the luminosity of SN\,2009ip at peak for \emph{any} value of 
mass and kinetic energy of the ejecta: the energy release of SN\,2009ip
is therefore \emph{not} powered by $^{56}\rm{Ni}$ radioactive decay. 
\cite{Fraser13} independently derived $M_{\rm{Ni}}<0.02\,\rm{M_{\sun}}$,
consistent with our findings.
In the following we explore a model where the major UV-bright peak is
powered by shock break-out from a dense shell ejected by the precursor bump, while continued
interaction with previously ejected material is responsible for the peculiar, bumpy 
light-curve that follows.
\subsection{Shock break-out plus continued interaction scenario for the 2012b explosion}
\label{SubSec:ShockBO}

\begin{figure}
\vskip -0.0 true cm
\centering
\includegraphics[scale=0.4]{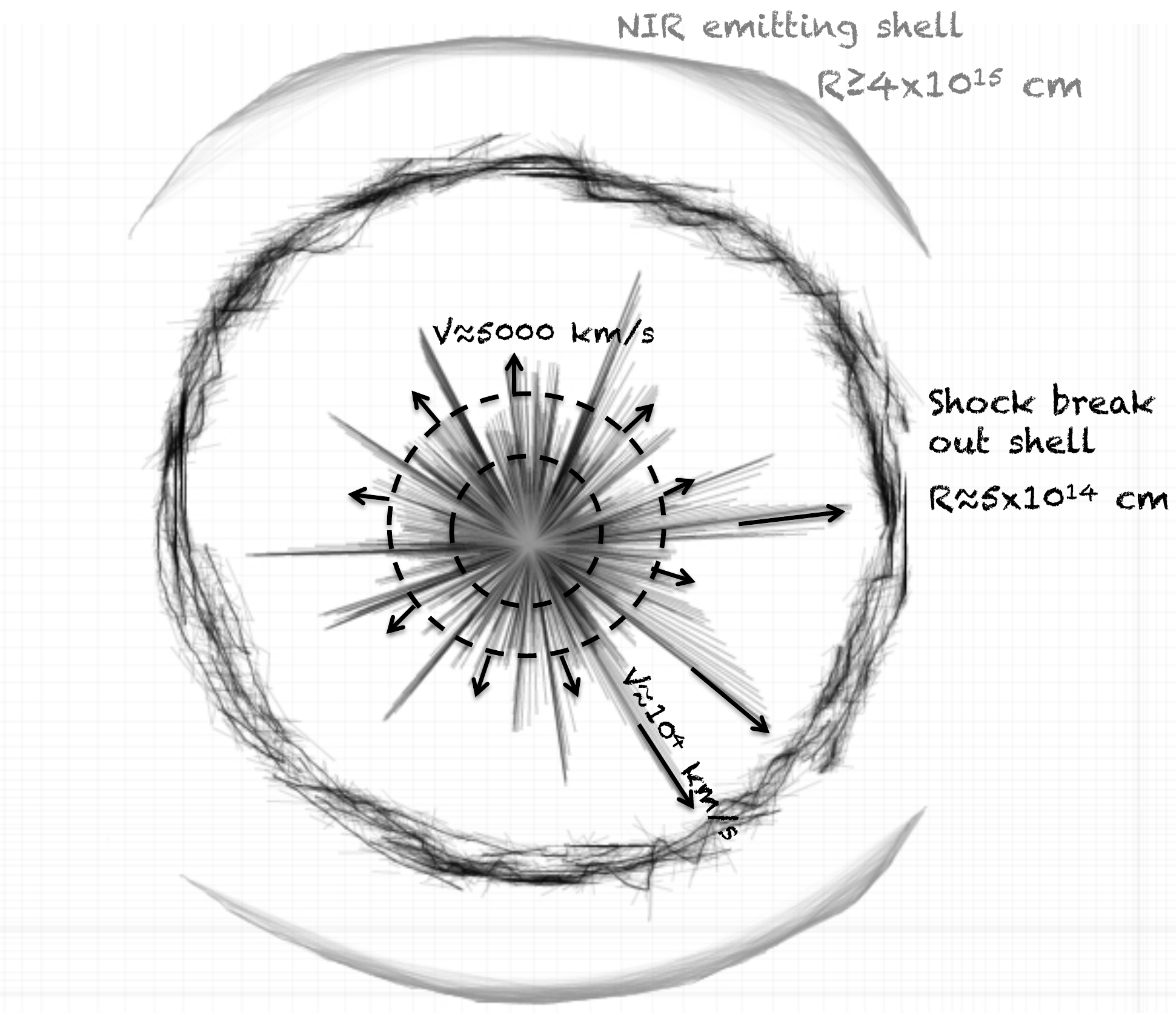}
\caption{Sketch of SN\,2009ip.
Its environment as well as its ejecta are likely to have an
extremely complex, potentially asymmetric structure. Here we show the basic components 
of the 2012b explosion and its environment.  In the ejecta we recognize the presence
of three velocity components at $v\sim2500\,\rm{km\,s^{-1}}$, $v\sim5000\,\rm{km\,s^{-1}}$
together with very fast material at $v>10000\,\rm{km\,s^{-1}}$ (Fig. \ref{Fig:absorptions}).  
Shock break-out from a dense shell of material left over by the 2012a outburst is
responsible for the major peak in the light-curve (Section \ref{SubSec:ShockBO}). 
Material sitting at larger distance,
connected with previous episodes of eruption, is 
responsible for partially re-processing the radiation into the NIR band, producing 
the NIR excess of Fig. \ref{Fig:SED}  (Sec. \ref{SubSec:NIRexcess}). Our analysis
requires this material to have a low filling factor and/or asymmetric distribution.}
\label{Fig:Sketch1.pdf}
\end{figure}

The rapid rise and decay times of the major 2012b explosion 
(Fig. \ref{Fig:AbsRbandRenorm}) suggest that the shock wave is interacting with 
a \emph{compact} shell(s) of material. The relatively fast fading of CSM-like features 
and subsequent emergence of Type IIP features shown in Fig. \ref{Fig:09ipV06bp} supports a similar 
conclusion.  The bumps in the light-curve further suggest an inhomogeneous medium.
We consider a model where the ejecta from the
2012b explosion initially interact with an optically \emph{thick} shell of material,
generating the UV-bright, major peak in the light-curve 
(Fig. \ref{Fig:Sketch1.pdf}). In our model, the light-curve is powered at later times by interaction with
optically \emph{thin} material.

In the shock break-out scenario the escape of radiation is delayed with respect to the onset of the explosion
until the shock is able to break-out from the shell at an optical depth $\tau_{\rm{w}}\approx c/v_{\rm{sh}}$.
This happens when the
diffusion time $t_{\rm{d}}$ becomes comparable to the expansion time.
Radiation is also released on the diffusion time scale, which implies that the observed bolometric
light-curve rise time is
$t_{\rm{rise}}\approx t_{\rm{d}}$. Following \cite{Chevalier11}, the radiated energy at break-out $E_{\rm{rad}}$,
the diffusion time $t_{\rm{d}}$ and the radius of the contact discontinuity at $t=t_{\rm{d}}$ 
($\equiv R_{\rm{bo}}$) depend on the explosion energy $E$, the ejecta mass $M_{\rm{ej}}$, the
environment density $\rho_{\rm{w}}$ (parametrized by the progenitor mass-loss rate) and opacity $k$. 
From our data we measure: $t_{\rm{rise}}\approx 10$ days;
$R_{\rm{bo}}\approx5\times 10^{14}$ cm; $E_{\rm{rad}}\approx 1.3 \times 10^{49}$ erg.
We solve the system of equations for our observables in Appendix \ref{App:ShockBreakOut} 
and obtain the following estimates for the properties of the explosion and its local environment.
Given the likely complexity of the SN\,2009ip environment, those should be treated as
order of magnitudes estimates. 

The onset of the 2012b explosion is around 20 days before peak  (2012 September 13).
Using  Eq. \ref{Eq:Dstar} and Eq. \ref{Eq:Mdot}, the progenitor mass-loss rate is 
$\dot M\approx 0.07 (v_{\rm{w}}/ 200\, \rm{km\,s^{-1}})\rm{M_{\sun} yr^{-1}}$. We choose
to renormalize the mass-loss rate to $200\, \rm{km\,s^{-1}}$, which is the FWHM of the
narrow emission component in the H$\alpha$ line (e.g. Fig. \ref{Fig:Halpha_evol}). 
The observed bolometric luminosity goes below the level of the luminosity expected from continued interaction
of Eq. \ref{Eq:Linteraction}  around $32$ days after the onset of the explosion or $t_{\rm{pk}}+12$ days.
By this time she shock must have reached the edge of the dense wind shell: $t_{\rm{w}}\lesssim 32$
days. This constrains the wind shell radius to be $R_{\rm{w}}\approx 1.2\times 10^{15}
\rm{cm}$ (Eq. \ref{Eq:Rwind}), therefore confirming the idea of a compact and dense shell of material,
while the total mass in the wind shell is $M_{\rm{w}}\approx
0.1 \rm{M_{\sun}}$  (Eq. \ref{Eq:Mwind}).\footnote{The mass
swept up by the shock by the time of break-out is $\approx 0.05 \rm{M_{\sun}}$.} The system of 
equations is degenerate for $M_{\rm{ej}}/E^2$.  Adopting our estimates of the observables above
and Eq. \ref{Eq:Mej} we find $M_{\rm{ej}}\approx 50.5 (E/ 10^{51}\rm{erg})^2 \rm{M_{\sun}}$.\footnote{Using
the line of reasoning of Section \ref{subsec:LimitNi}, the relation 
between $M_{\rm{ej}}$ and $E$ just found implies 
$M_{\rm{Ni}}<0.02\,\rm{M_{\sun}}$ for $E=10^{51}\,\rm{erg}$ and
$M_{\rm{Ni}}<0.07\,\rm{M_{\sun}}$ for $E=10^{50}\,\rm{erg}$, consistent
with the limits presented in Sec. \ref{subsec:LimitNi}.} 
The efficiency of conversion of kinetic energy into radiation depends
on the ratio of the total ejecta to wind shell mass  (e.g. \citealt{vanMarle10};
\citealt{Ginzburg12}; \citealt{Chatzopoulos12c}).
This  suggests $M_{\rm{ej}}\approx M_{\rm{w}}$ as order of magnitude estimate,
from which $E\sim10^{50}\,\rm{erg}$.

After $t_{\rm{w}}$ the bolometric luminosity starts to decay faster, especially at UV wavelengths 
(Fig. \ref{Fig:OptPhot}). By this time the shock has overtaken the dense thick shell and starts
to interact with less dense, optically \emph{thin} layers of material producing continued
power for SN\,2009ip. In this regime the observed luminosity tracks the energy deposition rate: 
$L=4\pi R^{2}\rho_{\rm{w}}(v_{\rm{fs}}-v_{\rm{w}})^3$, where $R$ is the radius of the cold 
dense shell that forms as a result of the loss of radiative energy from the shocked region;
$v_{\rm{fs}}$ is the forward shock velocity; $v_{\rm{w}}$ and $\rho_{\rm{w}}$ are the
velocity and density of the material encountered by the shock wave (\citealt{Chevalier11}
and references therein).
The presence of clearly detected bumps in the bolometric light-curve (with associated
rise in the effective temperature of the radiation, Fig. \ref{Fig:Lbol}) suggests that the medium has a complex 
structure and it is likely inhomogeneous. consequently
$\rho_{\rm{w}}$ might significantly depart from the $\propto R^{-2}$ profile expected in the
case of steady wind. The increasing FWHM with time measured for the narrow
component of the H$\alpha$ line in Fig. \ref{Fig:Halpha_evol} points to larger $v_{\rm{w}}$
at larger distances from the explosion, therefore deviating from the picture
of a steady wind with constant $v_{\rm{w}}$ (see Section \ref{SubSec:2012a}).
Given the complexity of the explosion environment, we adopt a simplified 
shock interaction model (see e.g. \citealt{Smith10a}) and parametrize the observed
luminosity as: $L=(\eta/2)wv^3$, where $\eta$ is the efficiency of conversion of
kinetic energy into radiation; $w(R)\equiv \dot M/ v_{\rm{w}}$ (hence $\rho_{\rm{w}}
=w(R)/4\pi R^2$); while $v$ is a measure of the expansion velocity of the shock 
into the environment. We estimate $v$ from the evolution of the black-body radius with 
time ($v\approx \dot R_{\rm{HOT}}$ of Fig. \ref{Fig:Lbol}), assuming that for $t_{\rm{pk}}+17$ days
the true shock radius continues to increase with  $v\approx 4500\,\rm{km\,s^{-1}}$
(while the measured $R_{\rm{HOT}}$ stalls and then decreases as the interaction 
shell progressively transitions to the optically thin regime, see Section \ref{Sec:SED}).
Using the bolometric luminosity of Fig. \ref{Fig:Lbol}, we can therefore constrain 
the properties (density and mass) of the environment as sampled by the 2012b
explosion.

We find that the total mass swept up by the shock from $t_{\rm{w}}=32$ days until the
end of our monitoring  (112 days since explosion) is $M_{\rm{w}}^{\rm{thin}}
\approx (0.05/\eta)\rm{M_{\rm{\sun}}}$. The total mass in the environment swept up
by the 2012b explosion shock is therefore $M_{\rm{tot}}=M_{\rm{w}}+M_{\rm{w}}^{\rm{thin}}
\approx (0.2-0.3) \rm{M_{\rm{\sun}}}$ for $\eta=50-30\%$. As a comparison,
\cite{Ofek13} derive a mass of $\sim 0.1\rm{M_{\rm{\sun}}}$.
Our analysis points to a steep density
profile with $\rho_{\rm{w}}\propto R^{-5.3}$ for $R>1.4\times 10^{15}\rm{cm}$. The 
mass-loss rate is $\dot M(R)=w(R)v_{\rm{w}}(R)$.
We estimate $v_{\rm{w}}$ from the evolution of the FWHM of the narrow H$\alpha$ 
component in Fig. \ref{Fig:Halpha_evol}. Combining this information with the expression 
above we find $\dot M(R)\propto R^{-2}$ for $(1.4<R<4.4)\times 10^{15}\rm{cm}$,
with $\dot M(R)\approx (0.08/\eta)\rm{M_{\sun}/yr}$ at $R=1.4\times 10^{15}\rm{cm}$,
declining to $(0.008/\eta)\rm{M_{\sun}/yr}$ at $R=4.4\times 10^{15}\rm{cm}$.
\subsection{Origin of the interacting material in the close environment}
\label{SubSec:2012a}

During the 2012b explosion, the shock interacts with an environment
which has been previously shaped by the 2012a explosion and previous
eruptions. 
In this section
we infer the properties of the pre-2012a explosion environment,
using the 2012a outburst as a probe. We look to: (i) understand the
origin of the compact dense shell with which the 2012b shock interacted,
whether it is newly ejected material by the 2012a outburst or material originating
from previous eruptions; (ii) constrain the nature of the slowly moving 
material ($v\approx $ a few $  100\,\rm{km\,s^{-1}}$) responsible for narrow line emission 
in our spectra.

We put an upper limit on the total amount of mass in the surroundings
of SN\,2009ip \emph{before} the 2012a explosion assuming that the
observed luminosity of the 2012a outburst is entirely powered\footnote{Any additional
source of power would lower the required interacting mass.} by optically thin shock interaction with
some previously ejected material of mass $M_{\rm{w}}^{\rm{12a}}$. 
As before: $L=(\eta/2)wv^3$.  $M_{\rm{w}}^{\rm{12a}}=\int w(R)dR$. 
The evolution of the black-body radius of Fig. \ref{Fig:Lbol} suggests
$v\approx 2500\,\rm{km\,s^{-1}}$ before 2012 August 21. We apply the same
line of reasoning as above and assume that the shock continues to 
expand with this velocity, while the photosphere transitions to the
thin regime and stalls at $\approx 0.4\times 10^{15}\,\rm{cm}$. 
In this picture, the 2012a shock sampled the environment on distances 
$R<1.2\times 10^{15}\rm{cm}$, sweeping up a total mass of 
$M_{\rm{w}}^{\rm{12a}}\leq (0.02/\eta)\rm{M_{\sun}}$. 
In Section \ref{SubSec:ShockBO} we estimated that the total mass 
of the dense wind shell from which the 2012b shock breaks out is
$M_{\rm{w}}\approx 0.1\,\rm{M_{\sun}}$ with radius
$R_{\rm{w}}\approx 1.2\times 10^{15}\rm{cm}$. The wind shell mass is $M_{\rm{w}}=
M_{\rm{w}}^{\rm{12a}}(R<R_{\rm{w}})+M_{\rm{ej}}^{\rm{12a}}(R<R_{\rm{w}})$,
where $M_{\rm{ej}}^{12a}(R<R_{\rm{w}})$ is the portion of the ejecta mass 
of the 2012a explosion within $R_{\rm{w}}$ at $t=t_{\rm{w}}$\footnote{Assuming MJD 56140 
(2012 August 1) as zero-time for the 2012a outburst, this would correspond to ejecta with
velocity $\leq 2000\,\rm{km\,s^{-1}}$ for free expansion.}. This implies 
$M_{\rm{ej}}^{12a}(R<R_{\rm{w}})>0.06\,\rm{M_{\sun}} (>0.09\,\rm{M_{\sun}})$
for our fiducial efficiency $\eta=30\%$ ($\eta=50\%$), comparable with
the mass of the dense wind shell. We conclude that the UV-bright 2012b explosion
results from shock break-out from a dense shell which 
mostly (if not entirely) originates from the ejecta mass of the 2012a
explosion, therefore establishing a direct connection between the properties of the
2012a-2012b episodes.


The previous result also implies a solid lower limit on the \emph{total} ejecta mass 
of the 2012a outburst: $M_{\rm{ej}}^{12a}>0.06\,\rm{M_{\sun}} (>0.09\,\rm{M_{\sun}})$
for $\eta=30\%$ ($\eta=50\%$). In Section  \ref{SubSec:ShockBO} we estimated that
the total mass collected by the 2012b shock by the end of our monitoring is
$M_{\rm{tot}}\approx 0.2-0.3\,\rm{M_{\sun}}$, which constrains $0.06<M_{\rm{ej}}^{12a}<0.3
\,\rm{M_{\sun}}$ for $\eta \geq 30$\%. In the following we use $M_{\rm{ej}}^{12a}\approx 0.1 
\rm{M_{\sun}}$ as an order of magnitude estimate for the mass ejected by the
2012a outburst.\footnote{Strictly speaking, we are only sensitive to the
2012a ejecta mass that has been overtaken by the 2012b explosion by the end
of our monitoring. However, the analysis by \cite{Pastorello12} shows evidence for 
strong deceleration of the 2012a ejecta by 2012 September 15, which suggests that most 
of $M_{\rm{ej}}^{12a}$ has been encompassed by the 2012b explosion $\sim100$ days after.}

Our spectra show evidence for narrow line emission (Section \ref{Sec:SpecLinesOpt})
typically observed in SNe IIn (and LBVs), which is 
usually interpreted as signature of the ejecta interaction with material deposited
by the progenitor wind before explosion. For SN\,2009ip we observe during the 2012b 
event a velocity gradient in the narrow emission from H$\alpha$ (Fig. \ref{Fig:Halpha_evol},
panel \emph{c}), with \emph{increasing} velocity with time. This
increase is consistent with being linear with time. This might suggest
a Hubble-like expansion for the CSM following the simple velocity profile $v\propto R$: 
as time goes by, the shock samples material at larger distances from the explosion 
(hence with larger velocity $v$). Our analysis indicates that episodes of mass ejection 
with approximate age 11-19 months before the 2012b explosion (roughly between
February and October 2011) might reasonably account for the observed velocity 
gradient. We suggest that  CSM material in the
surroundings of SN\,2009ip moving at velocities of  hundreds $\rm{km\,s^{-1}}$ originates
from this sudden episode(s) of mass ejection. Remarkably, 
SN\,2009ip has been reported to be in eruptive phase between May and 
October 2011 \citep{Pastorello12}, consistent with this picture.

The Hubble-like flow is \emph{not} consistent with a steady wind and points
instead to some mechanism leading to explosive mass ejections. Interestingly,
it is during the September 2011 outburst that SN\,2009ip showed
evidence for material with unprecedented velocity, reaching $v=12500 \,\rm{km\,s^{-1}}$
\citep{Pastorello12}.
Since no line-driven or continuum-driven wind mechanism is known to be able
to accelerate stellar surface material to these velocities (\citealt{Mauerhan12}),
stellar-core related mechanisms have to be invoked. The explosive mass
ejection suggested by our analysis might therefore be linked to
instabilities developing deep inside the stellar core. 

\subsection{The role of asymmetries in SN\,2009ip}
\label{SubSec:Asymmetry}

The analysis of Section \ref{SubSec:absorption} indicates the presence of
ejecta traveling at three distinct velocities:
$v\sim -12000\,\rm{km\,s^{-1}}$, $v\sim -5500\,\rm{km\,s^{-1}}$ and
$v\sim -2500\,\rm{km\,s^{-1}}$. These values correspond to the velocity
of material seen in \emph{absorption} (i.e. placed outside the photosphere). 
The radius of the hot blackbody $R_{\rm{HOT}}$ of Fig. \ref{Fig:Lbol} tracks the position of
the photosphere with time. Assuming free expansion of the ejecta 
and the explosion onset time ($t_{\rm{pk}}-20$ days) derived in the previous sections, we can predict at 
which time $t_{v}$  ejecta moving at a certain velocity $v$ will overtake the photosphere 
at $R_{\rm{HOT}}$. Only for $t\gtrsim t_{v}$ can the ejecta give rise to absorption 
features in the spectra. Spherical symmetry is an implicit assumption 
in the calculation of $R_{\rm{HOT}}$, so that comparing the predicted $t_{v}$
to the observed time of appearance of the absorption edges makes it possible to 
test the assumption of spherical symmetry of the explosion.

For $v\sim -2500\,\rm{km\,s^{-1}}$ we find $t_{\rm{v}}= t_{\rm{pk}}+55$ days (2012 November 27) in excellent
agreement with our observations, which constrain the $v\sim -2500\,\rm{km\,s^{-1}}$ absorption
edge to appear between 2012 November 23 and December 5 (Fig: \ref{Fig:absorptions}). 
No departure from spherical symmetry needs to be invoked for slow-moving 
ejecta, which likely includes most of the ejecta mass.\footnote{Note however that in no way this argument
can be used as a proof of spherical symmetry.} 
Spherical symmetry is instead clearly broken by the high-velocity material traveling at $v\sim -12000\,\rm{km\,s^{-1}}$.
For the 2012b explosion we detect high velocity material in absorption starting from $\sim1$  week
after peak. Around peak the spectrum of SN\,2009ip is optically thick and shows
no evidence for material with  $v\sim -12000\,\rm{km\,s^{-1}}$ (Fig. \ref{Fig:Xshooter1}
and Fig. \ref{Fig:Xshooter2}). However, in no way could a perfectly spherical photosphere 
traveling at $4000-5000\,\rm{km\,s^{-1}}$ mask the fast-moving ejecta at \emph{any} time
during the evolution, and in particular until the first week after peak,
as we observed. This indicates a departure of the high-velocity \emph{ejecta} from spherical geometry and might
suggest the presence of a preferred direction in the explosion.

Asymmetry can also have a role in the spatial distribution of the \emph{interacting material},
as supported by the observed co-existence of broad and (unresolved) intermediate 
components in the spectrum (Fig. \ref{Fig:Halpha}). 
In this respect, \cite{Chugai94} proposed the possibility of an enhanced
mass loss on the equatorial plane of SNe IIn to explain the intermediate
velocity component in SN 1988Z, other explanations being a clumpy 
circumstellar medium or, again, an asymmetric flow. In this context it is worth noting
that asymmetry is also a likely explanation for the discrepant mass-loss estimates
found by \cite{Ofek13}, as noted by the authors. A disk-like geometry
for SN\,2009ip was proposed by \cite{Levesque12} based on the H
Balmer decrement. Finally, the binary-star merger scenario proposed by 
\cite{Soker13} to interpret SN\,2009ip naturally leads to ejecta
with a bipolar structure.

\subsection{Hard X-rays and X-rays from shock break-out and continued interaction}
\label{SubSec:XraysBO}

\begin{figure}
\vskip -0.0 true cm
\centering
\includegraphics[scale=0.4]{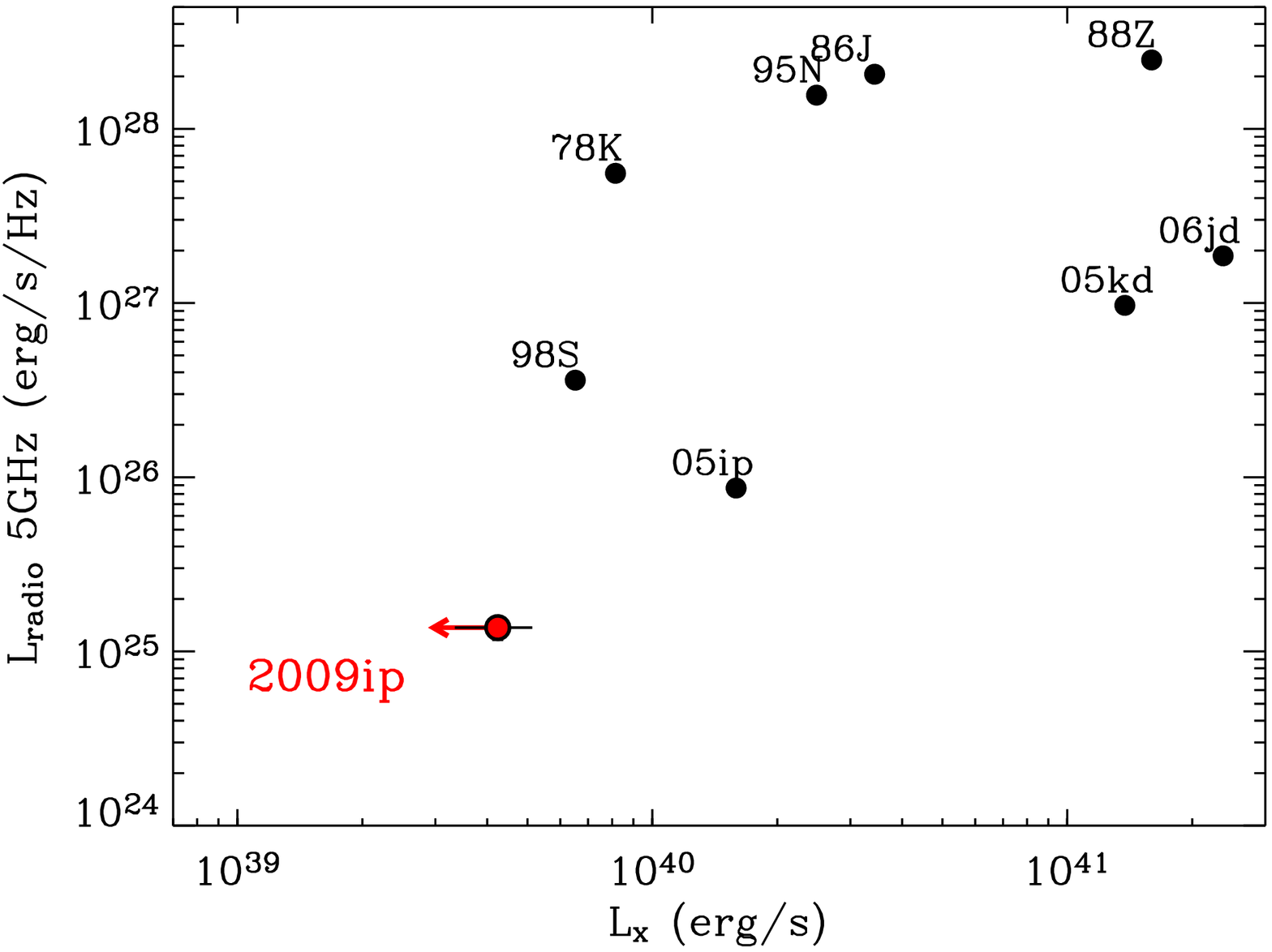}
\caption{Radio luminosity at peak vs. X-ray luminosity at the radio peak 
for a sample of Type IIn SNe (black dots). SN\,2009ip is shown in red.
Data have been collected from the literature. (See \citealt{Pooley02}; \citealt{Smith07};
\citealt{Stritzinger12}; \citealt{Chandra12a}; \citealt{Pooley07}; \citealt{Chandra07};
\citealt{Chandra09}; \citealt{Zampieri05}; \citealt{Houck98}; \citealt{Chevalier87};
\citealt{VanDyk93}; \citealt{Fabian96}; and references therein).}
\label{Fig:XrayRadio}
\end{figure}

\begin{figure*}
\vskip -0.0 true cm
\centering
\includegraphics[scale=0.65]{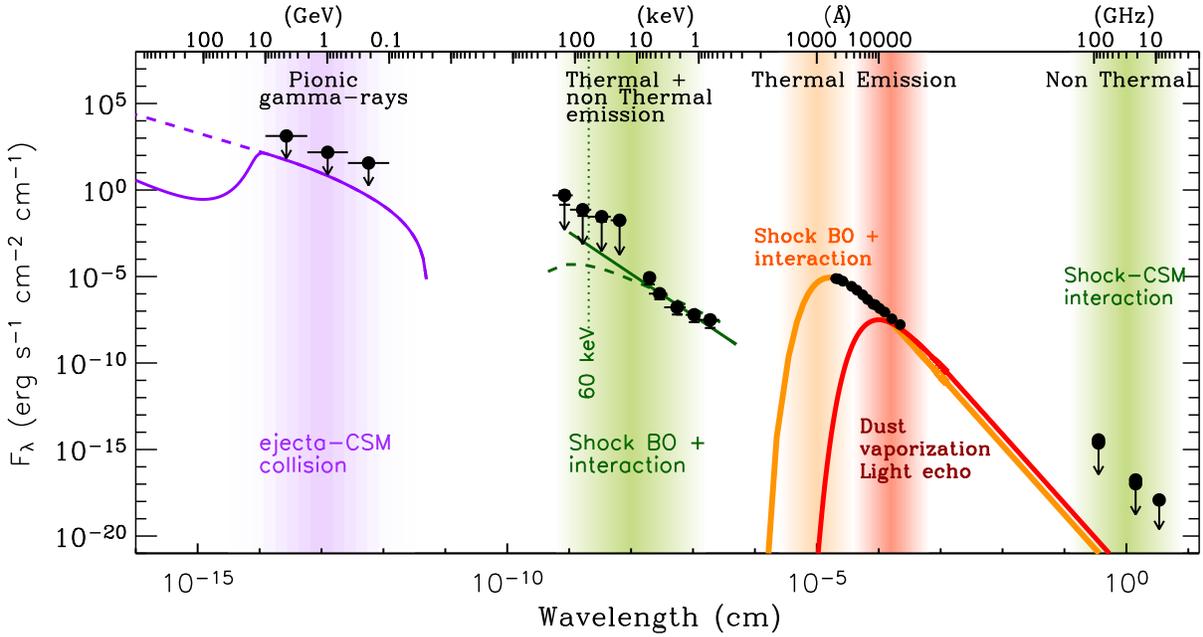}
\caption{Broad-band SED of SN\,2009ip around the optical peak. Shaded bands
highlight different components of emission that dominate at different
wavelengths. Neutral pion decay leads to $\gamma$-rays which are ultimately
powered by the collision between the ejecta and extremely dense CSM shells of 
material. We show here the limits obtained by \textit{Fermi}-LAT in the week around maximum light
(from $t=t_{\rm{pk}}-2$ days until $t=t_{\rm{pk}}+4$ days). Dashed
purple line: expected GeV emission based on the shock explosion parameters
derived in Section \ref{SubSec:ShockBO} and the model by \cite{Murase11} assuming no 
attenuation. The thick-line model includes partial attenuation as relevant for a clumpy 
and/or asymmetric medium (Sec. \ref{SubSec:Asymmetry}) 
with filling factor $<1$ that reduces the effects of  the Bethe-Heitler process. 
Soft and hard X-ray emission originates from the interaction of the shock with the CSM: 
both thermal and non-thermal emission might arise, mainly depending on the environment 
density. We show here a power-law (non-thermal emission, thick line with 
$F_{\lambda}\propto \lambda^{-2.4}$)  and a thermal bremsstrahlung model with $kT=60$ 
keV (dashed line) that fit the \emph{Swift}-XRT 0.3-10 keV data obtained between $t_{\rm{pk}}-2$
days and $t_{\rm{pk}}+4$ days. \emph{Swift}-BAT 15-150 keV data  acquired 
at $t_{\rm{pk}}\pm1$ days are also shown.  The dotted line marks 
the energy of 60 keV: this is the typical frequency of high-energy photons emitted by the
collisionless shock that forms at shock break-out (\citealt{Katz11}). Optical and UV emission 
is powered by the shock break-out plus continued shock interaction with the environment.
The orange thick line is the best-fitting "hot" black-body component with $T\approx
17000$ K obtained at maximum light. NIR and IR emission could originate from dust formation, dust vaporization but 
it might also be a light echo. Due to the high temperature we infer ($T\approx 3000$ K, red solid line)
we favor  dust vaporization here. Shock-CSM interaction is also the source
of millimeter and radio photons through synchrotron emission. At these times 
this emission was quenched by free-free absorption by the dense shell of material.}
\label{Fig:superSED}
\end{figure*}

SN\,2009ip is a weak X-ray and radio emitter (Fig. \ref{Fig:XrayRadio}). In the following two
sections  we connect the lack of high X-ray/radio luminosity to the 
shock break-out plus interaction scenario we developed to explain the optical
properties of SN\,2009ip.

The shock break-out plus continued interaction scenario gives rise to a two-component
spectrum, with a hard (X-rays) and a soft (UV/optical, at the break-out velocity of interest here) 
component (e.g. \citealt{Svirski12}; \citealt{Chevalier12}).
The hard component is generated by bremsstrahlung emission from hot electrons 
behind the shock. Theory predicts that 
$L_{\rm{hard,bo}}\sim 10^{-4}L_{\rm{bo}}$ (where $L_{\rm{bo}}$ is the break-out 
luminosity, resulting from the soft and hard component)
as long as: (i) Inverse Compton (IC) cooling dominates over bremsstrahlung; 
(ii) high-energy photons undergo Compton degradation in the unshocked wind during
their diffusion to the observer. We show in the following that both processes are
relevant for SN\,2009ip and provide a natural explanation for its very low X-ray to optical 
luminosity ratio ($L_{\rm{x}}/L_{\rm{optical}} \lesssim 10^{-4}$).


From our modeling
of Section \ref{SubSec:ShockBO}, we inferred a density parameter $D_{*}\approx 0.4$ 
(where $D_{*}$ is a measure of the density of the dense wind shell $\rho_{\rm{w}}$,
as explained in Appendix \ref{App:ShockBreakOut}). With a shock velocity $v\approx
4500\,\rm{km\,s^{-1}}$, the density measurement above implies that around break-out time
the main source of cooling of the hot electrons is IC (see \citealt{Chevalier12}). For our parameters, 
the IC to Bremsstrahlung (ff) emissivity ratio at break-out is (\citealt{Svirski12},
their Eq. 17) $\epsilon^{IC}/\epsilon^{ff} \approx0.01 (v/10^{9}\rm{cm\,s^{-1}})^2$ or
$\epsilon^{IC}/\epsilon^{ff} \approx 0.05$ for the observed $v\approx4500\,\rm{km\,s^{-1}}$.
IC is the dominant cooling source, suppressing the emission of hard photons in SN\,2009ip.
The calculations by \cite{Ofek13} instead assume negligible IC cooling. 

Comptonization of the hard photons as they propagate through the unshocked wind region
to the observer furthermore leads to a suppression of high-energy radiation. This
process can effectively suppress photons
with $\approx$keV energy  if $\tau_{\rm{es}}\gtrsim 15-20$, the photon energy
being limited by $\epsilon_{\rm{max}}=511/\tau_{\rm{es}}^2$ keV. Our modeling of 
Section \ref{SubSec:ShockBO} implies $\tau_{es}\approx 15$ for SN\,2009ip around 
shock break-out. This demonstrates that both the domination of IC over bremsstrahlung 
(i) and Compton losses (ii) are relevant to explain the weak X-ray emission in SN\,2009ip.
Identifying $L_{\rm{x}}$ with $L_{\rm{hard}}$ and $L_{\rm{bol}}$ with $L_{\rm{bo}}$,
the shock break-out scenario therefore naturally accounts for the
observed $L_{\rm{x}}/L_{\rm{optical}} \lesssim 10^{-4}$ ratio even in the absence of 
photo-absorption.

\cite{Chevalier12} calculate that full ionization (which gives minimal photo-absorption) 
is achieved for high-velocity shocks with $v\gtrsim 10^4\rm{km\,s^{-1}}$. For SN\,2009ip  
$v\approx 4500\,\rm{km\,s^{-1}}$ likely leads to incomplete ionization (i.e. potentially 
important photo-absorption), that will further reduce the escaping X-ray flux. 
Using the explosion observables and Eq. \ref{Eq:N} we constrain the total column density 
between the shock break-out radius and the observer to be $N\approx 2\times 10^{25}
\rm{cm^{-2}}$ which gives a bound-free optical depth  $\tau_{\rm{bf}}\approx2\times10^{3}$
at 1 keV (Eq. \ref{Eq:taubf}).\footnote{This calculation assumes a neutral medium and adopts the 
approximation by \cite{Ofek13} which is accurate within a factor of 2 with
respect to more detailed calculations of the cross section by \cite{Morrison83}.}
Since $\tau_{\rm{bf}}\propto R^{-1}$ and $R\propto t$, soft ($E\approx2$ keV)
X-ray emission would $\emph{not}$ be expected from SN\,2009ip until $R\approx 2\times10^{16}
\rm{cm}$ which happens for $\Delta t\gtrsim 440$ days since the explosion, or 
 $\gtrsim 44$ break-out time scales $ t_{\rm{d}}$
(where $t_{\rm{d}}\approx t_{\rm{rise}}$ in our scenario).\footnote{Adopting standard
parameters \cite{Svirski12} find
that under standard parameters the X-ray emission is expected to dominate the energy release
on time scales of the order of $10-50\,t_{\rm{d}}$.} This calculation assumes an extended
and spherically symmetric wind profile.  The Chandra \emph{detection} of SN\,2009ip
at much earlier epochs ($\Delta t \approx 4\,t_{\rm{d}}$) indicates that at least one 
of these assumptions is invalid, therefore pointing to a truncated and/or highly asymmetric
wind profile. This is consistent with the picture of a dense but compact wind shell of radius 
$R_{\rm{w}}\approx 1.2\times 10^{15}\rm{cm}$ followed by a steep density gradient 
$\rho_{\rm{w}}\propto R^{-5.3}$ we developed in Section \ref{SubSec:ShockBO}.
Asymmetry also plays a role, as independently suggested by observations
in the optical  (Section \ref{SubSec:Asymmetry}).

Finally, \cite{Katz11} predict that hard-X-ray emission with typical energy of $60$ keV
is also produced by the collision-less shock. The details of the spectrum are however unclear.
Bound-free absorption is less important at these energies giving the chance to detect
hard X-rays at earlier times. Our \emph{Swift}-BAT campaign in the 15-150 keV range
revealed a tentative detection. With these observations we put a solid upper limit on the hard X-ray to 
optical luminosity around maximum light which is $L_{\rm{X,hard}}/L_{\rm{bol}}<5\times 10^{-3}$
at $5\sigma$.  The broad band SED around maximum light is shown in 
Fig. \ref{Fig:superSED}.
\subsection{Radio emission from shock break-out and continued interaction}
\label{SubSec:RadioBO}

The shock-CSM interaction is a well known source of radio emission (e.g. \citealt{Chevalier82};
\citealt{Chevalier84}).
The limited shock velocity we infer for SN\,2009ip likely leads to a
partially ionized medium as discussed above, so that free-free absorption
plays a key role in suppressing the emitted radio flux. We quantify this statement below.
The high density of the wind
shell derived from our modeling of Section \ref{SubSec:ShockBO} implies
a free-free optical depth at radius $R$, $\tau_{\rm{ff}}\approx 3.8\times 10^{54}(\nu/\rm{GHz})
^{-2.1}(R/\rm{cm})^{-3}$  (Eq. \ref{Eq:tauff}). With $\tau_{\rm{ff}}\approx 10^5-10^8$
at $R\approx R_{\rm{bo}}$, no detectable radio emission is expected around
break-out, consistent with our lack of radio detection around these times.
\emph{If} the dense wind profile extends out to large distances, the calculation
above shows that no radio emission is expected until very late times, 
when the shock reaches $R\approx 7\times 10^{16}\rm{cm}$. Our radio
detection at much earlier epochs ($\Delta t\approx 60-80$ days since explosion)
demonstrates instead that the dense wind shell is \emph{not} extended
but  truncated and adds further, independent evidence for  a complex medium where
inhomogeneity, asymmetry and/or low wind filling factor might also be relevant.
This is consistent with the idea we suggested in Section \ref{SubSec:2012a}
that the dense shell was the outcome of a short eruption, 
since  eruptions are more likely to eject shells with sharp density edges as opposed
to a steady mass loss.

\subsection{GeV and neutrino emission at shock break-out}
\label{SubSec:GeVNeutrino}

\begin{figure}
\vskip -0.0 true cm
\centering
\includegraphics[scale=0.43]{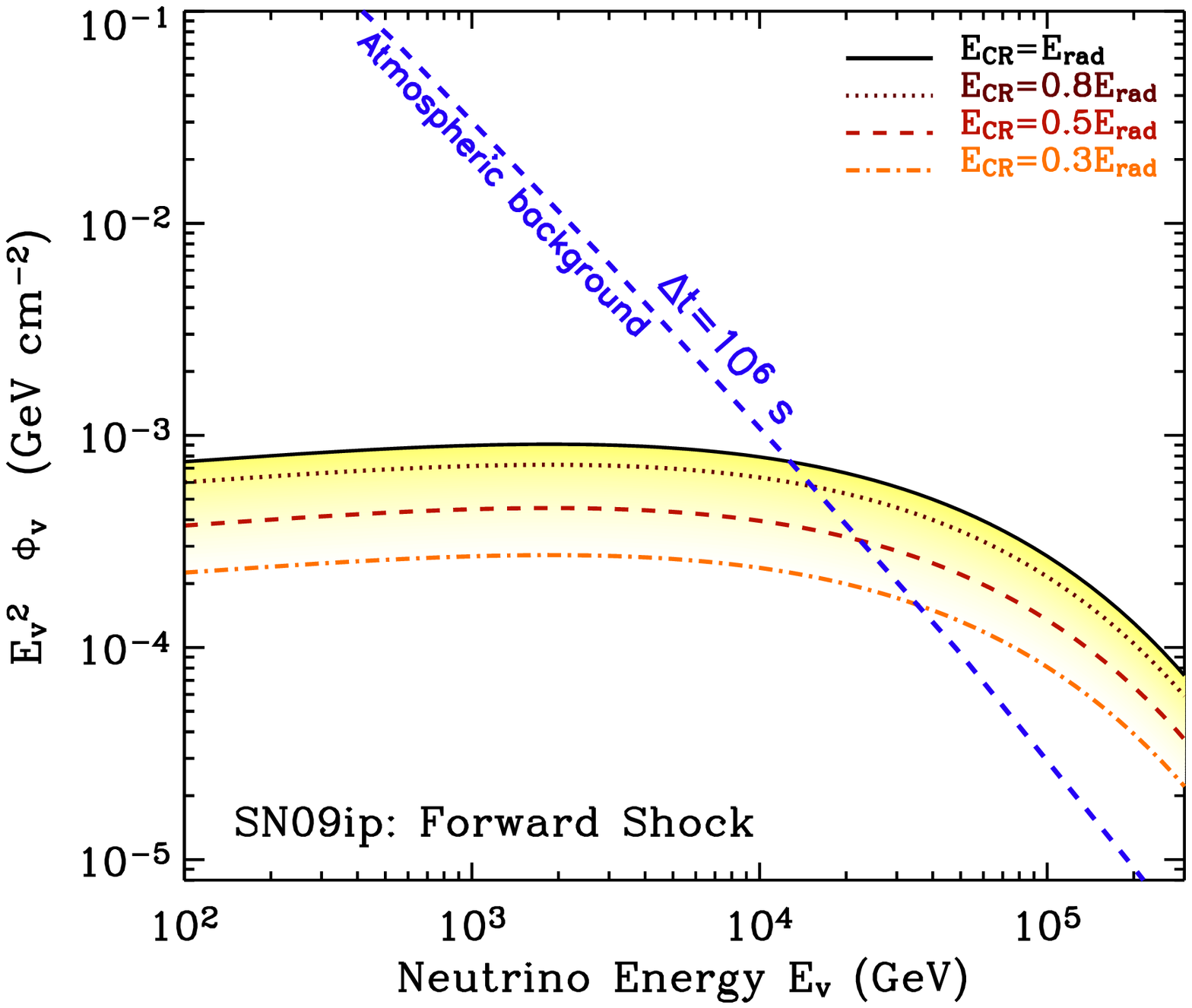}
\caption{Predicted muon and anti-muon neutrino fluence from SN\,2009ip using the
observables and explosion parameters of Sec. \ref{SubSec:ShockBO} according to 
the model by \cite{Murase11}. The shaded band corresponds to different
values for the total energy in cosmic rays $E_{\rm{CR}}$ here parametrized as a function
of the radiated energy at $t_{\rm rise}$. Note that, given $E_k \gg
E_{\rm rad}$, larger values of $E_{\rm CR}$ are also possible. Dashed blue line: 
atmospheric neutrino background integrated
over $\Delta t\approx t_{\rm{rise}}$ (zenith-angle averaged
within 1$\degr$), which is shown for comparison. For this event, the
atmospheric neutrino background is more severe since 
SN\,2009ip occurred in the southern hemisphere.
For better localized explosions, this plot shows how limits on the neutrino 
emission can be used to constrain $E_{\rm{CR}}$.}
\label{Fig:Neutrino}
\end{figure}

In the previous sections we suggested the presence of a dense ($n\approx 4\times 
10^{10}\rm{cm^{-3}}$ at the break-out radius) and compact shell in the close
environment of SN\,2009ip that we associate with material ejected during the 2012a outburst. 
Shock break-out from this shell powers the major re-brightening of SN\,2009ip 
at optical wavelengths (2012b explosion) and naturally explains the weak
radio and X-ray emission we observe.
The collision of the ejecta with massive shells is also 
expected to accelerate cosmic rays (CRs) and generate GeV gamma-rays
(\citealt{Murase11, Katz11}) with fluence
that depends both on the explosion and on the environment parameters. 
The GeV emission is expected to be almost simultaneous with the UV-optical-NIR emission.
The proximity of SN\,2009ip ($\approx 24$ Mpc) justifies the first search
for GeV emission from shock break-out. \emph{Fermi}-LAT observations
covering the period 2012 September 3 -- October 31  ($t_{\rm{pk}}-30$
days until $t_{\rm{pk}}+28$ days) yielded no detection (Sec. \ref{SubSec:GeVObs}). 
Following \cite{Murase11}, we predict the GeV emission from SN\,2009ip
(2012b outburst) using the explosion and environment parameters inferred in Sec.
\ref{SubSec:ShockBO} (shell density $\rho_{\rm{w}}$ or mass $M_{\rm{w}}$) 
together with the observables of the system ($t_{\rm{rise}}$, $R_{\rm{bo}}$). 
We take into account the attenuation due to $\gamma \gamma \rightarrow e^+ e^-$
pair production, assuming a black-body spectrum with $T=20000$ K.
The attenuation by extragalactic background light is also included.
Since $n$ is not too large, the Bethe-Heitler pair production will be irrelevant
in our case, as it happens for a clumpy  or
asymmetric CSM (Sec. \ref{SubSec:Asymmetry}) with filling factor $<1$.
The injected CR energy $E_{\rm CR}$ is assumed to be equal to that of
the radiation energy at $t_{\rm rise}$.
We compare the expected GeV fluence with observations in
Fig. \ref{Fig:superSED}: the \textit{Fermi}-LAT non-detection is consistent with our
picture of ejecta crashing into a compact and dense but low-mass ($\lesssim0.1-0.2
\,\rm{M_{\sun}}$) shell of material at small radius that we developed in the previous sections. 
For the detection of $\gamma-$rays, brighter SNe (closer SNe or SNe accompanying 
larger dissipation) are needed.

Using the set of parameters above we predict the muon and anti-muon
neutrino fluence from SN\,2009ip in Fig. \ref{Fig:Neutrino}, using the model by \cite{Murase11}. 
CRs produce mesons through inelastic proton-proton scattering, leading to neutrinos 
as well as $\gamma-$rays. As for the $\gamma-$rays, given that the explosion/ environment parameters are
constrained by observations at UV/optical/NIR wavelengths, the
neutrino fluence directly depends on the total energy in CRs, $E_{\rm{CR}}$ (or, equivalently
on the CR acceleration efficiency $\epsilon_{\rm{CR}}$, $E_{\rm{CR}}\equiv \epsilon_{\rm{CR}}E$,
being $E$ the explosion energy). Note that the injected CR energy $E_{\rm CR}$ is assumed to be the
radiation energy at $t_{\rm rise}$, but larger values of $E_{\rm CR}$ are also possible. 
The maximum proton energy is also set to $1.5$ PeV, corresponding to
$\varepsilon_B=0.01$. Our calculations show that SN\,2009ip was a bright source of neutrinos
if $E_{\rm{CR}}$ is comparable to the radiated energy.
We note however that its location in the souther sky was not optimal for searches for neutrino emission
by IceCube because of the severe atmospheric muon background. 
With this study we demonstrate how it is possible to constrain the CR acceleration efficiency if:
(i) the explosion/ environment parameters are constrained by observations at
UV/optical/NIR wavelengths; (ii) deep limits on muon neutrinos are available, as
 will be the case also for southern sky sources once KM3Net will be online in the near future.
\subsection{Origin of the NIR excess}
\label{SubSec:NIRexcess}

The time-resolved broad-band SED analysis of Section \ref{Sec:SED} identifies
the presence of a NIR excess of emission (Fig. \ref{Fig:SED}) and allows
us to constrain its temporal evolution (Fig. \ref{Fig:Lbol}).
Contemporaneous NIR spectroscopy from Section \ref{SubSec:NIRspec}
clearly shows that the NIR excess cannot be ascribed to line emission 
(Fig. \ref{Fig:NIRSpec}), therefore pointing to a physical
process producing NIR \emph{continuum} emission.
Adopting a black-body spectral model (with the implicit assumption of 
spherical symmetry) we find that: (i) the equivalent black-body radius is
$R_{\rm{COLD}}\sim4\times 10^{15}\,\rm{cm}$ with very limited evolution with time
over 30 days of monitoring.\footnote{Our observations imply $v_{\rm{COLD}}< 10^3
\,\rm{km\,s^{-1}}.$} 
This is in contrast with the hot component 
radius $R_{\rm{HOT}}$ which increases linearly with time \emph{inside} the cooler component.  (ii)
The cold black-body temperature is also stable, with $T_{\rm{COLD}}\sim 3000\,\rm{K}$
(while the hot black body cools from $19000$ K to $<10000$ K). 
Considering that our fits can overestimate the true dust temperature of 
hundreds of degrees (up to 20\% according to \citealt{Nozawa08}) the real dust
temperature might be close to $\sim 2500$ K.
(iii) The resulting NIR excess luminosity is $L_{\rm{COLD}}\sim4\times 10^{41}\rm{erg\,s^{-1}}$
which represents $(2-4)$\% of $L_{\rm{HOT}}$.
We use these properties to constrain the origin of the NIR excess below. 

A clear spectroscopic
signature of dust formation is the development of highly asymmetric and blue-shifted
line profiles (see e.g. \citealt{Smith08}, their Fig. 4) which is not observed. 
The temperature of $\sim 2500-3000$ K is also prohibitively high for dust to form. At this very early epochs
the shock radius is also $<R_{\rm{COLD}}$. Note that  $R_{\rm{COLD}}$, derived assuming
a black-body spectrum, represents a lower limit to the real size $R_{\rm{NIR}}$ of the NIR emitting region
(Fig. \ref{Fig:Sketch1.pdf}): $R_{\rm{NIR}}\propto R_{\rm{COLD}}/f^{0.5}$, 
where $f<1$ is the covering factor of the NIR emitting material. $f<1$ is indeed
required for the hot black-body radiation to be able to escape.
This clearly implies that the NIR emission cannot originate from dust created \emph{behind} the 
reverse shock.  
The dust creation scenario is therefore highly unlikely, leading to the conclusion
that the NIR excess originates from pre-existing material ejected by the progenitor before the 2012b explosion.
Since the geometry of the NIR emitting material can be non spherical (as we find in Section \ref{SubSec:Asymmetry}) and/or
have low filling factor (i.e. the material can be clumpy), we do not expect this material to 
necessarily produce absorption along our line of sight.

Material ejected during the 2012a outburst would be required to travel at an average
velocity $v> 8000\,\rm{km\,s^{-1}}$ (to reach $4\times 10^{15}\rm{cm}$ at the
time of the 2012b explosion) and then to decelerate to $v< 1000\,\rm{km\,s^{-1}}$ 
(to match the observed  evolution of $R_{\rm{COLD}}$). We consider this scenario
unlikely\footnote{Our SED fitting indicates the
presence of a NIR excess with similar radius during the 2012a 
eruption as well (Fig. \ref{Fig:Lbol}), adding further evidence that the 
NIR emitting material pre-existed the 2012a episode.} and suggest that the origin of the pre-existing material is rather linked
to the eruption episodes of the progenitor of SN\,2009ip in the \emph{years} before.
The same conclusion was independently reached by \cite{Smith13}.
We note that the size of the cool emitting region of SN\,2009ip is remarkably similar 
to the pre-outburst dust radii of other optical transients linked to eruption
episodes of their progenitor stars like NGC 300 OT2008-1 ($R\sim 5\times 10^{15}\,\rm{cm}$) 
and SN\,2008S ($R\sim 2\times 10^{15}\,\rm{cm}$, see e.g. \citealt{Berger09b}, their Fig. 28).
 
NIR emission in Type II SNe has also been connected to the extended atmosphere of the 
expanding star (e.g. \citealt{Dwek83}). However, for SN\,2009ip, the large radius we infer from our modeling 
($R_{\rm{COLD}}\sim4\times 10^{15}\,\rm{cm}$) and the lack of a clear evolution of the
temperature with time  argue against this interpretation.

A light echo from dust (i.e. pre-existing dust heated up by the UV and optical radiation from the explosion)
would require the dust grains to \emph{survive} the harsh environment.
At the high temperature of $T\sim 2500-3000\,\rm{K}$ this is however unlikely, while 
dust \emph{vaporization} is more likely to happen (see e.g. \citealt{Draine79}). 
We speculate on the dust vaporization scenario below.

In this picture a cavity is excavated by the explosion radiation out to a radius $R_{\rm{c}}$:
this radius identifies the position of the vaporized dust, while it does not track the outer
dust shell radius (which is instead likely to expand with time, see e.g. \citealt{Pearce92}). 
Being the dust shell created in the years before the 2012 explosion,
we expect the smaller grains to be located at the outer edge of the dust shell as
a result of forces acting on them (e.g. radiative pressure).  At smaller distances
we are likely dominated by the larger dust  grains. Following \cite{Dwek83}, their Eq.
8 (see also \citealt{Draine79})
the radius of the dust-free cavity for a UV-optical source with luminosity $L$ is:
$R_{\rm{c}}=23 (\langle Q \rangle (L/\rm{L_{\sun}})/(\lambda_0/\rm{\mu m}) T^{5})^{0.5}$,
where $R_{\rm{c}}$ is in units of pc; $T$ is the temperature at the inner boundary 
of the dust shell which we identify with the evaporation temperature $T_{\rm{ev}}$;
$\langle Q \rangle$ is the grain emissivity: $ Q=(\lambda_0/\lambda)^n$ for
$\lambda \geq \lambda_0$ while $Q=1$ for $\lambda < \lambda_0$. Here we adopt
$n=1$.\footnote{A value of $n$ between 1 and 2 is usually assumed. 
Using $n=2$ have no impact on our major conclusions. In particular,
our estimate of the dust grain radius would be $a\approx 1 \rm{\mu m}$.}
The dust grain radius is $a$, with $\lambda_0\sim 2a$ which implies 
$R_{\rm{c}}\propto a^{-0.5}$. For $R_{\rm{c}}= R_{\rm{COLD}}\sim 4\times 10^{15}
\rm{cm}$, $T_{\rm{ev}}\sim
3000$ K and $L=L_{\rm{pk}}\sim 1.7\times 10^{43}\rm{erg\,s^{-1}}$, we constrain
the radius of the vaporizing dust grains to be $a\lesssim(2-5)\rm{\mu m}$.

Grain radii of $0.2-2\,\rm{\mu m}$ are typically found in dust shells, suggesting that
we are possibly witnessing the vaporization of the larger grains at $R\sim R_{\rm{c}}$.
The very high vaporization temperature is only potentially compatible with
materials like graphite, silicates, corundum and carbide, that have binding
energy 
$U_0\gtrsim 6\,\rm{eV}$. Even in these cases
$T_{\rm{ev}}\sim2500-3000$ K requires extremely short vaporization time scales of the 
order of $\sim1$ day or less, and large grain dimensions of the order of $1\,\rm{\mu m}$
(see \citealt{Draine79}, their Eq. 24).

The passage of a SN explosion shock through a dust shell is one of the processes
believed to establish the radius distribution of dust particles in our Universe.  
The explosion shock will eventually interact with the NIR emitting "shell"  
likely causing a flattening in the light-curve decay.
This flattening should be then followed by a rapid decline once the shock reaches 
the edges of the NIR "shell". The timing of the interaction is however critically 
dependent on the velocity of the shock, the asymmetry of the explosion, the dust shell
filling factor and/or the asymmetry of the dusty material (but also on the expansion velocity of the
shell).\footnote{High-velocity ejecta with $v \sim 10000\,\rm{km\,s^{-1}}$  reached 
$R_{\rm{COLD}}$ at $t_{\rm{pk}}+26$ days, which is interestingly close
to the time of the second major peak in the bolometric light-curve in Fig. \ref{Fig:Lbol}.}
Allowing for a deceleration of the blast wave to $v \sim 2500\,\rm{km\,s^{-1}}$
and a low pre-existing dust filling factor $f=0.3$, we anticipate the shock - dust shell interaction 
to happen in the second half of 2013.  SN\,2009ip might be one of the rare cases 
where  it has been possible to map the properties of dust before and after the interaction with the 
explosion shock.  Future IR observations of  SN\,2009ip are of primary importance in this respect
and will clarify how newly condensed dust in the explosion ejecta mixes with pre-existing dust. 

Finally, we end by noting that a NIR echo from pre-existing \emph{gas} can possibly 
account for the high temperature of the NIR excess while 
naturally explaining the almost flat JHK light-curve around maximum light , compared to the
steeply decaying UV emission of Fig. \ref{Fig:OptPhot}. 
This possibility is further explored in a dedicated paper (Margutti et al. in preparation).
A complementary discussion of the NIR properties of SN\,2009ip can be found
in  \cite{Smith13} who favor the infrared echo hypothesis.
\section{Discussion}
\label{Sec:Discussion}

\begin{figure}
\vskip -0.0 true cm
\centering
\includegraphics[scale=0.55]{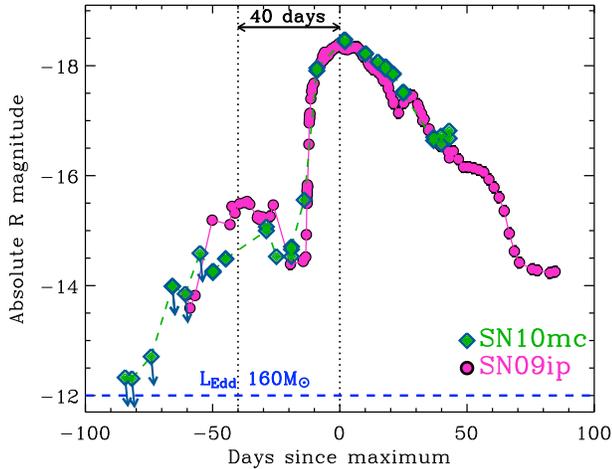}
\caption{The comparison of the absolute R band magnitude of SN\,2009ip and Type IIn
SN\,2010mc \citep{Ofek13b} reveals a striking similarity between the two explosions both 
during the precursor-bump and the major outburst. }
\label{Fig:09ipvs10mc}
\end{figure}

Our analysis  characterizes the 2012b episode as a low-energy, 
asymmetric explosion happening in a complex medium, previously shaped by multiple 
episodes of shell ejection by the progenitor at different times. Here we address three major questions:
\begin{itemize}
\item What is the nature of the SN\,2009ip double explosion in 2012? 
\item What is the underlying physical mechanism? 
\item What is the progenitor system of SN\,2009ip?
\end{itemize}

We address the first two questions by considering the close similarity of SN\,2009ip 
and SN\,2010mc (Section \ref{SubSec:09ip10mc}), the properties of the 
progenitor system of SN\,2009ip (Section \ref{SubSec:progenitor}),
the physical mechanisms
that can lead to sustained mass loss (Section \ref{SubSec:physicalmechanism}) and  the constraints
on the energetics we derived from our modeling (Section \ref{SubSec:nature09ip}).
\subsection{SN\,2009ip and SN\,2010mc}
\label{SubSec:09ip10mc}

In interpreting the fate and nature of SN\,2009ip we have to consider its
close likeness to SN\,2010mc \citep{Ofek13b}, the only other hydrogen-rich
explosion with clear signs of interaction and a detected precursor before
the major event. The similarity extends to the energetics, time scales 
(Fig. \ref{Fig:09ipvs10mc}) and spectral properties both during the precursor bump
and the major re-brightening, as noted by \cite{Smith13}.

The first important conclusion is that (i) the precursor 
and the major re-brightening are \emph{causally} connected events, being otherwise difficult
to explain the strictly similar phenomenology observed in SN\,2009ip and SN\,2010mc, 
two distinct and unrelated explosions. The same conclusion is independently 
supported by the very short time interval between the precursor and main
outburst when compared to the progenitor star lifetime, as pointed out by
\cite{Ofek13b} for SN\,2010mc.
A second conclusion is that (ii) whatever causes the precursor plus major outburst phenomenology,
this is \emph{not unique} to SN\,2009ip and might represent an important evolutionary channel
for massive stars. 

Furthermore, (iii) SN\,2009ip and SN\,2010mc must share some fundamental properties.
In particular their evolution through the explosive phase must be driven 
by \emph{few} physical parameters. A more complicated scenario would require
unrealistic fine tuning to reproduce the close similarity of SN\,2009ip and SN\,2010mc. 
This also suggests we are
sampling some fundamental step in the stellar evolution of the progenitor system.
Finally, (iv) whatever the physical mechanism behind, 
the time interval of $\sim40$ days (Fig. \ref{Fig:09ipvs10mc}) between the precursor explosion
and the main event must be connected to some physically important 
time scale for the system. 

We employ the fast $\chi^2$ technique for irregularly sampled
data by \cite{Palmer09} to search for periodicity and/or dominant
time-scales in the outburst history of SN\,2009ip \emph{before}\footnote{See
Martin et al., in prep. for a temporal analysis of SN\,2009ip during the
main episode of emission in 2012.} the major 2012 explosion. Details can be found in
Appendix \ref{App:tempan}. Applying the method above to the R-band data 
we find evidence for a dominant time-scale of $\sim38$ days, (with 
significant power distributed on time-scales between 30 and 50 days),
intriguingly similar to the $\Delta t\sim 40$ days between the precursor
and major explosion in 2012. We emphasize that this is not a
claim for periodicity, but the identification of a dominant variability time-scale 
of the signal.

Our analysis identifies the presence of a fundamental time-scale
which regulates both the progenitor outburst history and the major explosion, 
and that is also shared by completely independent events like SN\,2010mc.
This time-scale corresponds to a tiny fraction of a massive star lifetime: 
$\sim10^{-8}$ for $\tau\sim 4-6$ Myr, as appropriate for 
a $45-85\,\rm{M_{\sun}}$ star.
We speculate on the nature of the underlying physical process in the next
two sections.

\subsection{The progenitor system of SN\,2009ip}
\label{SubSec:progenitor}

In Section \ref{SubSec:Asymmetry} we showed that asymmetry plays a role 
in the 2012 explosion, which might point to the presence of a preferred direction in the
progenitor system of SN\,2009ip. This suggests either a (rapidly?) rotating single star or
an interacting binary as progenitor for SN\,2009ip. We first update previous estimates
of the progenitor mass of SN\,2009ip using the latest stellar evolutionary tracks 
and then discuss the effects of stellar rotation and the possibility of a binary progenitor. 

From HST pre-explosion images, employing the latest Geneva 
stellar evolutionary tracks \citep{Ekstrom12}  which include important updates 
on the initial abundances, reaction rates and mass loss prescriptions with respect
to  \cite{Schaller92}, we determine for $M_{\rm{V}}=-10.3\pm 0.3$ \citep{Foley11} 
a ZAMS  mass  $M\gtrsim 60\,\rm{M_{\sun}}$ assuming a 
non-rotating progenitor at solar composition\footnote{The
simulations extend up to ZAMS mass of $120\,\rm{M_{\sun}}$.}, consistent with 
previous estimates.  $M_{\rm{V}}=-10.3$ corresponds to 
$L_{\rm{V}}\sim 10^{6.1}L_{\rm{\sun}}$. This implies $L_{\rm{bol}}>2\times 10^{6}\,\rm{L_{\sun}}$,
thus rivaling the most luminous stars ever discovered (e.g. \citealt{Crowther10}). Luminosities of a few 
$10^6\,\rm{L_{\sun}}$ have been indeed associated to the group of LBV stars with typical
temperature of $\sim 15000-25000\,\rm{K}$. Adopting this range of temperature results in 
$L_{\rm{bol}}\sim 5\times10^{6}\,\rm{L_{\sun}}$, suggesting that any  progenitor with 
$M<160\,\rm{M_{\rm{\sun}}}$ was super Eddington at the time of the HST observations.
For $(2<L_{\rm{bol}}<5)\times10^{6}\,\rm{L_{\sun}}$ the allowed mass range is
$60\,\rm{M_{\sun}}<M_{\rm{ZAMS}}<300 \,\rm{M_{\sun}}$ (e.g. \citealt{Crowther10}).
Including the effects of axial rotation results instead in a more constrained range of 
allowed progenitor mass:  $45\,\rm{M_{\sun}}<M_{\rm{ZAMS}}<85 \,\rm{M_{\sun}}$
(for $\Omega/\Omega_{\rm{crit}}=0.4$).  

Rapid rotation strongly affects the evolution of massive stars, in particular by increasing
the global mass-loss rate (e.g. \citealt{Maeder2000}). More importantly, \cite{Maeder02}
showed that the mass loss in rapidly rotating massive stars does not remain isotropic, 
but it is instead enhanced at the polar regions, thus favoring bipolar stellar winds
(e.g. \citealt{Georgy11}, their Fig. 2). As a result, the formation of an asymmetric 
(peanut shaped) nebula around rapidly rotating stars is very likely. Additionally,
rapid rotation can induce mechanical mass loss, resulting in some matter to be
launched into an equatorial Keplerian disk. It is clear that any explosion/eruption of the 
central star will thus naturally occur in a non-isotropic medium, as we find for SN\,2009ip.
Rotation further leads to enhanced chemical mixing (e.g. \citealt{Chatzopoulos12b};
\citealt{Yoon12}).

HST pre-explosion images cannot, however, exclude the presence of a compact 
companion.\footnote{The variability argument allows us to conclude that the observed emission 
is however dominated by the progenitor of SN\,2009ip.} The massive progenitor of SN\,2009ip
might be part of a binary system and the repetitive episodes of eruption might be
linked to the presence of an interacting companion.  
The close periastron passages of a companion star in an eccentric binary system has been invoked by 
\cite{Smith11a} to explain the brightening episodes of $\eta$Car in 1838 and 1843. In the context 
of SN\,2009ip, \cite{Levesque12}  discussed the presence of a binary companion, while
\cite{Soker13} suggested that the 2012b explosion was the result of a merger of
two stars, an extreme case of binary interaction. 
A binary scenario was also proposed by \cite{Soker13b} for the cosmic twin of SN\,2009ip, 
SN\,2010mc.

A binary system would have a natural
asymmetry  (i.e., the preferred direction defined by the orbital plane)  
and a natural time scale (i.e., the orbital period) as indicated by the observations. 
A possible configuration suggested to lead to 
substantial mass loss is that of a binary system 
made of a compact object (NS) closely orbiting around a massive, extended star (\citealt{Chevalier12b}
and references therein). In this picture the mass loss is driven by the inspiral of the compact
object  in the common envelope (CE) evolution and the expansion velocity of the material 
is expected to be comparable to the escape velocity for the massive star.  For\footnote{This is the
velocity required to reach $R_{\rm{bo}}$ after $\sim 50$
days since the 2012a explosion.} $v\sim1000\,\rm{km\,s^{-1}}$ and $M\sim45-85\,M_{\sun}$
the radius of the extended star  is $R_{*}<(1-2)\times 10^{12}\rm{cm}$ (where the inequality 
accounts for the fact that we used the ZAMS mass).
In this scenario, the mass loss is concentrated on the orbital plane of
the binary (\citealt{Ricker12}), offering a natural explanation for the observed 
asymmetry (Section \ref{SubSec:Asymmetry}). However, it remains unexplained how this 
mechanism would be able to launch material with $v\sim 10^4\,\rm{km\,s^{-1}}$ as observed 
during the 2012a episode (\citealt{Mauerhan12}, \citealt{Pastorello12}).
A potential solution might be the presence of a high-velocity
accretion disk wind.
While the presence of a dominant time scale common to the eruptive and explosive 
phases makes the binary progenitor explanation particularly appealing\footnote{Adopting 
$t\sim 40$ days as orbital period implies an orbital radius 
$R<10^{13}\,\rm{cm}$.}, the common envelope
model in its present formulation seems to have difficulties in explaining the observed
phenomenology. Alternative scenarios are explored in Section \ref{SubSec:physicalmechanism}.

\subsection{Physical mechanisms leading to substantial mass loss in evolved single stars}
\label{SubSec:physicalmechanism}

The outer envelope of an evolved massive star contains $\sim95\%$ of the stellar radius
but only a tiny fraction of the stellar mass. The physical 
mechanism leading to the 2012a outburst must have been fairly deep seated
to unbind more than $\sim 0.1\,\rm{M_{\sun}}$. It is furthermore required to generate a large
amount of energy ($10^{48}$ erg plus the energy to lift the mass out of the
deep gravitational potential well) and to explain the presence of a dominant
time scale of $\sim 40$ days which is also shared with the observed
eruption history in the years before the explosion. 
The mass-loss rate of at least a few  $\sim 0.1\,\rm{M_{\sun}}/yr$
can only be sustained for an extremely small fraction of the life of a star and in principle
only for $\sim$ years, before inducing important adjustments to the stellar structure.
We explore here different physical mechanisms that 
have been proposed to lead to substantial mass loss in the late stages of evolution of massive 
single stars and discuss their relevance for SN\,2009ip. They can basically be grouped into 4 categories: 
(i) pulsational pair-instability; (ii) super-Eddington fusion
luminosity; (iii) Eddington-limit instabilities; (iv) shock heating of the stellar envelope.

Pulsational pair-instability (PPI, \citealt{Barkat67}) applies to 
massive stars with helium core of $\sim 40-60\,\rm{M_{\sun}}$. 
PPI leads to partial unbinding of the star, with a sequence of 
eruptions accompanied by the ejection of shells of material of the stellar envelope.
This mechanism has been considered as a plausible explanation for SN\,2009ip
by \cite{Pastorello12}, \cite{Mauerhan12} and \cite{Fraser13} 
but has been rejected by \cite{Ofek13b} and \cite{Ofek13} for both SN\,2010mc and SN\,2009ip
(the leading argument being that PPI  would lead to the ejection of much larger
amounts of mass than the $\sim10^{-1}-10^{-2}\,\rm{M_{\sun}}$ they infer).
According to \cite{Woosley07}, a non-rotating star with zero metallicity (Z=0) and 
ZAMS mass $95$ and $130\,\rm{M_{\sun}}$ meets the criteria for PPI.  However,
with updated prescriptions for the mass-loss rate, and assumption of solar metallicity 
(which more closely represents our conditions of $0.4\,\rm{Z_{\sun}}<Z<0.9\,\rm{Z_{\sun}}$, 
Section \ref{Sec:Metallicity}) \cite{Ekstrom12} predict
that even stars with $M=120\,\rm{M_{\sun}}$ will end the C-burning phase 
with  mass $\sim 31\,\rm{M_{\sun}}$, below the threshold of $\sim 40\,\rm{M_{\sun}}$ 
to trigger PPI. While our limits on the progenitor mass of  SN\,2009ip in Sec. \ref{SubSec:progenitor} 
do not formally rule out the PPI scenario in the non-rotating case, they definitely allow
for progenitors starting with much lower mass than required for the PPI to develop 
($M\gtrsim 120\,\rm{M_{\sun}}$).

Rapidly rotating progenitors ($\Omega/ \Omega_{\rm{crit}}>0.5$) enter the PPI regime starting with 
substantially lower ZAMS mass: $\sim 50\,\rm{M_{\sun}}$ for  $Z=0$ (\citealt{Chatzopoulos12};
\citealt{Yoon12}). On the other hand, for $Z=Z_{\sun}$ even very massive rotating $120\,\rm{M_{\sun}}$ stars
develop a core with $M\sim 19\,\rm{M_{\sun}}$, insufficient to trigger PPI  \citep{Ekstrom12}.  
In the previous sections we constrained the progenitor of SN\,2009ip with ZAMS mass $45\,\rm{M_{\sun}}<M<85\,\rm{M_{\sun}}$  
and metallicity $0.4\,\rm{Z_{\sun}}<Z<0.9\,\rm{Z_{\sun}}$. 
Following the prescriptions from  \cite{Chatzopoulos12}, adopting 
the most favorable conditions (i.e. $M=85\,\rm{M_{\sun}}$ and $Z=0.4\,Z_{\sun}$) and
starting with a very rapidly rotating star  with $\Omega/ \Omega_{\rm{crit}}=0.9$, we find the 
final oxygen core mass to have $M\sim35\,\rm{M_{\sun}}$, formally below the threshold of 
$40\,\rm{M_{\sun}}$ for PPI. This indicates some difficulties for rotating
progenitors to reach the PPI threshold. 
What remains difficult to interpret both for rotating and non-rotating progenitors is the
presence of a dominant time-scale of $\sim40$ days: depending on the pulse properties
and the physical conditions of the surviving star, the interval between pulses can 
be \emph{anywhere} between days to decades \citep{Woosley07}.

Alternatively, convective motions stimulated by the super-Eddington fusion luminosity in the core
of a massive star can excite internal gravity waves (g-mode non-radial oscillations,
\citealt{Quataert12}). An important
fraction of the energy in gravity waves can be converted into sound waves, with
the potential to unbind several $M_{\sun}$. For a $40\,M_{\sun}$ star,  \cite{Quataert12}
estimate that $10^{47}-10^{48}\,\rm{erg}$ will be deposited into the stellar envelope. 
If this mechanism is responsible for the ejection of material by the 2012a explosion,
the velocity $v\sim 1000\,\rm{km\,s^{-1}}$ of the unbound material inferred from our 
observations implies an ejecta mass $\lesssim 0.1 \,\rm{M_{\sun}}$,
consistent with our estimate of the mass of the compact shell encountered by the
2012b explosion in Section \ref{SubSec:ShockBO}. However, this mechanism is likely 
to lead to a steady mass-loss  (E. Quataert private communication) as opposed to
shell ejection and does not offer a natural explanation for the 40-day time scale involved
in the process. We therefore consider the super-Eddington luminosity mechanism unlikely
to apply to SN\,2009ip and SN\,2010mc,  (see however \citealt{Ofek13b}).

With $R$-band observed magnitudes between $\sim18$ mag and $\sim21$ mag reported by \cite{Pastorello12},
SN\,2009ip oscillates between $\sim$Eddington and super-Eddington
luminosity episodes in the years preceding the double explosion 
in 2012. Depending on the effective temperature of the emission, $R\sim21$ mag
corresponds to the Eddington luminosity of a star with mass $40-80\,\rm{M_{\sun}}$.
$R\sim18$ would correspond to an Eddington limit for a star of $M=600-1200\,\rm{M_{\sun}}$. 
A number of instabilities have been shown to develop in stars approaching
and/or exceeding the Eddington limit, actually allowing the super-Eddington luminosity
to persist (e.g. \citealt{Owocki12}), potentially powering LBV-like eruptions. The eruption time scale,
repetition rate and ejected mass are however loosely predicted, so that it is unclear 
if any of these mechanisms would apply to the eruption history of SN\,2009ip and, even more
importantly, how these are connected with the 2012 double explosion (which
clearly differs from super-Eddington powered winds). With luminosity 
between $\sim 10^6$ and $\sim10^7\,\rm{L_{\sun}}$ the progenitor of SN\,2009ip falls
into the region where radiation pressure starts to have a major role in supporting the star 
against gravity (see e.g. \citealt{Owocki12}, their Fig. 12.1): we speculate that the
dominance of radiation pressure over gas pressure in the envelope might have an 
important role in determining the repeated ejection of shells of SN\,2009ip.

Indeed, the outer layers of massive stars close to the Eddington limit and/or critical rotation 
are only loosely bound and might be easily ejected if enough energy is suddenly deposited.
A potential source of energy deposition has been identified in thermonuclear flashes associated
with shell burning \citep{Dessart10}. Differently from the PPI, this scenario does not require the 
progenitor to be extremely massive. 
Alternatively, non-radial gravity mode oscillations
above the core or near to the H-burning shell could provide fresh fuel, triggering a burst
of energy (and subsequent shell ejection) in massive stars like $\eta$Car \citep{Guzik05}.
Both scenarios have the advantage to be basically driven by two 
parameters: the deposited energy and the envelope binding energy, naturally satisfying the 
"simplicity" criterion of Section \ref{SubSec:09ip10mc}. 

Finally we mention that  direct numerical simulations of pre-collapse hydrodynamics
by \cite{Meakin06}, \cite{Meakin07} and \cite{Arnett11a,Arnett11b} found eruptive instabilities due to the 
interaction of oxygen and silicon burning shells. The instabilities are related to turbulent 
convection (and being inherently nonlinear, are invisible to conventional linear stability analysis).
These simulations specifically predicted mass ejection prior to core collapse,
suggesting that pre-collapse evolution is far more eventful than previously thought. 
However, as for all the other mechanisms analyzed in this section, it is at the moment unclear how
to explain the 40-day time scale.
\subsection{The nature of SN\,2009ip double-explosion in 2012: the explosive ejection 
of the stellar envelope}
\label{SubSec:nature09ip}

\begin{figure*}
\vskip -0.0 true cm
\centering
\includegraphics[scale=0.65]{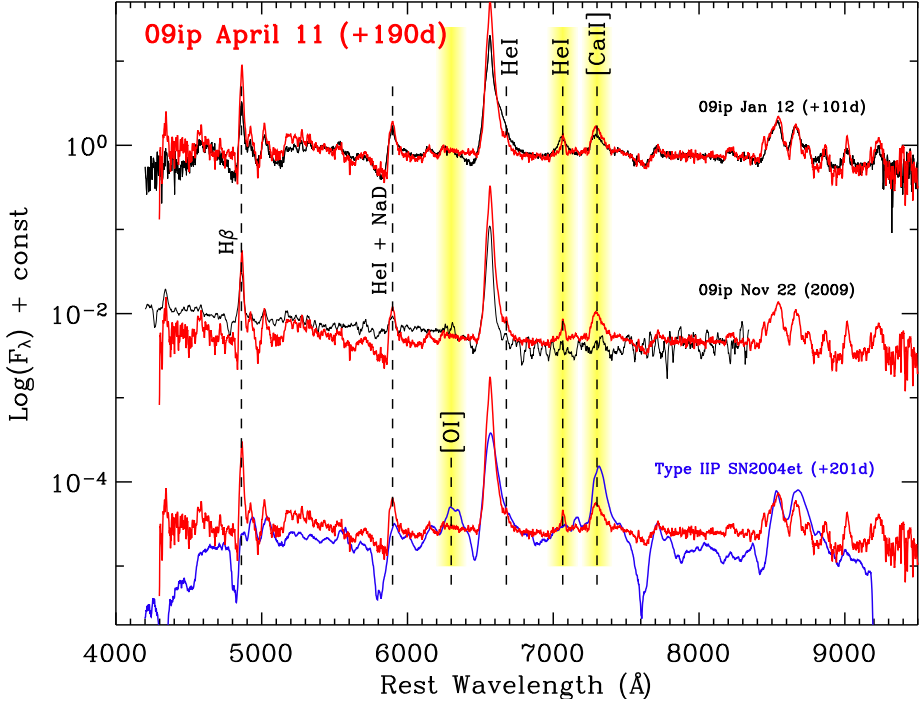}
\caption{Our spectrum taken in April 2013 ($t_{\rm{pk}}+190$ days) shows minimal evolution with respect to 
$\sim100$ days  before.  The most notable difference with respect to 
spectra acquired after or during the LBV-like eruptions is the presence of intermediate \ion{He}{1} and 
[CaII] in emission.  We still do not find evidence for SN-synthesized material (e.g. 
[OI] 6300 \AA\,).  We refer to  \cite{Sahu06} for details on Type IIP SN\,2004et.}
\label{Fig:latetimeSpec}
\end{figure*}

Our modeling shows that the 2012b explosion is \emph{not} powered by Ni radioactive
decay (Section \ref{subsec:LimitNi}) and demonstrates that a shock breaking out from 
a shell of material previously ejected by the 2012a outburst can reasonably account 
for the observed properties of the 2012b explosion, constraining the mass of the ejecta to  
$M_{\rm{ej}}=50.5(E/10^{51}\rm{erg})^2\,M_{\rm{\sun}}$ (Section \ref{SubSec:ShockBO}).
This strongly suggests an explosion energy well below $10^{51}\rm{erg}$ (likely
around $10^{50}$ erg), and brings to
question the fate of SN\,2009ip: a terminal SN explosion (\citealt{Mauerhan12})
or a SN "impostor" (\citealt{Pastorello12}; \citealt{Fraser13})?

A key prediction of the SN scenario is the presence of chemical elements produced
by the SN nucleosynthesis (like oxygen) in late-time spectra, which 
is not expected in the case of a non-terminal explosion.  Our
latest spectrum acquired on 2013 April 11 ($t_{\rm{pk}}+190$ days) still shows no evidence for
SN-synthesized material  (Fig. \ref{Fig:latetimeSpec}).
This non-detection is consistent with (but surely not
a proof of) the non-terminal explosion scenario. This spectrum, dominated
by H emission lines and where the high-velocity absorption features have
completely disappeared, shares some similarities with pre-explosion spectra of SN\,2009ip.
A notable difference, however, is the presence of  (intermediate width) \ion{He}{1} and [\ion{Ca}{2}].
The present data set therefore does not yet offer compelling evidence for a standard (i.e.
resulting from the collapse of a degenerate core) SN-explosion scenario.

Interaction seems to be still dominating the emission at this time.
Our late-time observations indicate a flattening in the light-curve 
(our latest V-band photometry acquired at $t_{\rm{pk}}+190$ days
implies a decay of $\Delta V=+0.86 \pm0.22$ over 98.5 days, or 
$0.0087\pm0.0022\,\rm{mag/day}$), which might be caused by the interaction of the
shock with the NIR emitting material.
Our model predicts that the ``flat'' phase should be followed by a rapid decline,
once the shock has reached the edge of the ``NIR shell''.  

We propose that SN\,2009ip was the consequence of the \emph{explosive ejection of the envelope of a massive
star}, which later interacted with shells of material ejected during previous eruption episodes.
While $E<10^{51}\rm{erg}$ is insufficient to fully unbind a massive star, its
outer envelope is only loosely bound, and can be easily ejected by a lower-energy explosion.
The origin of the energy deposition is not constrained as unclear is its 
potential relation with the possible binary nature of the system. 
The final fate of SN\,2009ip depends on the properties of the remaining ``core'', and
in particular, on its mass and rotation rate: if the star managed to explode its entire 
H envelope still retaining a super-critical core mass, it might re-explode in the future as a genuine
H-poor (Ib/c) SN explosion; if instead the star partially retained its envelope, it will 
possibly give rise to other SN-impostor displays, on time scales which are difficult
to predict.\footnote{Note that SN\,2009ip might still undergo PPI in the future even if
it managed to eject its entire envelope during the 2012b explosion.}.
The core might also have directly collapsed to a black hole: in this
case this would mark the ``end''.
Only close inspection of the explosion site after SN\,2009ip has appreciably
faded will reveal if the star survived the ejection of its outer layers.
Given the impressive similarity with SN\,2010mc, our interpretation extends to this event as well.
\section{Summary and Conclusions}
\label{Sec:Conclusions}

The ``2012 double-explosion'' of SN\,2009ip brought to light the limits of our current 
understanding of massive star evolution, pointing to the existence of new channels for
sustained mass-loss in evolved stars, whose origins has still to be identified.  Our extensive
follow-up campaign and modeling allow us to identify the properties of
a complex ejecta structure and an explosion 
environment shaped by repeating outbursts during the previous years. We find that:
\begin{itemize}
\item SN\,2009ip is embedded in a sub-solar metallicity environment ($0.4\,Z_{\sun}<Z<0.9\,Z_{\sun}$)
in the outskirts of its host galaxy ($d\sim5$ kpc), radiated $\sim3\times10^{49}$ erg
during the 2012b explosion, reaching a peak luminosity of $\sim2\times 10^{43}\,\rm{erg\,s^{-1}}$. 
The 2012a precursor bump released $\sim2\times10^{48}$ erg.
\item The explosion is not powered by $^{56}\rm{Ni}$ radioactive decay. From late-time photometry,
the total $^{56}\rm{Ni}$ mass is $M_{\rm{Ni}}<0.08\,\rm{M_{\sun}}$. Narrow emission lines in the optical 
spectra require interaction of the explosion shock with the CSM at some level. We 
suggest this interaction is also responsible for mediating the conversion of the shock kinetic energy 
into radiation, powering the observed light-curve.
\item Spectroscopy at optical to NIR wavelengths further identifies three distinct velocity components
in the ejecta with $v\sim 12000\,\rm{km\,s^{-1}}$, $v\sim 5500\,\rm{km\,s^{-1}}$ and
$v\sim 2500\,\rm{km\,s^{-1}}$, arguing against a continuous velocity distribution. 
No departure from spherical symmetry needs to be invoked for the slow-moving 
($v\leq 5500\,\rm{km\,s^{-1}}$) ejecta. Instead, spherical symmetry is clearly broken by the 
high-velocity material, possibly pointing to the presence of a preferred direction in the explosion.
Broad and intermediate components in the optical/NIR spectra are consistent with the 
view that asymmetry might also have a role in the spatial distribution of the interacting material.
\item CSM material in the region surrounding SN\,2009ip, with a velocity of a few to several
$100\,\rm{km\,s^{-1}}$, originates from abrupt episode(s) of mass ejection in the previous
years. Mass loss is unlikely to have occurred in the form of a steady wind: instead,
our analysis favors explosive mass ejections, likely linked to instabilities developing 
deep inside the stellar core.
\item We interpret the major 2012 re-brightening to be caused by the explosion shock 
breaking out through a compact and dense CSM shell previously ejected in the 2012a outburst. Our
analysis constrains the onset of the explosion to be $\sim 20$ days before peak. The break-out
radius is $R_{\rm{bo}}\sim 5\times 10^{14}\rm{cm}$. The shell extends to $R_{\rm{w}}\sim1.2\times
10^{15}\rm{cm}$, with a total mass of $M_{\rm{w}}\sim 0.1\,\rm{M_{\sun}}$. The presence
of a compact and dense shell is independently supported by (i) the detection of X-ray radiation
with $L_{\rm{x}}/L_{\rm{opt}}<10^{-4}$ and (ii) by the late-time rise of radio emission.
After break-out the optical-UV light-curve is powered by continued interaction with
optically thin material characterized by a steep density profile.
\item A shock breaking out through a dense shell is expected to produce hard X-rays with a
typical energy of 60 keV. We report a tentative detection of hard X-rays (15-150 keV range) 
around the optical peak and put a solid upper limit on the hard X-ray to optical emission from
SN\,2009ip: $L_{\rm{X,hard}}/L_{\rm{bol}}<5\times 10^{-3}$. 
\item The collision of the ejecta with a massive shell(s) is expected to accelerate cosmic rays and
generate GeV gamma-rays. 
Using \textit{Fermi}-LAT data we detect no GeV
radiation, consistent with the picture of ejecta crashing into a compact and dense but low-mass 
shell of material at small radius. 
\item Our calculations indicate that the latest outburst from SN\,2009ip was a bright source of neutrinos
if the CR energy is comparable to the radiated energy.
We demonstrate with SN\,2009ip how to constrain the
cosmic-ray acceleration efficiency using broad band electromagnetic data and deep limits 
on muon neutrinos that will be available in the near future when new facilities will come
online. 
\item We speculate that the NIR excess of emission is a signature of vaporization of dust grains with
radius $(2-5)\rm{\mu m}$ in a shell of material located at $R>4\times 10^{15}\rm{cm}$. The very high
temperature associated with the NIR excess remains however difficult to explain.
The shell was ejected by the progenitor in the years preceding 2012, and will
be overtaken by the explosion shock during 2013. 
\item Finally, our modeling of the 2012b explosion implies a small ejecta mass $M_{\rm{ej}}\sim 0.5\,\rm{M_{\sun}}$,
with an explosion energy well below $10^{51}$ erg (likely around $10^{50}$ erg), thus raising 
questions about the nature of SN\,2009ip: was it a terminal SN explosion or a SN impostor?  
\end{itemize}

This analysis constrains the 2012b re-brightening of SN\,2009ip 
to be a low-energy,  asymmetric explosion  in a complex medium.
We interpret this 2012b episode to be the \emph{explosive ejection of the envelope 
of the massive progenitor star} that later interacted with shells of material ejected during
previous eruption episodes.

To unravel the nature of the physical mechanism behind the complex phenomenology of SN\,2009ip,
two key observational findings stand out:
(i) its extreme similarity with SN\,2010mc
both in terms of time scales and energetics of the precursor bump and main
explosion; (ii) the presence of a dominant time scale of $\sim 40$ days, which
regulates both the progenitor outburst history and major explosion
(and which is also shared by completely independent events like SN\,2010mc).
While it is clear that the physical process leading to the 2012a eruption must
have been fairly deep-seated to unbind $\sim0.1\,\rm{M_{\sun}}$, it is unclear if
any of the proposed mechanisms for substantial episodic mass loss (i.e.
pulsational pair instability, super-Eddington fusion luminosity, Eddington-limit
instabilities and shock heating of the stellar envelope) would be able
to fully account for the observed properties, and in particular for the
presence of dominant time scales. Indeed, the presence of
dominant time scales might suggest a binary progenitor.

Finally, the extreme similarity to SN\,2010mc allows us to conclude that 
the mechanism behind  the precursor plus major outburst is \emph{not unique}
to SN\,2009ip and is likely driven by \emph{few} physical parameters. Moreover, this eruptive dynamic
might represent an important evolutionary channel for massive stars. Future observations 
will reveal if SN\,2009ip was able to survive.

\acknowledgments
We are grateful to E. Dwek, E. Quataert, H. Krimm, S. Barthelmy, I. Czekala, G. Chincarini for many interesting 
discussions and helpful suggestions. We would also like to thank 
the entire \emph{Swift} team for their hard work and excellent support in scheduling the observations.

C.C.C.\ was supported at NRL by a Karles' Fellowship and NASA DPR S-15633-Y.
J.V. and T.S. are supported by the Hungarian OTKA Grant NN 107637.
C.G. is supported by the NASA Postdoctoral Program (NPP).
M.I. and C.C. were supported by the Creative Research Initiative program of the Korea Research Foundation (KRF) grant No. 2010-000712.
R.C.  acknowledges support  from the National Science Foundation under grant AST-0807727
M. S. gratefully acknowledges generous support provided by the Danish Agency for Science and Technology and Innovation realized through a Sapere Aude Level 2 grant.

The Dark Cosmology Centre is funded by the Danish National Science Foundation.
The research of JCW, the Texas Supernova Group and EC is supported in part by NSF 
AST-1109801 and by StScI grant HST-AR-12820. E.C. wishes to thank the University 
of Texas Graduate School for the William C. Powers fellowship given in support of his studies.
The work of the Carnegie Supernova Project is supported by the National Science Foundation under grant AST1008343.

Based on observations made with ESO Telescopes at the La Silla Paranal Observatory under 
programme ID  090.D-0719.
Observations reported here were obtained at the MMT Observatory, a joint facility of 
the Smithsonian Institution and the University of Arizona. This paper includes data gathered 
with the 6.5 meter Magellan Telescopes located at Las Campanas Observatory, Chile.

Observations were obtained with the JVLA operated by the
National Radio Astronomy Observatory. The National Radio Astronomy Observatory is a facility of 
the National Science Foundation operated under cooperative agreement by Associated Universities, Inc.

Support for CARMA construction was derived from the Gordon and Betty Moore Foundation, the Kenneth T. and Eileen L. Norris Foundation, the James S. McDonnell Foundation, the Associates of the California Institute of Technology, the University of Chicago, the states of California, Illinois, and Maryland, and the National Science Foundation. Ongoing CARMA development and operations are supported by the National Science Foundation under a cooperative agreement, and by the CARMA partner universities.

The \textit{Fermi} LAT Collaboration acknowledges generous ongoing support
from a number of agencies and institutes that have supported both the
development and the operation of the LAT as well as scientific data analysis.
These include the National Aeronautics and Space Administration and the
Department of Energy in the United States, the Commissariat \`a l'Energie Atomique
and the Centre National de la Recherche Scientifique / Institut National de Physique
Nucl\'eaire et de Physique des Particules in France, the Agenzia Spaziale Italiana
and the Istituto Nazionale di Fisica Nucleare in Italy, the Ministry of Education,
Culture, Sports, Science and Technology (MEXT), High Energy Accelerator Research
Organization (KEK) and Japan Aerospace Exploration Agency (JAXA) in Japan, and
the K.~A.~Wallenberg Foundation, the Swedish Research Council and the
Swedish National Space Board in Sweden. Additional support for
science analysis during the operations phase is gratefully
acknowledged from the Istituto Nazionale di Astrofisica in Italy and 
the Centre National d'\'Etudes Spatiales in France.

This paper made use of the SUSPECT database (http://www.nhn.ou.edu/ suspect/).

\appendix
\section{Shock break-out in a dense wind shell: direct constraints from observables }
\label{App:ShockBreakOut}

We provide here the set of equations we used to constrain the ejecta mass $M_{\rm{ej}}$, energy 
of the explosion $E$, the environment density $\rho_{\rm{w}}$ and mass-loss rate $\dot M$, 
starting from three observables: the radiated energy at break-out $E_{\rm{rad}}$, the bolometric light-curve
rise-time $t_{\rm{rise}}$ and the radius at shock break-out $R_{\rm{bo}}$. This work is based on
\cite{Chevalier11} . We consider here their model (a)  (see their Fig.  1), where the break-out 
happens \emph{inside} the dense wind shell of radius
$R_{\rm{w}}$ so that $R_{\rm{bo}}<R_{\rm{w}}$. For SN\,2009ip it turns out that 
$R_{\rm{bo}}\lesssim R_{\rm{w}}$ which leads to comparable estimates of the explosion and
the environment parameters even if one were to use their model (b).
The density in the wind shell is $\rho_{\rm{w}}=\dot M/(4\pi r^2 v_{\rm{w}})$, or $\rho_{\rm{w}}=
D r^{-2}$ with $D\equiv5.0\times 10^{16}(\rm{g/cm})D_{*}$ and $r$ is in cgs units.  
We follow the parametrization by \cite{Chevalier11} and solve their Eq. 1, Eq. 3 and Eq. 5 for the
three observables. The system of equation is degenerate for $M_{\rm{ej}}/E^2$. We obtain: 
\begin{equation}
	M_{\rm{ej}}=10 \Big( \frac{R_{\rm{bo}}}{4.0\times10^{14}\rm{cm}}\Big)^{-2}
	\Big(\frac{E_{\rm{rad}}}{0.44\times 10^{50}\rm{erg}} \Big)^{-1}
	\Big (\frac{t_{\rm{rise}}}{6.6\,\rm{days}}\Big)^2
	\Big (\frac{E}{10^{51}\rm{erg}}\Big)^2M_{\sun}
	\label{Eq:Mej}
\end{equation}

\begin{equation}
	D_{*}=\Big (\frac{t_{\rm{rise}}}{6.6\,\rm{days}}\Big)^2\Big( \frac{R_{\rm{bo}}}{4.0\times10^{14}\rm{cm}}\Big)^{-3}
	\Big(\frac{E_{\rm{rad}}}{0.44\times 10^{50}\rm{erg}} \Big)
	\label{Eq:Dstar}
\end{equation}

\begin{equation}
	\dot M = D_{*}\Big (\frac{v_{\rm{w}}}{1000\,\rm{km\,s^{-1}}}\Big) \rm{M_{\sun}yr^{-1}}
	\label{Eq:Mdot}
\end{equation}

\begin{equation}
	k=0.34
	\Big( \frac{R_{\rm{bo}}}{4.0\times10^{14}\rm{cm}}\Big)^{3}
	\Big(\frac{E_{\rm{rad}}}{0.44\times 10^{50}\rm{erg}} \Big)^{-1}
	\Big (\frac{t_{\rm{rise}}}{6.6\,\rm{days}}\Big)^{-1}
	\rm{cm^2/g}
	\label{Eq:k}
\end{equation}

where $v_{\rm{w}}$ is the wind velocity and $k$ is the opacity. The mass of the wind shell enclosed within a radius $R$ is:
\begin{equation}
	M_{\rm{w}}(r<R)=6.3\times10^{17} 
	\Big(\frac{R}{\rm{cm}} \Big)
	\Big (\frac{t_{\rm{rise}}}{6.6\,\rm{days}}\Big)^2
	\Big( \frac{R_{\rm{bo}}}{4.0\times10^{14}\rm{cm}}\Big)^{-3}
	\Big(\frac{E_{\rm{rad}}}{0.44\times 10^{50}\rm{erg}} \Big)
	\rm{g}
	\label{Eq:Mwind}
\end{equation}

The luminosity from continued shock -wind interaction (\citealt{Chevalier11}, their Eq. 9)  after break-out can be written as: 
\begin{equation}
	L(t)=7.1\times 10^{43} 	
	\Big(\frac{E_{\rm{rad}}}{0.44\times 10^{50}\rm{erg}} \Big)
	\Big (\frac{t_{\rm{rise}}}{6.6\,\rm{days}}\Big)^{-0.4}
	\Big (\frac{t}{10\,\rm{days}}\Big)^{-0.6}
	\rm{erg\,s^{-1}}
	\label{Eq:Linteraction}
\end{equation}

The radius of the wind shell is:
\begin{equation}
	R_{\rm{w}}=8.3\times 10^{13}
	\Big (\frac{t_{\rm{rise}}}{6.6\,\rm{days}}\Big)^{-0.8}
	\Big( \frac{R_{\rm{bo}}}{4.0\times10^{14}\rm{cm}}\Big)
	\Big (\frac{t_{\rm{w}}}{\rm{days}}\Big)
	\rm{cm}
	\label{Eq:Rwind}
\end{equation}
where $t_{\rm{w}}$ is the time when the shock reaches the edge of the shell.  In this scenario, the onset of the explosion
is at: 
\begin{equation}
	t_{0}\approx	t_{\rm{peak}}-2 t_{\rm{rise}}
\end{equation}

Assuming  hydrogen rich material with $\langle \mu_{\rm{p}}\rangle=1$, the column density
from radius $R$  to the observer can be written as (see \citealt{Ofek13}):
\begin{equation}
	N(R)\approx3.02\times 10^{25}
	\Big (\frac{t_{\rm{rise}}}{6.6\,\rm{days}}\Big)^2
	\Big( \frac{R_{\rm{bo}}}{4.0\times10^{14}\rm{cm}}\Big)^{-3}
	\Big(\frac{E_{\rm{rad}}}{0.44\times 10^{50}\rm{erg}} \Big)
	\Big (\frac{R}{10^{15}\,\rm{cm}}\Big)^{-1}
	\rm{cm^{-2}}	
	\label{Eq:N}
\end{equation}
The bound-free optical depth at radius $R$ as a function of the energy of the photons, $E$ is $\tau_{\rm{bf}}=N\sigma(E)$ or 
\begin{equation}
	\tau_{\rm{bf}}(E)\approx 3\times 10^3
	\Big (\frac{t_{\rm{rise}}}{6.6\,\rm{days}}\Big)^2
	\Big( \frac{R_{\rm{bo}}}{4.0\times10^{14}\rm{cm}}\Big)^{-3}
	\Big(\frac{E_{\rm{rad}}}{0.44\times 10^{50}\rm{erg}} \Big)
	\Big (\frac{R}{10^{15}\,\rm{cm}}\Big)^{-1}
	\Big (\frac{E}{\rm{keV}}\Big)^{-2.5}
	\label{Eq:taubf}
\end{equation}
adopting the approximated cross-section as in \cite{Ofek13}. The free-free optical depth $\tau_{\rm{ff}}$ at
radius $R$ as a function of photon frequency $\nu$ is:
\begin{equation}
	\tau_{\rm{ff}}(\nu)\approx 2.6\times 10^8
	\Big (\frac{T_{e}}{10^4\,\rm{K}}\Big)^{-1.35}	
	\Big (\frac{\nu}{\rm{GHz}}\Big)^{-2.1}	
	\Big (\frac{R}{10^{15}\,\rm{cm}}\Big)^{-3}
	\Big (\frac{t_{\rm{rise}}}{6.6\,\rm{days}}\Big)^4
	\Big( \frac{R_{\rm{bo}}}{4.0\times10^{14}\rm{cm}}\Big)^{-6}
	\Big(\frac{E_{\rm{rad}}}{0.44\times 10^{50}\rm{erg}} \Big)^2
	\label{Eq:tauff}
\end{equation}
where $T_{e}$ is the electron temperature.
\section{Temporal Analysis}
\label{App:tempan}

\begin{figure}
\vskip -0.0 true cm
\centering
\includegraphics[scale=0.35,angle=270]{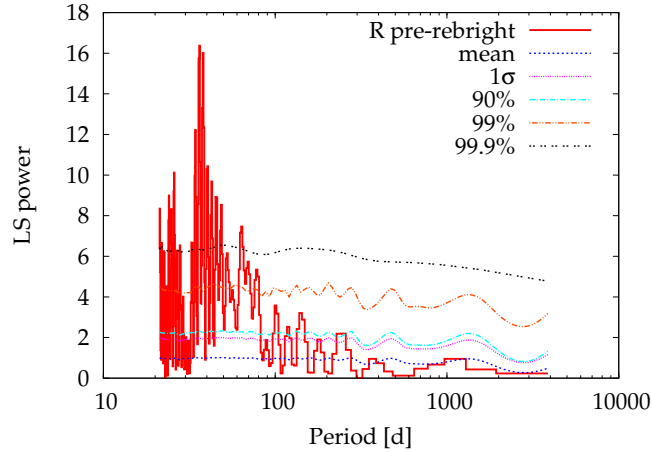}
\caption{Lomb-Scargle periodogram of the R-band magnitude time series for the time interval
August 2009 - April 2012 (i.e. \emph{before} the 2012a and 2012b explosions). This plot shows
significant power on time-scales around $\sim40$ days. We emphasize that this is \emph{not} a claim for periodicity.
Contours at various significance levels are also shown.}
\label{Fig:tempan}
\end{figure}

We identify the presence of a dominant time-scale of $\sim 40$ days by applying the 
the Fast $\chi^2$ algorithm\footnote{http://public.lanl.gov/palmer/fastchi.html} developed by \cite{Palmer09}
to the R-band photometry obtained for SN\,2009ip by \cite{Pastorello12} in the period August 2009 - April 2012.
This algorithm  is suitable for irregularly sampled data with nonuniform errors and it is 
designed to identify the presence of
periodic signals by minimizing the $\chi^2$ between the data and the model.
We adopt a linear model for the trend. The best period was found to be $T_0=115$~days.
However, the reduction in the $\chi^2$ obtained for this period is comparable over a broad
range of period candidates, thus proving that there it is \emph{no} periodic or quasi--periodic
signal. Assuming the Fourier decomposition with the fundamental period $T_0$,
among the first seven harmonics most (52\%) of the temporal power is carried by the third harmonic
($T_0/3=38$~days),  followed by the fundamental (19\%), the
fifth harmonic ($T_0/5=23$~days, 10\%), the second and fourth harmonics ($T_0/2=58$ and $T_0/4=29$~days,
5\% each). This is evidence for a dominant time-scale of about 40~days, with significant power
distributed on time-scales between 30 and 50 days.
Replicating the same analysis on time series obtained by randomizing the observed
magnitudes has the effect of washing out the excess of power on those time scales.
This demonstrates that the excess of power around 40 days is not due to the data sampling.

Additionally and independently, we calculated the Lomb--Scargle (LS; \citealt{Lomb76}; \citealt{Scargle82})
periodogram of the same time series. This is particularly suitable to unevenly sampled
data sets and measures the power contributed by the different frequencies to the total
variance. This reduces to the Fourier power spectrum in the uniform sampling limit.
As shown in Fig. \ref{Fig:tempan}, most power is concentrated between $\sim30$ and
$\sim$50~days, with a peak around 40~days. To evaluate the impact of the aliasing and to quantify how significant
the power is with respect to the white noise case (i.e. no preferred time scale), we carried
out the following Monte Carlo simulation. We randomized the magnitudes among the same
observation times and generated $10^4$ synthetic profiles with the same variance as the
real one. For each period we derived the corresponding confidence levels of $1\sigma$,
90\%, 99\%, and 99.9\%, as shown in Fig. \ref{Fig:tempan}. Comparing the real LS periodogram with
the MC generated confidence levels, we conclude that power in excess of white noise
around 40~days is significant at 99.9\% confidence
and we identify this time scale as the dominant one in the overall variance.

In addition, we applied both methods to the time profiles obtained for the V, H and R filters
around the 2012 outburst. We applied both a third and a fourth degree polynomial de-trending.
In all cases, we obtained similar
results with dominant time scales in the range 30--50~days. This is a robust result
which does not crucially depend on the polynomial degree used for de-trending the series.
\section{Observing Logs}

\begin{deluxetable}{lllcccccc}
\tabletypesize{\scriptsize}
\tablecolumns{9} 
\tablewidth{0pc}
\tablecaption{Log of observed \emph{Swift}-UVOT spectra}
\tablehead{ \colhead{Date (UT)} & \colhead{Time range (UT)} & \colhead{Obs ID} & 
\colhead{Exposure (ks)} & \colhead{Roll Angle}}
\startdata
2012-09-27  &    09:03:41-14:37:27   &   31486008  & 5.1& 212.0 \\
2012-09-28  &   13:41:01-16:53:23    &   31486011  & 3.4& 215.0 \\
2012-09-29  &  10:27:52-12:24:46     &   31486012   & 3.3 & 215.0 \\
2012-09-30   &   05:48:22-17:19:31   &   32570001  & 4.9 & 216.0\\
2012-10-01  &    01:13:23-19:11:08    &   32570002 & 4.0&  216.0 \\
2012-10-02  &   14:08:21-22:05:35    &   31486020  &  5.3 &   229.0 \\
2012-10-03   &  02:49:01-15:52:51    &   32570003  &  4.4 & 220.0 \\
2012-10-04   &   09:10:01-14:19:17    &   32570004 & 5.1 & 220.0\\
2012-10-05    &  09:12:55-15:33:06     &   32570005& 7.2& 220.0\\
2012-10-06    &  04:26:57-11:09:55     & 32570006  & 6.7 &220.0\\
2012-10-07    &  07:41:54-08:05:27     &  32570007 &6.2 & 220.0\\
2012-10-08    &  07:44:55-12:57:21     &  32579001 &6.2& 243.0\\
2012-10-10    &  03:01:55-12:56:01      & 32579002  &6.5 & 243.0\\ 
2012-10-12    &  06:20:54-11:31:19      & 32579003 & 5.7 & 243.0\\
2012-10-14     & 04:50:54-06:54:14      & 32579004  & 5.8 & 243.0 \\  
2012-10-16     & 06:32:56-11:35:19      & 32579005  &5.7 & 243.0\\
2012-10-20    &   08:30:16-13:41:40     &  32579007 & 5.3 & 243.0\\
2012-10-22    &  09:59:14-13:30:15       &32579008   & 3.5 & 243.0\\
2012-10-24     &13:19:14-17:00:26     &  32605001  & 4.3 & 249.0\\
2012-10-26     & 10:10:14-13:53:54     &  32605002  & 3.7& 249.0\\
2012-10-28     & 08:41:21-09:00:03     &  32605003  & 1.1 & 249.0\\
2012-11-02    &  17:09:16-19:03:52      &   32605004  & 11.3   &  249.0\\
\enddata
\label{Tab:UVOTObsLog}
\end{deluxetable}

\begin{deluxetable}{lllcccccc}
\tabletypesize{\scriptsize}
\tablecolumns{9} 
\tablewidth{0pc}
\tablecaption{Log of observed optical spectra}
\tablehead{ \colhead{Date (UT)} & \colhead{Telescope/Inst.} & \colhead{Range (\AA)} & 
\colhead{R ($\lambda / \Delta\lambda$)} & \colhead{Grating (l/mm)} &  
\colhead{Aperture (arcsec)} & \colhead{Airmass} &  \colhead{Exp. Time (s)}}
\startdata
Aug 26.83&SALT/RSS& 3500$-$10000 &  300     & 300 & 1.25 & 2.84 &  900\\ 
Sep 26.75&SALT/RSS& 3500$-$10000 &  300     & 300 & 1.25 & 1.16 & 1200\\
Sep 27.99&Magellan/MagE\footnote{Observations were obtained under poor $2''$ seeing conditions.}        &3150.0$-$9400.0  & 3400     &  175 &  1.5 & 1.25 & 600\\
Sep 28.99& Magellan/MagE\footnote{$1.5''$ seeing.}        &3150$-$9400  & 3400     & 175&  1.5 & 1.23   &1200\\ 
Sep 30.02&VLT/X-shooter &3000$-$25000 & 5100$-$8800 &-- & 1.0/0.9/0.9\footnote{Refers to UBV, VIS and NIR ranges, respectively} & 1.10 & 744/556/800$^{\rm{c}}$ \\
Oct 11.15 &KPNO/RCSpec& 3080$-$8760 & 1200 & 316 & 1.5 & 2.15 & 1200 \\
Oct 11.93 &SALT/RSS& 3500$-$10000& 300      & 300 & 1.25 &1.24&  300\\
Oct 12.16 &KPNO/RCSpec& 3080$-$8760 & 1200 & 316 & 1.5 & 2.09 & 1200 \\
Oct 12.94 &SALT/RSS& 3500$-$10000 & 300      & 300 & 1.25 & 1.29& 300\\
Oct 13.16 &KPNO/RCSpec& 5380$-$8290 & 2500 & 632 & 1.5 & 2.07 & 6000 \\
Oct 13.18 & FLWO/FAST & 3470$-$7414 & 2700 &  300 &  3.0 &  2.04 &      1800 \\ 
Oct 14.15 &KPNO/RCSpec& 5380$-$8290 & 2500 & 632 & 1.5 & 2.10 & 3600 \\
Oct 14.17 &MMT/Blue Channel & 3350$-$8570 & 1200 & 300 & 1.0 & 2.03 & 300 \\
Oct 14.18 &MMT/Blue Channel & 5860$-$7160 & 5000 & 1200 & 1.0 & 2.03 & 600 \\
Oct 14.21 &MMT/Blue Channel & 3800$-$5120 & 3000 & 1200 & 1.0 & 2.12 & 900 \\
Oct 14.21 & FLWO/FAST & 3469$-$7413 & 2700 &  300 &  3.0 &  2.07 &      1800 \\ 
Oct 15.21 & FLWO/FAST & 3473$-$7417 & 2700 &  300 &  3.0 &  2.08 &      1800 \\ 
Oct 15.14 &KPNO/RCSpec& 5380$-$8290 & 2500 & 632 & 1.5 & 2.11 & 4800 \\
Oct 15.14 &MMT/Blue Channel & 3350$-$8570 & 1200 & 300 & 1.0 & 2.03 & 300 \\
Oct 16.19 & FLWO/FAST & 3472$-$7416 & 2700 &  300 &  3.0 &  2.05 &      1500 \\ 
Oct 17.19 & FLWO/FAST & 3474$-$7418 & 2700 &  300 &  3.0 &  2.04 &      1602 \\ 
Oct 20.25 & FLWO/FAST & 3481$-$7422 & 2700 &  300 &  3.0 &  2.20 &      1500 \\ 
Oct 21.25 & FLWO/FAST & 3475$-$7419 & 2700 &  300 &  3.0 &  2.80 &      1500 \\ 
Oct 22.18 & FLWO/FAST & 3474$-$7418 & 2700 &  300 &  3.0 &  2.04 &      1500 \\ 
Oct 23.90 &SALT/RSS& 3200-9000   & 1500  & 900 & 1.25 & 1.20 & 600 \\
Oct 27.89 &SALT/RSS& 3200-9000   & 1500  & 900 & 1.25 & 1.21 & 600 \\
Oct 31.13 & VLT/X-shooter &3000$-$25000 &5100$-$8800 &-- & 1.0/0.9/0.9$^{\rm{c}}$ & 1.23 &  744/556/800$^{\rm{c}}$\\
Nov 10.85 &SALT/RSS&3300$-$10500 &    300   & 300 & 1.25&1.25 &600  \\
Nov 14.06 &MMT/Blue Channel & 3320$-$8540 & 1200 & 300 & 1.0 & 2.11 & 900 \\
Nov 14.08 &MMT/Blue Channel & 5840$-$7140 & 5000 & 1200 & 1.0 & 2.06 & 1000 \\
Nov 14.11 &MMT/Blue Channel & 3780$-$5110 & 3000 & 1200 & 1.0 & 2.04 & 1500 \\
Nov 17.04 &Magellan/IMACS& 4200 - 9300  & 1300  & 300  & 0.7  & 1.02 & 2700 \\
Nov 23.81 &SALT/RSS& 3200 - 9000  & 1500  & 900  & 1.25 & 1.25 & 600  \\
Dec 05.01 &Magellan/IMACS & 3500 - 9400&1300       &300  &0.7 &1.25 & 27000 \\
Dec 14.04 &SOAR/GHTS    & 3930 - 7985  & 300   & 1390 & 0.84 & 1.44 & 1800  \\
Dec 21.07 & MMT/Blue Channel    & 3300 - 8500  & 300   & 740  & 1.0  & 2.46 & 2000 \\
Jan 12.04 &Magellan/LDSS3 & 3780$-$10500 & 650 & VPH-all & 1.0 & 2.75 & 1200 \\
Apr 11.39 &Magellan/IMACS& 4000$-$10200 & 1600 & 300 & 0.9 & 1.87 & 1200 \\
\enddata
\label{Tab:OptSpecObsLog}
\end{deluxetable}

\begin{deluxetable}{lllcccccc}
\tabletypesize{\scriptsize}
\tablecolumns{9} 
\tablewidth{0pc}
\tablecaption{Log of observed NIR spectra}
\tablehead{ \colhead{Date (UT)} & \colhead{Inst.} & \colhead{Range (\AA)} & 
\colhead{R ($\lambda / \Delta\lambda$)} & \colhead{Grating (l/mm)} &  
\colhead{Aperture (arcsec)} & \colhead{Airmass} &  \colhead{Exp. Time (s)}}
\startdata
Sep 27.21 &MDM& 9700 $-$ 18000 &  720  & \nodata  & 0.7 & 2.00  & 1600  \\
Sep 29.19 &MDM& 9700 $-$ 24300 &  720  & \nodata  & 0.7 & 2.00  & 1600  \\
Sep 30.20 &MDM& 9700 $-$ 24300 &  720  & \nodata  & 0.7 & 2.00  & 1600  \\
Nov 05.11 &FIRE&  8000 $-$ 27400&   500 & \nodata& 0.6 & 1.24  &  761  \\ 
Nov 19.09 &FIRE&  8200 $-$ 25000 & 6000 &  & 0.6  & 1.28 & 1200  \\ 
Nov 25.11 &FIRE&  8000 $-$ 27200&   500 & \nodata& 0.6 & 1.67 &  1014  \\ 
Dec 03.04 &FIRE&  8000 $-$ 27200&  500 & \nodata& 0.6  & 1.31 & 1014  \\ 
\enddata
\label{Tab:NIRSpecObsLog}
\end{deluxetable}

\section{Photometry Tables}

\begin{deluxetable}{cccccccccccc}
\tabletypesize{\scriptsize}
\tablecolumns{9} 
\tablewidth{0pc}
\tablecaption{\emph{Swift}-UVOT photometry}
\tablehead{ \colhead{Date} & \colhead{v} & \colhead{Date} & \colhead{b} & \colhead{Date} &  \colhead{u} & \colhead{Date} &  \colhead{w1} & \colhead{Date} &  \colhead{w2 }& \colhead{Date} &  \colhead{m2}
\\ \colhead{(d)}  & \colhead{(mag)} & \colhead{(d)} & \colhead{(mag)} & \colhead{(d)} &  \colhead{(mag)} & \colhead{(d)} &  \colhead{(mag)} & \colhead{(d)} &  \colhead{(mag) }& \colhead{(d)} &  \colhead{(mag)}}
\startdata
84.68\footnote{Dates are in MJD-55000 (days).} & 20.50(0.30) & 84.68 & 20.76(0.19) & 84.68 & 20.08(0.17) & 84.67 & 20.48(0.19) & 84.68 & 20.77(0.17) & 84.69 & 20.87(0.23) \\ 
1174.88 & 17.10(0.21) & 1174.90 & 17.29(0.05) & 1174.90 & 16.48(0.06) & 1174.89 & 17.09(0.07) & 1174.90 & 17.85(0.09) & 1176.64 & 17.73(0.09) \\ 
1176.64 & 17.52(0.15) & 1176.64 & 17.24(0.07) & 1176.64 & 16.46(0.07) & 1176.63 & 16.96(0.08) & 1176.64 & 17.86(0.10) & 1183.44 & 18.92(0.12) \\ 
1183.44 & 17.74(0.15) & 1183.44 & 17.62(0.08) & 1183.44 & 17.23(0.09) & 1183.43 & 18.08(0.11) & 1183.44 & 19.09(0.14) & 1192.74 & 20.25(0.22) \\ 
1192.77 & 18.16(0.10) & 1190.75 & 18.39(0.15) & 1190.75 & 17.77(0.12) & 1190.75 & 19.08(0.21) & 1192.77 & 20.31(0.18) & 1196.35 & 12.78(0.04) \\ 
1196.24 & 14.91(0.05) & 1192.76 & 18.13(0.06) & 1192.76 & 17.88(0.07) & 1192.76 & 19.23(0.16) & 1196.34 & 12.83(0.05) & 1198.48 & 12.30(0.04) \\ 
1196.31 & 14.95(0.05) & 1196.23 & 14.81(0.05) & 1196.23 & 13.46(0.05) & 1196.34 & 12.98(0.04) & 1197.46 & 12.66(0.05) & 1200.99 & 12.10(0.04) \\ 
1196.37 & 14.86(0.05) & 1196.30 & 14.75(0.05) & 1196.30 & 13.44(0.05) & 1197.47 & 12.75(0.04) & 1197.71 & 12.52(0.05) & 1201.99 & 12.12(0.04) \\ 
1196.45 & 14.83(0.05) & 1196.36 & 14.77(0.05) & 1196.36 & 13.39(0.05) & 1197.70 & 12.62(0.04) & 1198.48 & 12.42(0.04) & 1202.75 & 12.17(0.04) \\ 
1198.48 & 14.38(0.04) & 1196.45 & 14.74(0.05) & 1196.45 & 13.36(0.05) & 1198.47 & 12.51(0.04) & 1198.59 & 12.43(0.04) & 1202.26 & 12.14(0.04) \\ 
1199.66 & 14.26(0.05) & 1197.70 & 14.39(0.05) & 1197.70 & 12.99(0.05) & 1198.61 & 12.59(0.04) & 1199.65 & 12.32(0.04) & 1203.07 & 12.13(0.04) \\ 
1200.99 & 14.14(0.05) & 1198.48 & 14.28(0.04) & 1198.47 & 12.90(0.04) & 1199.47 & 12.53(0.04) & 1200.99 & 12.20(0.04) & 1204.06 & 12.06(0.04) \\ 
1201.99 & 14.07(0.05) & 1199.65 & 14.11(0.05) & 1199.65 & 12.78(0.04) & 1199.65 & 12.41(0.04) & 1201.99 & 12.19(0.04) & 1206.13 & 12.04(0.04) \\ 
1202.26 & 14.11(0.05) & 1200.98 & 13.94(0.05) & 1200.98 & 12.65(0.04) & 1200.98 & 12.30(0.04) & 1201.38 & 12.23(0.04) & 1208.68 & 12.30(0.04) \\ 
1203.06 & 14.10(0.05) & 1201.98 & 13.98(0.05) & 1201.98 & 12.61(0.04) & 1201.98 & 12.31(0.04) & 1202.26 & 12.21(0.04) & 1210.07 & 12.47(0.04) \\ 
1204.05 & 13.91(0.05) & 1202.25 & 13.96(0.05) & 1200.47 & 12.72(0.04) & 1202.25 & 12.31(0.04) & 1203.06 & 12.23(0.04) & 1212.20 & 12.58(0.04) \\ 
1206.13 & 13.88(0.05) & 1203.06 & 13.96(0.05) & 1202.25 & 12.64(0.04) & 1203.06 & 12.30(0.04) & 1200.49 & 12.26(0.04) & 1214.14 & 12.71(0.04) \\ 
1208.67 & 13.92(0.05) & 1203.32 & 13.96(0.05) & 1203.06 & 12.63(0.04) & 1204.05 & 12.22(0.04) & 1201.43 & 12.23(0.04) & 1216.21 & 13.16(0.04) \\ 
1210.06 & 13.88(0.05) & 1204.05 & 13.84(0.05) & 1203.12 & 12.67(0.04) & 1206.12 & 12.17(0.04) & 1203.27 & 12.27(0.04) & 1216.41 & 13.20(0.06) \\ 
1212.20 & 13.95(0.05) & 1204.52 & 13.76(0.13) & 1203.32 & 12.63(0.06) & 1208.67 & 12.44(0.04) & 1204.05 & 12.16(0.04) & 1218.83 & 13.88(0.04) \\ 
1214.14 & 13.93(0.05) & 1205.52 & 13.73(0.11) & 1203.39 & 12.67(0.06) & 1210.06 & 12.55(0.04) & 1204.50 & 12.13(0.04) & 1220.55 & 14.22(0.04) \\ 
1216.21 & 14.13(0.05) & 1206.12 & 13.75(0.05) & 1204.05 & 12.60(0.04) & 1212.20 & 12.69(0.04) & 1205.53 & 12.13(0.04) & 1222.29 & 14.49(0.04) \\ 
1218.83 & 14.30(0.05) & 1208.67 & 13.86(0.05) & 1204.38 & 12.56(0.04) & 1214.13 & 12.77(0.04) & 1206.12 & 12.14(0.04) & 1222.41 & 14.49(0.05) \\ 
1220.55 & 14.30(0.05) & 1210.06 & 13.91(0.05) & 1205.38 & 12.57(0.04) & 1218.82 & 13.71(0.04) & 1206.33 & 12.18(0.04) & 1224.49 & 14.82(0.04) \\ 
1222.29 & 14.43(0.05) & 1212.20 & 13.91(0.05) & 1205.52 & 12.54(0.04) & 1220.55 & 13.95(0.04) & 1208.67 & 12.47(0.04) & 1226.49 & 15.20(0.05) \\ 
1224.49 & 14.46(0.05) & 1214.14 & 13.97(0.05) & 1205.65 & 12.56(0.04) & 1222.28 & 14.18(0.04) & 1207.44 & 12.30(0.04) & 1226.23 & 15.16(0.05) \\ 
1226.23 & 14.75(0.06) & 1216.27 & 14.18(0.07) & 1206.12 & 12.50(0.04) & 1224.49 & 14.40(0.04) & 1208.44 & 12.45(0.04) & 1228.71 & 15.64(0.05) \\ 
1228.71 & 15.01(0.06) & 1218.83 & 14.39(0.05) & 1206.25 & 12.51(0.04) & 1226.23 & 14.75(0.05) & 1210.34 & 12.70(0.04) & 1230.38 & 15.92(0.06) \\ 
1230.38 & 15.23(0.07) & 1220.55 & 14.44(0.05) & 1208.39 & 12.63(0.04) & 1228.70 & 15.30(0.05) & 1210.06 & 12.65(0.04) & 1232.86 & 16.21(0.06) \\ 
1232.86 & 15.09(0.06) & 1222.29 & 14.57(0.05) & 1208.52 & 12.60(0.06) & 1230.37 & 15.48(0.06) & 1212.20 & 12.80(0.04) & 1234.41 & 16.21(0.05) \\ 
1233.98 & 14.85(0.09) & 1224.49 & 14.68(0.05) & 1208.52 & 12.61(0.04) & 1232.85 & 15.64(0.06) & 1214.14 & 12.94(0.04) & 1234.98 & 16.18(0.08) \\ 
1234.98 & 14.86(0.09) & 1226.23 & 14.94(0.05) & 1208.67 & 12.63(0.04) & 1234.48 & 15.52(0.06) & 1216.21 & 13.44(0.04) & 1236.18 & 16.13(0.06) \\ 
1236.21 & 14.95(0.05) & 1228.71 & 15.25(0.06) & 1210.06 & 12.80(0.04) & 1236.21 & 15.53(0.05) & 1216.47 & 13.50(0.04) & 1238.09 & 16.31(0.05) \\ 
1238.09 & 15.16(0.05) & 1230.37 & 15.50(0.06) & 1210.19 & 12.75(0.04) & 1238.08 & 15.72(0.05) & 1218.83 & 14.22(0.04) & 1240.06 & 16.58(0.06) \\ 
1240.06 & 15.34(0.06) & 1232.85 & 15.38(0.06) & 1210.46 & 12.75(0.04) & 1240.09 & 16.01(0.05) & 1212.38 & 12.82(0.04) & 1242.30 & 16.93(0.06) \\ 
1242.29 & 15.68(0.06) & 1234.48 & 15.16(0.06) & 1212.20 & 12.76(0.04) & 1242.29 & 16.27(0.05) & 1214.25 & 12.97(0.04) & 1242.30 & 16.93(0.06) \\ 
1244.43 & 15.88(0.06) & 1236.18 & 15.25(0.06) & 1212.33 & 12.78(0.06) & 1244.46 & 16.63(0.09) & 1216.39 & 13.47(0.04) & 1244.43 & 17.28(0.07) \\ 
1246.91 & 16.00(0.06) & 1238.08 & 15.40(0.05) & 1212.33 & 12.80(0.04) & 1246.93 & 16.94(0.08) & 1220.47 & 14.59(0.04) & 1246.94 & 17.39(0.08) \\ 
1248.05 & 15.83(0.06) & 1240.05 & 15.73(0.05) & 1212.46 & 12.81(0.04) & 1248.04 & 16.88(0.07) & 1222.50 & 14.84(0.04) & 1248.05 & 17.57(0.07) \\ 
1250.05 & 16.04(0.07) & 1242.29 & 16.07(0.05) & 1214.14 & 12.84(0.04) & 1250.04 & 17.05(0.08) & 1220.55 & 14.58(0.04) & 1250.05 & 17.85(0.08) \\ 
1252.59 & 16.17(0.07) & 1244.46 & 16.23(0.08) & 1216.20 & 13.09(0.04) & 1252.59 & 17.34(0.08) & 1222.29 & 14.86(0.04) & 1252.59 & 18.32(0.10) \\ 
1254.09 & 16.33(0.06) & 1246.91 & 16.42(0.06) & 1216.27 & 13.05(0.06) & 1254.09 & 17.50(0.07) & 1224.49 & 15.19(0.05) & 1254.09 & 18.32(0.08) \\ 
1256.70 & 16.32(0.09) & 1248.04 & 16.33(0.06) & 1216.27 & 13.07(0.04) & 1256.66 & 17.83(0.09) & 1226.23 & 15.54(0.06) & 1256.70 & 18.57(0.13) \\ 
1258.53 & 16.42(0.07) & 1250.01 & 16.47(0.07) & 1216.40 & 13.06(0.06) & 1258.52 & 17.93(0.09) & 1224.64 & 15.21(0.04) & 1258.54 & 18.78(0.10) \\ 
1260.34 & 16.45(0.07) & 1252.59 & 16.80(0.06) & 1216.41 & 13.09(0.04) & 1260.33 & 18.04(0.09) & 1226.44 & 15.56(0.06) & 1260.34 & 18.65(0.09) \\ 
1262.21 & 16.41(0.06) & 1254.09 & 16.85(0.06) & 1218.82 & 13.48(0.04) & 1262.20 & 18.42(0.11) & 1228.71 & 16.10(0.06) & 1262.21 & 18.95(0.11) \\ 
1264.61 & 16.60(0.07) & 1256.66 & 17.00(0.06) & 1220.55 & 13.66(0.04) & 1264.60 & 18.32(0.11) & 1230.37 & 16.32(0.06) & 1264.61 & 18.99(0.12) \\ 
1266.15 & 16.71(0.07) & 1258.53 & 17.16(0.06) & 1220.45 & 13.69(0.04) & 1266.14 & 18.41(0.11) & 1232.86 & 16.55(0.06) & 1266.15 & 19.28(0.13) \\ 
1268.66 & 17.00(0.08) & 1260.30 & 17.20(0.08) & 1222.29 & 13.75(0.04) & 1268.66 & 18.73(0.13) & 1233.79 & 16.48(0.05) & 1268.66 & 19.43(0.14) \\ 
1270.35 & 17.28(0.10) & 1262.20 & 17.11(0.06) & 1224.62 & 14.02(0.04) & 1270.34 & 18.70(0.14) & 1228.38 & 15.94(0.07) & 1271.32 & 19.77(0.15) \\ 
1272.26 & 17.41(0.11) & 1264.61 & 17.34(0.07) & 1224.49 & 14.02(0.04) & 1272.25 & 18.85(0.14) & 1234.48 & 16.52(0.07) & 1275.20 & 20.00(0.16) \\ 
1274.56 & 17.94(0.14) & 1266.14 & 17.55(0.07) & 1226.23 & 14.29(0.04) & 1274.55 & 19.05(0.16) & 1234.26 & 16.50(0.04) & 1280.35 & 19.84(0.18) \\ 
1276.23 & 18.18(0.17) & 1268.66 & 18.04(0.09) & 1228.71 & 14.85(0.05) & 1276.22 & 19.16(0.17) & 1236.21 & 16.51(0.05) & 1286.55 & 20.00(0.15) \\ 
1278.40 & 18.29(0.27) & 1270.35 & 18.00(0.09) & 1228.36 & 14.81(0.04) & 1278.39 & 19.42(0.30) & 1238.08 & 16.70(0.05) & 1292.40 & 20.31(0.16) \\ 
1280.31 & 18.46(0.17) & 1272.25 & 18.35(0.10) & 1230.37 & 15.04(0.05) & 1280.30 & 19.64(0.21) & 1240.05 & 17.00(0.06) &  1387.30 & >21.90 \\ 
1282.15 & 18.51(0.20) & 1274.55 & 18.95(0.16) & 1232.85 & 14.99(0.05) & 1282.14 & 19.78(0.27) & 1242.29 & 17.36(0.06) &  &  \\ 
1284.24 & 18.70(0.20) & 1276.23 & 19.05(0.17) & 1233.98 & 14.90(0.08) & 1285.14 & 19.78(0.19) & 1244.43 & 17.71(0.07) &  &  \\ 
1286.08 & 18.67(0.25) & 1279.40 & 19.35(0.15) & 1234.98 & 14.77(0.07) & 1290.68 & 20.10(0.24) & 1246.94 & 17.82(0.10) &  & \\ 
1290.69 & 18.72(0.18) & 1283.17 & 19.15(0.13) & 1236.18 & 14.87(0.05) & 1294.46 & 19.66(0.18) & 1248.05 & 17.87(0.08) &  &  \\ 
1294.47 & 18.84(0.22) & 1287.48 & 19.48(0.18) & 1238.08 & 15.06(0.05) &  1387.30&  21.12(0.33)& 1250.04 & 18.18(0.09) &  & \\ 
1387.30 &  20.18(0.42)& 1291.36 & 19.44(0.20) & 1240.05 & 15.43(0.05) &  &  & 1252.59 & 18.49(0.10) &  &  \\ 
 &  & 1294.46 & 19.61(0.20) & 1242.29 & 15.82(0.05) &  &  & 1254.09 & 18.72(0.10) &  &  \\ 
 &  & 1387.30 &  20.32(0.20)& 1244.43 & 16.13(0.06) &  &  & 1256.66 & 18.86(0.14) &  &  \\ 
 &  &  &  & 1246.91 & 16.23(0.06) &  &  & 1258.53 & 19.01(0.12) &  &  \\ 
 &  &  &  & 1248.04 & 16.18(0.06) &  &  & 1260.34 & 19.27(0.13) &  &  \\ 
 &  &  &  & 1250.04 & 16.44(0.07) &  &  & 1262.20 & 19.17(0.12) &  &  \\ 
 & &  &  & 1252.59 & 16.75(0.07) &  &  & 1264.61 & 19.27(0.14) &  &  \\ 
 &  &  &  & 1254.09 & 17.07(0.07) &  &  & 1266.14 & 19.46(0.15) &  &  \\ 
 &  &  &  & 1256.66 & 17.30(0.09) &  &  & 1268.66 & 19.45(0.15) &  &  \\ 
 &  &  &  & 1258.53 & 17.24(0.09) &  &  & 1271.32 & 19.82(0.17) &  &  \\ 
 &  &  &  & 1260.33 & 17.30(0.08) &  &  & 1275.19 & 20.11(0.18) &  &  \\ 
 &  &  &  & 1262.20 & 17.26(0.08) &  &  & 1280.34 & 20.33(0.18) &  &  \\ 
 &  &  &  & 1264.61 & 17.41(0.09) &  &  & 1285.14 & 20.40(0.20) &  &  \\ 
 &  &  &  & 1266.14 & 17.67(0.10) &  &  & 1290.69 & 20.55(0.21) &  &  \\ 
 &  &  &  & 1268.66 & 18.11(0.12) &  &  & 1294.46 & 20.54(0.22) &  &  \\ 
  &  &  &  & 1387.30 & 20.88(0.38) &  &  & 1387.30& 21.56(0.30) &  &  \\ 
\enddata
\label{Tab:UVOTphot}
\end{deluxetable}

\begin{deluxetable}{cccc}
\tabletypesize{\scriptsize}
\tablecolumns{6} 
\tablewidth{0pc}
\tablecaption{R and I band photometry}
\tablehead{ \colhead{Date} & \colhead{R} & \colhead{Date} & \colhead{I} \\ 
		\colhead{(MJD)} 	&  \colhead{(mag)} & \colhead{(MJD)} &  \colhead{(mag)} }
\startdata
56181.25 & 16.54(0.04) & 56181.25 & 16.55(0.07) \\ 
56194.25 & 17.08(0.06) & 56194.25 & 17.82(0.09) \\ 
56199.50 & 14.01(0.01) & 56197.00 & 14.43(0.13) \\ 
56201.50 & 13.86(0.01) & 56198.00 & 14.17(0.07) \\ 
56204.50 & 13.70(0.01) & 56199.25 & 13.94(0.06) \\ 
56205.50 & 13.66(0.01) & 56200.00 & 13.97(0.06) \\ 
56206.50 & 13.65(0.01) & 56201.00 & 13.89(0.04) \\ 
56209.00 & 13.70(0.04) & 56202.00 & 13.79(0.08) \\ 
56209.50 & 13.70(0.01) & 56203.00 & 13.86(0.08) \\ 
56210.50 & 13.71(0.01) & 56204.00 & 13.77(0.07) \\ 
56215.50 & 13.75(0.01) & 56205.00 & 13.69(0.07) \\ 
56218.50 & 13.95(0.01) & 56206.00 & 13.65(0.05) \\ 
56219.50 & 14.01(0.01) & 56208.00 & 13.67(0.07) \\ 
56220.50 & 14.04(0.01) & 56209.00 & 13.66(0.07) \\ 
56221.50 & 14.04(0.03) & 56209.25 & 13.60(0.09) \\ 
56223.50 & 14.17(0.02) & 56210.00 & 13.69(0.07) \\ 
56224.50 & 14.22(0.01) & 56211.00 & 13.59(0.07) \\ 
56226.50 & 14.43(0.01) & 56212.00 & 13.65(0.06) \\ 
56226.50 & 14.40(0.03) & 56213.00 & 13.56(0.10) \\ 
56227.50 & 14.54(0.02) & 56214.00 & 13.56(0.07) \\ 
56228.50 & 14.65(0.02) & 56215.25 & 13.69(0.09) \\ 
56234.50 & 14.59(0.04) & 56216.00 & 13.63(0.08) \\ 
56238.50 & 14.78(0.05) & 56217.00 & 13.74(0.09) \\ 
56238.50 & 14.78(0.01) & 56218.00 & 13.89(0.08) \\ 
56239.50 & 14.87(0.02) & 56219.00 & 13.83(0.06) \\ 
56240.50 & 14.98(0.04) & 56220.00 & 13.83(0.08) \\ 
56240.50 & 14.98(0.02) & 56221.00 & 13.90(0.10) \\ 
56242.50 & 15.15(0.02) & 56222.00 & 14.00(0.07) \\ 
56245.50 & 15.34(0.02) & 56223.00 & 14.04(0.13) \\ 
56246.50 & 15.36(0.03) & 56224.00 & 14.16(0.10) \\ 
56247.50 & 15.36(0.08) & 56225.00 & 14.14(0.06) \\ 
56247.50 & 15.37(0.02) & 56226.00 & 14.19(0.07) \\ 
56248.00 & 15.37(0.04) & 56229.00 & 14.44(0.12) \\ 
56248.50 & 15.38(0.03) & 56230.00 & 14.77(0.09) \\ 
56250.50 & 15.68(0.06) & 56231.00 & 14.55(0.09) \\ 
 &  & 56233.00 & 14.52(0.10) \\ 
 &  & 56234.00 & 14.44(0.08) \\ 
 &  & 56235.00 & 14.53(0.08) \\ 
 &  & 56236.00 & 14.36(0.06) \\ 
 &  & 56237.00 & 14.43(0.07) \\ 
 &  & 56238.00 & 14.60(0.12) \\ 
 &  & 56239.00 & 14.74(0.08) \\ 
 &  & 56240.00 & 14.70(0.08) \\ 
 &  & 56241.00 & 14.90(0.07) \\ 
 &  & 56242.00 & 14.98(0.14) \\ 
 &  & 56243.00 & 14.97(0.19) \\ 
 &  & 56244.00 & 14.99(0.08) \\ 
 &  & 56245.00 & 15.14(0.14) \\ 
 &  & 56246.00 & 15.26(0.08) \\ 
 &  & 56247.00 & 15.19(0.08) \\ 
 &  & 56248.00 & 15.20(0.06) \\ 
 &  & 56248.00 & 15.10(0.10) \\ 
 &  & 56249.00 & 15.00(0.09) \\ 
 &  & 56250.00 & 15.26(0.09) \\ 
 &  & 56251.00 & 15.19(0.09) \\ 
\enddata
\label{Tab:amateurphot}
\end{deluxetable}

\begin{deluxetable}{ccccccc}
\tabletypesize{\scriptsize}
\tablecolumns{6} 
\tablewidth{0pc}
\tablecaption{NIR photometry from PAIRITEL}
\tablehead{ \colhead{Date} & \colhead{J} & \colhead{Date} & \colhead{H} &   \colhead{Date} & \colhead{K} \\ 
		\colhead{(MJD)} 	&  \colhead{(mag)} & \colhead{(MJD)} &  \colhead{(mag)} & \colhead{(MJD)} &  \colhead{(mag)}}
\startdata
56213.19 & 13.53(0.01) & 56213.19 & 13.37(0.02) & 56213.19 & 13.25(0.05) \\ 
56214.19 & 13.60(0.01) & 56214.19 & 13.45(0.03) & 56214.19 & 13.20(0.06) \\ 
56215.20 & 13.56(0.01) & 56215.20 & 13.41(0.02) & 56215.20 & 13.24(0.05) \\ 
56216.21 & 13.61(0.02) & 56216.21 & 13.51(0.04) & 56216.21 & 13.23(0.07) \\ 
56219.18 & 13.75(0.03) & 56219.18 & 13.60(0.06) & 56219.18 & 13.44(0.12) \\ 
56223.21 & 13.88(0.01) & 56223.21 & 13.85(0.03) & 56226.14 & 13.76(0.10) \\ 
56225.14 & 13.99(0.02) & 56225.14 & 13.90(0.04) & 56227.15 & 13.66(0.08) \\ 
56226.14 & 14.08(0.02) & 56226.14 & 13.90(0.03) & 56228.15 & 14.02(0.12) \\ 
56227.15 & 14.18(0.02) & 56227.15 & 14.08(0.04) & 56230.17 & 14.11(0.12) \\ 
56228.15 & 14.33(0.02) & 56228.15 & 14.22(0.05) & 56231.17 & 14.19(0.17) \\ 
56230.17 & 14.43(0.03) & 56230.17 & 14.38(0.05) & 56232.17 & 14.10(0.15) \\ 
56231.17 & 14.41(0.02) & 56231.17 & 14.32(0.05) & 56236.11 & 14.01(0.15) \\ 
56232.17 & 14.43(0.03) & 56232.17 & 14.26(0.07) & 56237.11 & 13.86(0.12) \\ 
56236.11 & 14.26(0.03) & 56236.11 & 14.11(0.05) & 56238.12 & 13.71(0.16) \\ 
56237.11 & 14.35(0.02) & 56237.11 & 14.12(0.04) & 56243.10 & 14.35(0.14) \\ 
56238.12 & 14.34(0.02) & 56238.12 & 14.22(0.05) & 56255.10 & 14.93(0.29) \\ 
56243.10 & 14.74(0.04) & 56245.13 & 14.65(0.06) & 56266.07 & 14.82(0.25) \\ 
56245.13 & 14.80(0.03) & 56248.09 & 14.66(0.11) & 56267.08 & 14.53(0.36) \\ 
56246.14 & 14.88(0.10) & 56253.11 & 14.88(0.10) &  &  \\ 
56248.09 & 14.88(0.04) & 56255.10 & 14.82(0.14) &  &  \\ 
56253.11 & 15.07(0.05) & 56256.07 & 15.09(0.19) &  &  \\ 
56255.10 & 15.03(0.05) & 56257.08 & 14.96(0.12) &  &  \\ 
56256.07 & 15.15(0.09) & 56266.07 & 14.94(0.13) &  &  \\ 
56257.08 & 15.15(0.05) & 56267.08 & 15.27(0.15) &  &  \\ 
56266.07 & 15.21(0.08) &  &  &  &  \\ 
56267.08 & 15.22(0.06) &  &  &  &  \\ 
 \enddata
\label{Tab:nirphot}
\end{deluxetable}

\end{document}